\documentclass[graybox]{svmult}

\usepackage{type1cm}        
\usepackage{makeidx}         
\usepackage{graphicx}        
\graphicspath{{./figures/}}

\usepackage{multicol}        
\usepackage[bottom]{footmisc}

\usepackage{newtxtext}       %
\usepackage[varvw]{newtxmath}       

\usepackage[utf8]{inputenc}
\usepackage{xcolor}
\usepackage{bm,bbm}
\usepackage{physics}

\usepackage[sort&compress,square,comma,numbers]{natbib}

\usepackage[colorlinks,bookmarks=true,linktocpage=true]{hyperref}
\usepackage[nameinlink]{cleveref}

\usepackage{esvect}

\usepackage[a4paper, margin=2.5cm]{geometry}

\hypersetup{
    colorlinks,
    citecolor=blue,
    filecolor=blue,
    linkcolor=blue,
   urlcolor=blue,
   linktoc=page
}

\def\llangle{\left\langle}
\def\rrangle{\right\rangle}

\begin{document}

\title*{Quantum Gravity from dynamical metric fluctuations}

\author{Jan M.~Pawlowski and Manuel Reichert}

\institute{Jan M.~Pawlowski  \at Institut f\"ur Theoretische Physik, Universit\"at Heidelberg, Philosophenweg 16, 69120 Heidelberg, Germany, \email{j.pawlowski@thphys.uni-heidelberg.de}
	\and Manuel Reichert \at Department of Physics and Astronomy, University of Sussex, Brighton, BN1 9QH, U.K. \email{m.reichert@sussex.ac.uk}}
%
%
\maketitle

\abstract{
In this contribution, we discuss the asymptotic safety scenario for quantum gravity by evaluating the correlation functions of dynamical metric fluctuations. This is done with a functional renormalisation group approach that disentangles dynamical metric fluctuations from the background metric. We detail the derivation of the respective flow equations on space-time manifolds with Euclidean and Lorentzian signatures and discuss the diffeomorphism symmetry constraints on the flow as well as the convergence of systematic vertex expansion schemes. We then proceed with a comprehensive review of results of momentum-dependent correlation functions at vanishing cutoff scale, the phase structure of the asymptotically safe Standard Model, and spectral properties of asymptotically safe gravity from direct computations in space-times with Lorentzian signatures such as the graviton spectral function. 
}



\section{Introduction}
The unification of general relativity with quantum field theory remains one of the key fundamental problems of modern physics. Despite decades of research the question of whether the metric field can be used as a fundamental degree of freedom in a quantum field theory has not been conclusively answered yet. While perturbative quantisations of metric quantum gravity run into problems of predictivity or unitarity, it has been pointed out by Weinberg \cite{Weinberg:1980gg} that gravity might possess a non-trivial ultraviolet (UV) fixed point of the renormalisation group (RG) flow. This UV fixed point would make the theory non-perturbatively renormalisable. 

The method of choice for respective investigations is the functional renormalisation group (fRG) in its form for the effective action \cite{Wetterich:1992yh}. The fRG approach to quantum gravity has been initiated by the seminal paper \cite{Reuter:1996cp}, where the UV fixed point has been studied in the Einstein-Hilbert truncation. In this approximation, one retains only two couplings, the Newton coupling $G_\text{N}$ and the cosmological constant $\Lambda$. Already this basic truncation exhibits a UV fixed point in four dimensions, see \cite{Reuter:1996cp, Souma:1999at}. This exciting finding has triggered a plethora of works for asymptotically safe gravity with and without matter, and we refer the reader to textbooks \cite{Percacci:2017fkn, Reuter:2019byg} and reviews \cite{Niedermaier:2006wt, Litim:2011cp, Reuter:2012id, Ashtekar:2014kba, Eichhorn:2017egq, Bonanno:2017pkg, Eichhorn:2018yfc, Pereira:2019dbn, Reichert:2020mja, Knorr:2022dsx, Eichhorn:2022gku, Platania:2023srt}. For very recent accounts of the challenges for asymptotically safe gravity see \cite{Bonanno:2020bil, Donoghue:2019clr}. For generic reviews on the fRG we refer to~\cite{Berges:2000ew, Aoki:2000wm, Polonyi:2001se, Pawlowski:2005xe, Gies:2006wv, Delamotte:2007pf, Kopietz:2010zz, Rosten:2010vm, Braun:2011pp, Dupuis:2020fhh}.

In this contribution, we discuss the fluctuation approach to metric quantum gravity. The key property of this approach is that the metric background and the dynamical fluctuations about it are strictly disentangled. The RG running is then set up in terms of correlation functions of the fluctuation and background field. The approach has been reviewed in \cite{Pawlowski:2020qer} to which  we refer for extended details and more background information. Here we concentrate on new developments in the approach, which includes the first results on Lorentzian backgrounds within the spectral fRG, full 1PI quantum correlation functions, and the UV and infrared (IR) physics of asymptotically safe matter-gravity systems including the full Standard Model (SM).

\section{Quantum field theory approach to quantum gravity}
\label{sec:qft-approach}
The quantum field theory approach to metric quantum gravity is based on the path integral over the metric fluctuations for a given classical gravity action. This reads schematically
\begin{align}
	\int\!\! \mathcal{D}  \hat g_{\mu\nu}\, e^{\imag \,S_\text{grav}[\hat g_{\mu\nu}]}\,, 
	\label{eq:path-integral}
\end{align}
where the hat indicates fields that are integrated over, which is here only the metric field $\hat g_{\mu\nu}$. In general, the gravitational action in \labelcref{eq:path-integral} also includes matter fields and an integration over these fields. A typical choice for the classical gravitational action is the Einstein-Hilbert action  
\begin{align} 	\label{eq:EH-Action} 
	S_\text{EH}[g_{\mu\nu}] &= \frac{1}{16 \pi G_\text{N}} \int\! \mathrm d^4 x \sqrt{-g} \left(R - 2 \Lambda\right)\,,
	&
	g&=\det g_{\mu\nu}\,,
\end{align}
where $G_\text{N}$ is the classical Newton coupling, $\Lambda$ is the cosmological constant, and $R$ is the Ricci scalar with
\begin{align}
	R &= R^\mu{}_{\mu} \, , 
	&
	R_{\mu\nu} &= R^\lambda{}_{\mu\lambda\nu} \, ,
	\label{eq:Curvatures}
\end{align}
computed from the Riemann tensor 
\begin{align}
	R^\rho{}_{\sigma\mu\nu} = \partial_\mu \Gamma^\rho{}_{\nu\sigma} -
	\partial_\nu \Gamma^\rho{}_{\mu\sigma} +
	\Gamma^\rho{}_{\mu\lambda}\Gamma^\lambda{}_{\nu\sigma} -
	\Gamma^\rho{}_{\nu\lambda}\Gamma^\lambda{}_{\mu\sigma} \,,
\end{align}
with the Levi-Civita connection, 
\begin{align}
	\Gamma^\sigma{}_{\mu\nu}= \tfrac{1}{2}g^{\sigma\rho}\left( \partial_\mu g_{\nu\rho} +
	\partial_\nu g_{\rho\mu} - \partial_\rho g_{\mu\nu} \right) \,.  
	\label{eq:LeviCivita}
\end{align}
Quantum gravity based on the Einstein-Hilbert action \labelcref{eq:EH-Action} is perturbatively non-renormalisable, see  \cite{tHooft:1974toh, Goroff:1985sz, Goroff:1985th, vandeVen:1991gw}, which is related to the negative mass dimension of the Newton constant in four spacetime dimensions, $[G_\text{N}] = -2$. 

One can construct theories of quantum gravity that are perturbatively renormalisable when higher-order curvature invariants are included in the classical action. A prominent example is provided by the Stelle action \cite{Stelle:1976gc, Stelle:1977ry}
\begin{align} 
	\label{eq:Stelle-Action} 
	S_\text{Stelle}[g_{\mu\nu}]=  S_\text{EH} +\int\! \mathrm d^4 x \sqrt{-g} \, \left( a \,R^2 + b \, C_{\mu\nu\rho\sigma}^2 + c \, E\right),
\end{align}
where $C_{\mu\nu\rho\sigma}$ is the Weyl tensor and $E$ is the Gau\ss-Bonnet density, given by $E= \frac{1}{32\pi^2}( R^2-4 R^{\mu\nu}R_{\mu\nu}+R^{\mu\nu\rho\sigma}R_{\mu\nu\rho\sigma})$, which is a topological term in four dimensions. This classical action is perturbatively renormalisable due to the dimensionless couplings $a$, $b$, and $c$. However, it contains a massive ghost state that might spoil unitarity \cite{Stelle:1976gc, Stelle:1977ry, Antoniadis:1986tu}.

The perturbative failure of making sense out of the gravitational path integral \labelcref{eq:path-integral} does not imply that quantum gravity cannot be described by a fundamental metric field. Instead, it emphasises the need for non-perturbative methods such as lattice and functional approaches. The key idea is that the existence of an interacting UV fixed point makes the theory non-perturbatively renormalisable \cite{Weinberg:1980gg}. This UV fixed point exists in $2+\varepsilon$ dimensions and might extend up to $d=4$ dimensions \cite{Kawai:1989yh, Jack:1990ey, Falls:2015qga, Falls:2017cze, Martini:2021slj, Martini:2021lcx, Martini:2022sll}. The four-dimensional case was tested with lattice computations \cite{Ambjorn:1991pq, Bilke:1996zd, Ambjorn:1998xu, Ambjorn:2004qm, Ambjorn:2005qt, Laiho:2011ya, Loll:2019rdj, Loll:2023hen}, and fRG \cite{Wetterich:1992yh, Ellwanger:1993mw, Morris:1993qb}, see \cite{Dupuis:2020fhh} for a review on the fRG. The first indications for the UV fixed point in four dimensions were found with the fRG in \cite{Reuter:1996cp, Souma:1999at}.

In the fRG approach pursued in the present work, the key object is the full quantum effective action $\Gamma[g_{\mu\nu}]$ with $g_{\mu\nu}=\langle \hat g_{\mu\nu}\rangle$, and the UV closure of the theory is obtained through a UV fixed point with finitely many relevant directions. The respective fixed-point effective action is a highly non-trivial functional of the (mean) metric and the fixed-point analysis in such a framework only depends on the field content: the underlying theory is fixed by the chosen UV fixed point or rather its respective UV-relevant directions. It is this choice and the selected trajectory emanating from the fixed point that determines the theory and hence the underlying classical or rather IR-effective action. For example, one can find fixed points that relate to an IR action that only features the cosmological-constant term, which is a non-dynamical IR theory. Furthermore one can find fixed points that contain the spin-two ghost of Stelle gravity \labelcref{eq:Stelle-Action}, indicating the absence of unitarity, and others that do not. We are interested in finding a dynamical IR effective theory without apparent loss of unitarity.

The fRG approach as well as the metric path integral in \labelcref{eq:path-integral} is based on the existence of the metric propagator. For the latter to exist, the diffeomorphism invariance of gravity has to be factored out, which is done within the Faddeev-Popov procedure. Then, the classical action in  \labelcref{eq:path-integral} is supplemented with a gauge-fixing and ghost term. The latter terms (as also do all source terms) require the introduction of a background metric $\bar g_{\mu\nu}$. The expansion of the full metric $g_{\mu\nu}$ about this background introduces the fluctuation field $h_{\mu\nu}$ as the dynamical field in this expansion. Accordingly, it is the fluctuation field that is integrated over in the path integral. Throughout this contribution, we only consider the linear metric split
\begin{align}
	\label{eq:LinSplit} 
	g_{\mu\nu} =\bar g_{\mu\nu} + h_{\mu\nu}\,, 
\end{align} 
or variations thereof with a scale-dependent prefactor multiplying the fluctuation field.  
This split is by no means unique, and various metric splits have been considered in the literature, for example the exponential split \cite{Kawai:1992np, Nink:2014yya, Falls:2015qga, Demmel:2015zfa, Percacci:2015wwa, Gies:2015tca, Ohta:2015efa, Labus:2015ska, Ohta:2015fcu, Ohta:2015zwa, Dona:2015tnf, Falls:2016msz, Ohta:2016npm, Ohta:2016jvw, Ohta:2017dsq, Alkofer:2018fxj, deBrito:2018jxt}. More elaborate choices are typically based on the geometrical properties of gravity as pioneered by Vilkovisky and DeWitt \cite{DeWitt:1980jv, Fradkin:1983nw, Vilkovisky:1984st, DeWitt:2003pm}, which was also used in asymptotic safety with the fRG, see \cite{Branchina:2003ek, Pawlowski:2003sk, Pawlowski:2005xe, Donkin:2012ud, Demmel:2014hla, Falls:2020tmj}. For a more detailed discussion on the splits, we refer to \cite{Pawlowski:2020qer}. 

A typical choice for the gauge-fixing action is a linear gauge-fixing condition for the fluctuation field $h_{\mu\nu}$,
\begin{align}
	S_{\text{gf}}[\bar g, h]=\frac{1}{2 \alpha} \int \!\mathrm{d}^4x
	\sqrt{-\bar{g}}\; \bar{g}^{\mu \nu} F_\mu F_\nu \,.
	\label{eq:gf} 
\end{align}
A common gauge-fixing condition $F_\mu$ is given by
\begin{align}
	F_\mu[\bar g, h] =
	\bar{\nabla}^\nu h_{\mu \nu} -\frac{1+ \beta}{4} \bar{\nabla}_\mu h^{\nu}_{~\nu} \,, 
	\label{eq:gf-condition}
\end{align}
where $\bar\nabla$ is the covariant derivative with the connection \labelcref{eq:LeviCivita} of the background metric $\bar g_{\mu\nu}$. The gauge fixing \labelcref{eq:gf-condition} is introduced in the path integral with the Faddeev-Popov trick and the Jacobi determinant of the respective reparameterisation. The Faddeev-Popov determinant can be rewritten in terms of a fermionic path integral with the ghost fields $c_\mu$ and $\bar c_\mu$. The ghost action related to \labelcref{eq:gf-condition} reads 
\begin{align}
	\label{eq:Sghost}
	S_{\text{gh}}[\bar g, h,c,\bar c]=\int \!\mathrm{d}^4x
	\sqrt{-\bar{g}}\; \bar c^\mu M_{\mu\nu} c^\nu\,, 
\end{align}
with the Faddeev-Popov operator 
\begin{align}
	\label{eq:OpFP}
	M_{\mu\nu}= \bar\nabla^\rho\! \left(g_{\mu\nu} \nabla_\rho +g_{\rho\nu} \nabla_\mu\right) -\frac{1+\beta}{2} \bar g^{\sigma\rho} \bar\nabla_\mu g_{\nu\sigma} \nabla_\rho\,,
\end{align}
which contains both, the background covariant derivative $\bar\nabla$ and the covariant derivative $\nabla$ with the connection \labelcref{eq:LeviCivita} of the full metric $g_{\mu\nu}$. Note that $M_{\mu\nu}$ is linear in the fluctuation field $h_{\mu\nu}$. In order to achieve this property, it was necessary to introduce the background metric $\bar g_{\mu\nu}$. Both, gauge-fixing and ghost action, depend explicitly on the background metric, as do the source terms, if they are kept linear in the fluctuation field. This implies that the quantum effective action also depends on both metrics, the background metric $\bar g_{\mu\nu}$ and the full metric $g_{\mu\nu}(\bar g,h)$. Note, however, that the correlation functions of diffeomorphism-invariant operators as well as the solutions to the quantum equations of motion do not depend on the gauge fixing. Hence, they are background-independent as explained below.

In summary, we arrive at the generating functional of metric quantum gravity,  
\begin{align}
	Z[\bar g, J] \simeq \int\! \mathcal{D} \hat \phi\,
	e^{\imag (S+ S_\text{gf}+S_\text{gh}+\int\!\mathrm{d}^4x \sqrt{-\bar g} \, J^a \hat\phi_a)}\,.
	\label{eq:ZpathI}
\end{align}
Here, $\hat \phi$ includes all fluctuation fields, the graviton fluctuation field, the ghost fields and potential matter fields, and $J$ is composed of the respective currents, 
\begin{align}
	\phi &=(h_{\mu\nu},c_\mu, \bar c_\mu,\phi_\text{mat})\,,
	&
	& \textrm{and}
	&
	J&=(J_{h_{\mu\nu}},J_{c_\mu},J_{\bar c_\mu},J_\text{mat})\,, 
	\label{eq:superfield}
\end{align}
where we have dropped the hat. We have not introduced a normalisation of the path integral in \labelcref{eq:ZpathI}, as it drops out anyway if performing the Legendre transformation to the effective action $\Gamma$, the quantum analogue of the classical action. We get 
\begin{align}
	\Gamma[\bar g, \phi] = - \imag  \log Z[\bar g, J]- \int\!\mathrm{d}^4 x \sqrt{-\bar g} \, J^a \phi_a \,, 
	\label{eq:Gamma}
\end{align}
where the extremisation of the currents is implied. The index $a$ in $J^a$ and $\phi_a$ comprises all Lorentz, Dirac, gauge group and further internal indices as well as the species of fields, and is summed over in \labelcref{eq:Gamma}. In terms of the superfield and super current in \labelcref{eq:superfield}, this yields 
\begin{align}
	J^a \phi_a = J_{h_{\mu\nu}} h_{\mu\nu} +J_{c_\mu} c_\mu - \bar c_\mu J_{\bar c_\mu} +J_\text{mat} \phi_\text{mat}\,. 
	\label{eq:ExplicitContraction}
\end{align}
So far, we formulated all equations in Lorentzian signature, which is apparent from the factors of $\imag$ in front of the classical action and the minus sign in $\sqrt{-g}$. Within non-perturbative approaches, such as the fRG or lattice, it is almost always necessary to start from an Euclidean formulation of the theory. This is achieved by a Wick rotation where the time component and the four-vector product transform as
\begin{align}
	\label{eq:Wick-rotation}
	t  &\to -\imag t_E\,,
	&
	x^2 &\to -x_E^2 \,.
\end{align}	
Most notably, this turns the $\imag$ in the exponentials in \labelcref{eq:path-integral,eq:ZpathI} into minus signs, which makes them applicable as probability density for Monte-Carlo lattice simulations. In the fRG, this allows for an ordering principle in the momentum-squared for the coarse graining. However, there has been significant recent progress in applying the fRG to Lorentzian settings \cite{Horak:2020eng, Fehre:2021eob, Braun:2022mgx, Banerjee:2022xvi, DAngelo:2022vsh, DAngelo:2023tis}. In the next section, we will set up the Euclidean and Lorentzian fRG equations. The majority of the results reviewed here are obtained within Euclidean signature, and only \Cref{sec:LorentzianGravity} focuses on Lorentzian results. This is owed to the fact that the latter approach has only been developed very recently in \cite{Fehre:2021eob, Braun:2022mgx}. 

We close this section with a brief discussion of the properties and physics content of the effective action. To begin with, the physics of a given system is very conveniently expressed in the effective action $\Gamma$, see \labelcref{eq:Gamma}, that by its definition generates one-particle-irreducible correlation functions. Hence it carries less redundancies as the generating functional $Z$ that generates all correlation functions including disconnected ones, or the Schwinger functional $\log Z$ that generated connected correlation functions. $\Gamma$ has a rather direct connection to physical processes as $S$-matrix elements can be constructed from $\Gamma$ in terms of tree-level diagrams with full 1PI vertices and full propagators. Moreover, the gauge-fixed effective action \labelcref{eq:Gamma} also encodes a diffeomorphism-invariant effective action of one field, the background effective action 
\begin{align}
	\bar \Gamma[g,\phi_\textrm{mat}]= \Gamma[g,\phi]\Big|_{h,c,\bar c=0}\,. 
	\label{eq:BackGamma}
\end{align}
\Cref{eq:BackGamma} only depends on the full metric as well as the matter fields. It is obtained from the full effective action \labelcref{eq:Gamma} with $h=0$. Moreover, also \labelcref{eq:BackGamma} can be linked to the $S$-matrix, even though this relation is less direct. Still, its diffeomorphism invariance allows us to extract many observables directly from the effective action $\bar \Gamma$, and it would be most convenient to only work with \labelcref{eq:BackGamma} from the onset. As we shall see, this is only possible within approximation whose applicability has to be assessed very carefully, and such an analysis is rather difficult and has not been done yet. It is the latter intricacies which led to the more direct fluctuation approach discussed in the following three sections: There we first set up the general flow equation approach to metric quantum gravity, see \Cref{sec:fRG-QG}, discuss the ensuing diffeomorphism symmetry constraints, see \Cref{sec:symmetry-identities}, and finally the fluctuation approach, see \Cref{sec:fluctuation-approach}.

\section{Functional renormalisation group approach to metric quantum gravity}
\label{sec:fRG-QG}
Based on the quantum field theory approach to metric quantum gravity outlined in the last section, we will now set up fRG equations for gravity, for reviews see \cite{Niedermaier:2006wt, Litim:2011cp, Reuter:2012id, Ashtekar:2014kba, Eichhorn:2017egq, Bonanno:2017pkg, Eichhorn:2018yfc, Pereira:2019dbn, Reichert:2020mja, Bonanno:2020bil, Donoghue:2019clr} and for generic fRG reviews see \cite{Berges:2000ew, Aoki:2000wm, Polonyi:2001se, Pawlowski:2005xe, Gies:2006wv, Delamotte:2007pf, Kopietz:2010zz, Rosten:2010vm, Braun:2011pp, Dupuis:2020fhh}.

The fRG approach to gravity is based on a successive integrating-out of quantum fluctuations. A general regularisation of metric quantum gravity is implemented by augmenting the kinetic term of the action $S$ with an IR cutoff term, $S\to S+\Delta S_k$, with 
\begin{align}
	\Delta S_k[\bar g, \phi] = \frac12 \int_{x,y} \phi_a(x) \,R^{ab}[\bar g](x,y)\, \phi_a(y)\,, 
	\label{eq:Cutofffunction}
\end{align}
where we have used the abbreviation 
\begin{align}
	\int_x = \int \!\mathrm d^4 x \sqrt{\pm \bar g(x)}\,, 
\end{align}
for the integrals with the volume factor $\sqrt{-\bar g}$ in a Lorentzian background and $\sqrt{\bar g}$ in a Euclidean one. Common Euclidean regulator classes implement a regularisation of covariant momentum modes, more details are provided in \Cref{sec:flowEffActEuclidean}. In turn, Minkowski regulators either break Lorentz symmetry and only suppress low spatial (covariant) momentum modes, or they are formulated as spectral cutoffs, more details are provided in \Cref{sec:flowEffActLorentzian}. We also note that while these are the common choices, the set-up is far more general and interesting other choices are cutoffs for the field amplitude \cite{Polonyi:2001se} or temporal cutoffs for the time evolution \cite{Gasenzer:2008zz}. 

Irrespective of its nature of being a momentum, spectral, amplitude, temporal or other  regularisation, the cutoff term \labelcref{eq:Cutofffunction} is conveniently introduced into the generating functional of metric quantum gravity \labelcref{eq:ZpathI}, which leads to  
\begin{align}
	\label{eq:Zhk} 
	Z_k[\bar g, J]= \exp \left(-c_g\int_{x,y}\frac{\delta}{\delta J^a(x)} R^{ab}_k(x,y) \frac{\delta}{\delta J^b(y)} \right) Z[\bar g, J]\,. 
\end{align}
The prefactor $c_g$ depends on the signature of the background metric: we have $c_g =1$ for a Euclidean background and  $c_g =\imag$ for a Lorentzian background. Furthermore, we have defined the functional derivative to include a factor of $1/\sqrt{\pm \bar g(x)}$, which makes them background-covariant. For example, for a tensor $J$ of rank $n$, the derivative reads
\begin{align}
	\label{eq:funDerGen} 
	\frac{\delta J_{\mu_1\cdots \mu_n}(x)}{\delta J_{\nu_1\cdots \nu_n}(y)} = \frac{1}{\sqrt{\pm \bar g}} \delta(x-y)\, \delta^{(\nu_1}_{\mu_1} \cdots \delta^{\nu_n)}_{\mu_n} \,.
\end{align}
This notation was also used in \cite{Pawlowski:2020qer}.

\subsection{Euclidean flow equation for the effective action} 
\label{sec:flowEffActEuclidean}

Here we review the standard fRG approach to metric quantum gravity: as mentioned above, the commonly used regulator function is a function of background covariant derivatives and hence ultra-local in position space, $R^{ab}(x,y) \propto \delta(x-y)$. Here we consider such cutoffs and elucidate their use at the example of Euclidean quantum gravity. The implementation of such a covariant momentum cutoff necessitates the choice of a background metric $\bar g$, and the covariant momenta are those related to the covariant Laplacian in the background metric, $\Delta_{\bar g}$, with the spectral values $p^2_{\bar g}$. 

Most of the computations in the fluctuation approach discussed here are done in a flat background with the notable exception of the computations in \cite{Christiansen:2017bsy, Burger:2019upn}. In these works the theory was expanded about a curved background $\bar g_{\mu\nu}\neq \delta_{\mu\nu}$, which allows us to compute background-metric dependent couplings and also provides direct access to the diffeomorphism invariant background effective action. In a curved background, the regulator depends on the Laplace-Beltrami operator in this background as well as background curvature terms. The background can be chosen such that it is the dynamical solution of the full quantum equation of motion in which case the flow equation takes into account fluctuations about this background. While this may facilitate the convergence of expansion schemes it is by no means a necessary choice.  

In the following, we concentrate on a flat Euclidean background 
\begin{align}
	\bar g_{\mu\nu} = \delta_{\mu\nu}\,.
	\label{eq:FlatEuclideanMetric}
\end{align}
With \labelcref{eq:FlatEuclideanMetric}, the eigenvalues of the Laplacian are just momentum squared, $p^2$, and the regulator suppresses the IR-momentum modes with $p^2 \lesssim k^2$. In turn, UV-momentum modes with $p^2 \gtrsim k^2$ propagate freely and the generating functional includes all quantum contributions generated by these modes. Then, one RG step with $k\to k-\Delta k$ constitutes the integration of momentum modes $p^2 \approx k^2$. 

It is convenient to write the regulator $R_k$ in terms of the classical or quantum dispersion of the field at hand,
\begin{align}
	\label{eq:RegSplits} 
	R_k^{ab}(p) &= T^{ab}_k(p)\, r_k(x)\,,
	&
	\text{with} \qquad x&=\frac{p^2}{k^2}\,, 
\end{align}
where momentum-squared is counted in cutoff units. In these units, the IR regime is given by $x\lesssim 1$ and the UV regime by $x\gtrsim 1$. The tensor part $T^{ab}_k$ of the regulator is proportional to the classical or quantum dispersion of the field. In the former case, it is the second derivative of the classical action with respect to the fields $\phi^a$ and $\phi^b$, i.e., $T^{ab}_k(p) = (S^{(2)})^{ab}(p)$. In the latter case, the regulator is proportional to the quantum dispersion of the fluctuation fields or parts of it. In both cases, this carries the kinetic information about the field whose propagation is regularised. In turn, the dimensionless shape function $r_k$ specifies how the propagation is regularised. The shape function $r_k$ has to be chosen such that the IR suppression of momentum modes as well as the UV decay of the regularisation is guaranteed. In the flow equation approach to Lorentzian quantum gravity set up in \cite{Fehre:2021eob}, we will employ a slightly different strategy in regularising the spectral UV modes, see \Cref{sec:flowEffActLorentzian}. For Euclidean quantum gravity, the regularisation of UV and IR modes leads to the following asymptotics of the regulator shape function, 
\begin{align}
	\lim_{x\to 0} r_k(x) &\to \infty \,,
	&
	\lim_{x\to \infty} x^{d/2} r_k(x) &= 0 \,, 
	\label{eq:rkLimits}
\end{align} 
where $d$ is the dimension of space-time. The first limit in \labelcref{eq:rkLimits} guarantees the IR suppression of momentum modes and also entails that for $k\to\infty$ all momentum modes are suppressed and the theory approaches the UV-scaling regime. For asymptotically free theories this is the classical theory, for asymptotically safe theories this is the non-trivial UV theory. The second limit in \labelcref{eq:rkLimits}, guarantees that the UV behaviour of the theory is unchanged by the IR regularisation. Moreover, the decay with more than $x^{d/2}$ guarantees that a change of the IR RG scale does not require an adaptation of the UV-renormalisation. This limit also has another implication: for $k\to 0$ the limit holds for all momenta and the cutoff is removed from the theory. 

The flow equation for the scale dependent effective action $\Gamma_k[\bar g,\phi]$ is then given by \cite{Wetterich:1992yh, Ellwanger:1993mw, Morris:1993qb}
\begin{align}
	\label{eq:flow} 
	\partial_t \Gamma_k[\bar g,\phi]= \frac12 \text{Tr}\, G_k[\bar g,\phi]\,\partial_t R_k\,,
\end{align}
where $t=\log(k/k_0)$ with some reference scale $k_0$, and $G_k[\bar g, \phi]$ is the full field-dependent propagator
\begin{align}
	\label{eq:Gk}
	G_k[\bar g,\phi] = \frac{1}{\Gamma_k^{(0,2)}[\bar g,\phi]+ R_k} \,.
\end{align}
The trace in \labelcref{eq:flow} sums over space-time, internal and gauge group indices as well as species of fields, including the symplectic metric for fermions. In \labelcref{eq:Gk}, we have used the notation $\Gamma_k^{(0,2)}$ for the second derivative of the effective action with respect to the dynamical fluctuation fields $\phi$. In general, we use the following notation for derivatives, 
\begin{align}
	\Gamma_{k,\bar g^n \phi_{a_1} \cdots\phi_{a_m}}^{(n,m)}[\bar g,\phi](\boldsymbol x_{n+m}) &=\frac{\delta}{\delta \bar g(x_1)}\cdots \frac{\delta}{\delta \bar g(x_n)} \frac{\delta}{\delta \phi_{a_1}(x_{n+1})} \cdots \frac{\delta}{\delta \phi_{a_m}(x_{n+m})} \Gamma_k[\bar g,\phi]\,, 
	\label{eq:def-derivates}
\end{align}
for general functionals of $\bar g$ and $\phi$. On the left-hand side, we used the abbreviation 
\begin{align}
	\boldsymbol{x}_{m}=(x_1,...,x_m)\,,
	\label{eq:xnpn}
\end{align}
which we will use in straight analogy for sets for four-momenta, $\boldsymbol{p}_{m}=(p_1,...,p_m)$.

In \labelcref{eq:def-derivates}, the subscript indicates the $k$-dependence and the fields indicate which field-derivative is taken and in which order. The latter is important as the fluctuation field includes also fermionic Grassmannian fields such as the gravity ghosts. Moreover, we shall also use the abbreviations 
\begin{align}
	\Gamma_k^{(n,m)}\,,\,	\Gamma_k^{(n,\phi_{a_1} \cdots\phi_{a_m})}=\Gamma_{k,\bar g^n \phi_{a_1} \cdots\phi_{a_m}}^{(n,m)} \,, \qquad\qquad \Gamma_k^{\phi_{a_1} \cdots\phi_{a_m}}=\Gamma_k^{(0,\phi_{a_1} \cdots\phi_{a_m})}\,,
	\label{eq:GnmShort}
\end{align} 
for \labelcref{eq:def-derivates} with or without the subscript ${}_k$ for the sake of brevity. This concludes our derivation of the functional flow equation in metric quantum gravity.

\subsection{Lorentzian flow equation for the effective action} 
\label{sec:flowEffActLorentzian}

In this section, we review the \textit{spectral} fRG approach to Lorentzian quantum gravity that has been put forward in \cite{Fehre:2021eob}. It constitutes the first \textit{direct} Lorentzian fRG approach and was used for the computation of the spectral function of the graviton propagator. 

The standard Euclidean setting allows for an ordering of fluctuations according to the size of the Euclidean momentum squared, $p^2$, or more generally in terms of some covariant spectral value as the spectral values of the Laplace-Beltrami operator. The respective numerical results have to be Wick-rotated in order to use them for Lorentzian quantum gravity. This is already a challenging task in a standard quantum field theory, where the Wick rotation is mathematically well-defined. There, the intricacy is solely given by the task of reconstructing real-time data from imaginary data, which is an ill-conditioned problem. Moreover, in gauge theories it is further complicated due to the fact, that at least the gauge field is not a physical field and its spectral representation is at stake, for a respective discussion see \cite{Cyrol:2018xeq}. In asymptotically safe quantum gravity the reconstruction problem is softened by the \textit{close to perturbativeness} of asymptotically safe gravity, see \cite{Eichhorn:2018ydy} for a brief description and respective references. This has been used for the reconstruction of the graviton propagator from its Euclidean data in \cite{Bonanno:2021squ}. 

Evidently, a direct Lorentzian renormalisation group approach is much wanted, and has been set up in quantum gravity in \cite{Fehre:2021eob} within the \textit{spectral} fRG approach \cite{Braun:2022mgx}, based on the general spectral functional approach put forward in \cite{Horak:2020eng}. The spectral fRG utilises the spectral representation of correlation functions, and most prominently the propagators $G_{\phi_a \phi_b}(p)$. Here we briefly present the key idea, for more details, we defer the reader to \cite{Fehre:2021eob, Braun:2022mgx}. For the discussion of spectral representations we use the Källén-Lehmann spectral representation of a scalar Euclidean propagator with
\begin{align} 
	G(p_0,|\vv{p}|) &= \int_{0_-}^\infty \frac{\mathrm d\lambda}{2 \pi}\frac{\lambda\,\rho(\lambda,|\vv{p}|)}{\lambda^2+p_0^2} \,, 
	\label{eq:KL-Rep}
\end{align}
where $p_0\in\mathbbm{R}$ singles out the Euclidean frequency axis. However, the frequency can take any complex value, $p_0\in\mathbbm{C}$, including Lorentzian frequencies $\omega$ with $p_0=\imag \omega$. The lower bound of the spectral integral, $0_-$, takes into account that for massless particles the spectral function contains a delta function at vanishing spectral values, $\delta(\lambda)$. \Cref{eq:KL-Rep} can be inserted into loop diagrams and, upon integration of the analytic frequency dependence, one is led to numerically accessible spectral and spatial momentum integrals without poles. This is the key idea in a nutshell. 

It is left to define the class of cutoffs that preserve the spectral representation  \labelcref{eq:KL-Rep}. Moreover, we would like to maintain causality and unitarity, Lorentz invariance as well as finiteness of the flow. The causality and unitarity constraint is tightly related to the existence of the spectral representation and constrains the position of poles in the complex frequency plane. For this discussion, we restrict ourselves to the propagator of a scalar field. This argument is readily generalised if writing a given propagator as the dimensionless tensor structure times the scalar part of the propagator. In contradistinction to \labelcref{eq:RegSplits} we use the parametrisation 
\begin{align}
	R_k(p) = k^2 r(x)\,,
	\label{eq:ParaRegSpectral}
\end{align}
also used in \cite{Braun:2022mgx}, where the spectral RG is introduced. \Cref{eq:ParaRegSpectral} entails, that the shape function $r(x)$ carries the whole momentum dependence of the regulator and the prefactor $k^2$ carries the dimension. Then, the causality and unitarity requirements only constraint the shape function and this constraint is satisfied for regulators with shape functions $r_k$ defined in \labelcref{eq:ParaRegSpectral} with 
\begin{align}
	r_k=r_k (x_{\vv{p}} )\,,\qquad \textrm{with}\qquad x_{\vv{p}} = \frac{\vv{p}^{\,2}}{k^2}\,. 
	\label{eq:rkCausal} 
\end{align}
However, \labelcref{eq:rkCausal} breaks Lorentz invariance. The Lorentz invariance constraint is only satisfied for regulators with shape functions $r_k$ depending on the Lorentz invariant four-momentum squared,
\begin{align}
	r_k=r_k(x_p )\,,\qquad \textrm{with}\qquad 	x_p = \frac{p^2}{k^2}\,,
	\label{eq:rkLorentz} 
\end{align}
which in turn do not satisfy the causality constraint. Finally, both shape functions can guarantee the  finiteness of the flow if they satisfy 
\begin{align}
	\lim_{x\to 0} r_k(x) &> 0 \,,
	&
	\lim_{x\to \infty} x^{d/2-1} r_k(x) &= 0 \,, 
	\label{eq:rkLimitsSpec}
\end{align} 
the analogue of the limits \labelcref{eq:rkLimits} for the parametrisation \labelcref{eq:RegSplits}.  

\begin{figure}[t]
	\centering
	\includegraphics[width=0.4\textwidth]{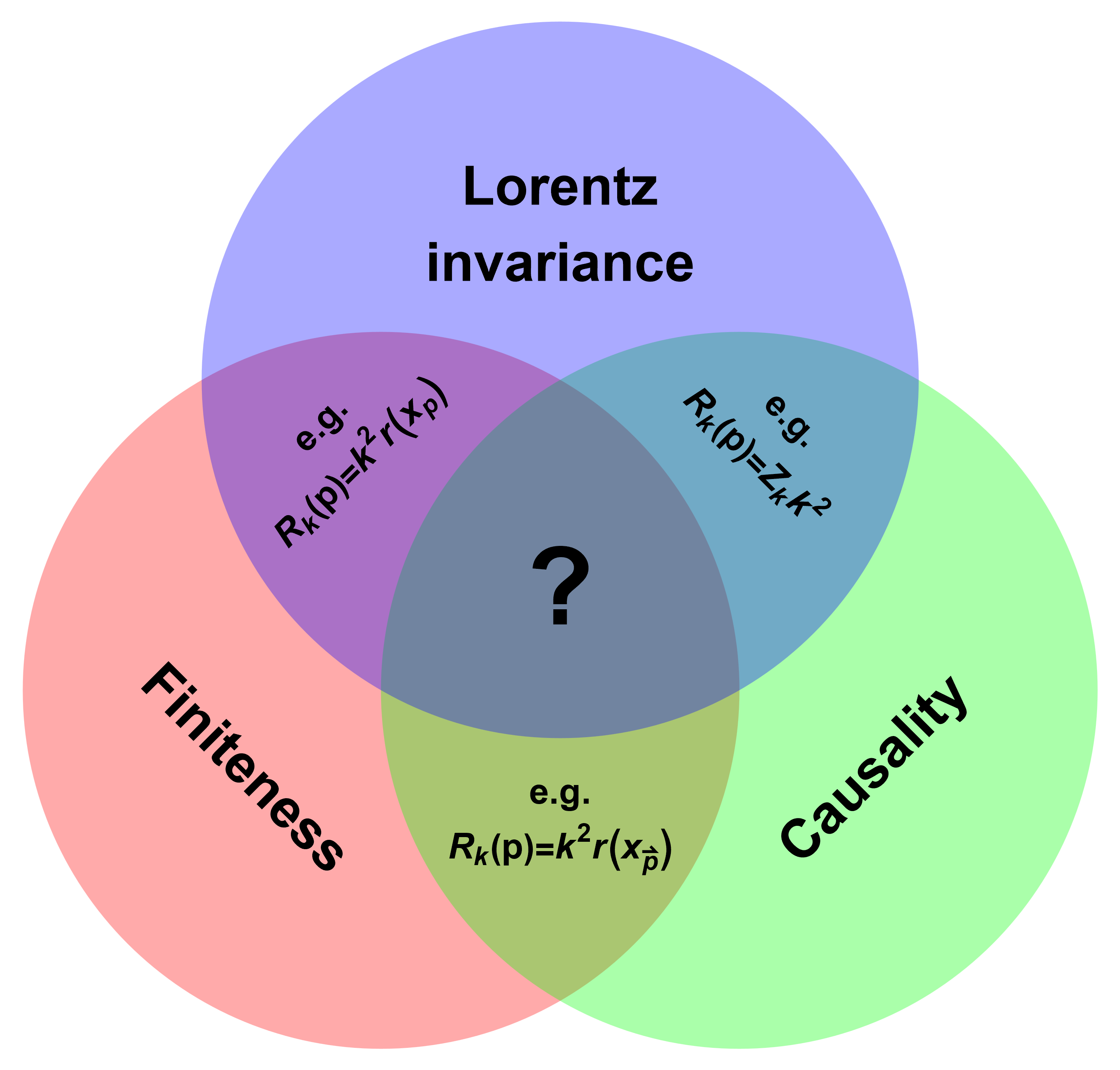}
	\caption{Sketch of the competing requirements for regulators: finiteness of the flow, Lorentz invariance and causality of regulators. Examples of regulators with two of the properties are given. A fully systematic construction of regulators with all three properties in the flow is lacking to date. Figure adapted from \cite{Braun:2022mgx}.}
	\label{fig:RegTrinity}
\end{figure}

The different demands for the shape function $r(x)$ in \labelcref{eq:ParaRegSpectral} are depicted in \Cref{fig:RegTrinity}, a detailed discussion can be found in \cite{Braun:2022mgx}. To date, the construction of a regulator with all the above properties is lacking. As the spectral fRG approach rests upon the existence of a spectral representation, or at least is facilitated by its prevalence, we rely on regulators that fulfil the causality constraint. Still, we note that Lorentz-invariant regulators with momentum argument \labelcref{eq:rkLorentz} can be constructed, that also satisfy \labelcref{eq:rkLimitsSpec}. The respective real-time flows are then based on the Keldysh contour and satisfy the causality constraint in a given frequency domain, see \cite{Pawlowski:2015mia}, for recent works see e.g.~\cite{Huelsmann:2020xcy, Tan:2021zid, Roth:2021nrd, Roth:2023wbp}, for a review see \cite{Berges:2012ty}.

For general regulators with \labelcref{eq:rkCausal}, the price to pay is the breaking of Lorentz invariance, which is restored in the limit $k\to0$. The only regulator that preserves causality and Lorentz invariance without any restriction is the Callan-Symanzik (CS) cutoff 
\begin{align}
	\label{eq:CS-Cutoff}
	R_k = Z_\phi\, k^2\,.
\end{align}
These properties make the CS cutoff best suited to extract spectral data. The price to pay is that the flow requires additional renormalisation because the standard UV divergences and counter terms resurface \cite{Symanzik:1970rt}. These local divergences of the flow must be absorbed in the parameters of the cutoff-dependent effective action. The CS regulator was later also employed in similar Lorentzian flow equations in algebraic QFT \cite{DAngelo:2022vsh, DAngelo:2023tis}.

Using the CS cutoff in the general flow equation \labelcref{eq:flow} leads us to the functional CS equation, first put down in \cite{Symanzik:1970rt}. Indeed it constitutes the first functional flow equation. Typically it is renormalised order by order in perturbation theory with the introduction of renormalised fields, couplings, and masses. This procedure cannot be applied within non-perturbative applications and in \cite{Braun:2022mgx} a general functional flow equation for the effective action $\Gamma_k$ with \textit{flowing} renormalisation has been derived. It reads for both Euclidean or Lorentzian signature 
\begin{align}
	\partial_t\Gamma_k[\phi] =  \frac12 \Tr\, G_k[\phi]\,\partial_t R_k - \partial_t S_{\text{ct},k}[ \phi] \,,
	\label{eq:RenFlow}
\end{align}
with the \textit{local} flowing counter-term action $S_{\text{ct},k}$. Importantly, \labelcref{eq:RenFlow} allows for a \textit{flowing} renormalisation, which is readjusted at each RG step. In \labelcref{eq:RenFlow}, the CS limit for momentum cutoffs with \labelcref{eq:rkCausal,eq:rkLorentz} was taken in a manifestly finite way as the flowing counter term action guarantees finiteness of the flow.  We emphasise that in the CS limit only the combination of the two terms on the right-hand side of \labelcref{eq:RenFlow} are finite and strictly speaking \labelcref{eq:RenFlow} only provides the structure of the flow. 

Importantly, the analytic nature of the momentum integrals in a spectral functional approach allows us to also define and use \textit{spectral dimensional regularisation} non-perturbatively and hence preserve gauge and diffeomorphism invariance within a non-perturbative numerical set-up. Of course, also other regularisation schemes such as the \textit{spectral BHPZ regularisation} can be used, for a detailed discussion see  \cite{Horak:2020eng,Braun:2022mgx}, for applications see \cite{Horak:2020eng,Horak:2021pfr,Horak:2022myj,Horak:2022aza,Horak:2023hkp}. In the explicit numerical computations of the graviton propagator in \cite{Fehre:2021eob}, spectral dimensional regularisation was used since it respects diffeomorphism invariance. 
More details and results of the spectral fRG approach to asymptotically safe metric quantum gravity are presented in \Cref{sec:LorentzianGravity}.

\subsection{Flow equations of correlation functions} 
\label{sec:flowCorFuncs}
The functional flow equations, \labelcref{eq:flow,eq:RenFlow}, for the effective action cannot be solved in a closed form. In most quantum field theories, the effective action has to be expanded in a systematic expansion scheme, and the flow is then solved in a given order of this expansion scheme. Here, we discuss this approach in terms of an expansion in terms of $n,m$-point correlation functions, where $n$ refers to background derivatives and $m$ to fluctuation derivatives, see \labelcref{eq:def-derivates}. They constitute the expansion coefficients of an expansion of the effective action $\Gamma_k[\bar g,\phi]$ in terms of powers of the fluctuation field and the background metric. The effective action takes the form 
\begin{align}
	\Gamma_k[\bar g,\phi] = \sum_{m=0}^\infty \int_{\boldsymbol{x}_m}\Gamma_k^{\phi_{a_1} \cdots\phi_{a_m}}[\bar g,0](\boldsymbol{x}_{m}) \phi_{a_1}(x_1)\cdots \phi_{a_m}(x_m)\,.
	\label{eq:GammaExpand}
\end{align}
In \labelcref{eq:GammaExpand}, we have used the abbreviation in \labelcref{eq:GnmShort} and a contraction of the indices $a_i$ as in \labelcref{eq:ExplicitContraction} is implied. Moreover, for the sake of simplicity, we have restricted ourselves to an expansion about $\phi=0$ but the expression is readily extended to one about a general background $\bar\phi$. We emphasise that the expansion coefficients $\Gamma_k^{\phi_{a_1} \cdots\phi_{a_m}}[\bar g,0]$ in  \labelcref{eq:GammaExpand} also comprise all $\Gamma_k^{(n,m)}$ upon $n$ further derivatives with respect to $\bar g$. 

The set of derivatives $\Gamma_k^{(n,m)}$ is the set of all one-particle irreducible parts of correlation functions of the fluctuation as well as the background field. As such it provides the full information about metric quantum gravity. More precisely, this information is comprised in the dressings of the $\Gamma_k^{(n,m)}$, which are evaluated on the equations of motion or another expansion point. We expand the correlation functions $\Gamma_k^{(n,m)}$ in complete tensor bases
\begin{align}
	\Gamma_k^{(n,m)}[\bar g,0] =\sum_i^{N_{n,m}}  \lambda^{(i)}_{n,m} {\cal T}^{(i)}_{n,m} \,, 
	\label{eq:GammanmTensors}
\end{align}
where $i=1,...,N_{n,m}$ labels the tensors in the basis $\{{\cal T}^{(i)}_{n,m}\}$ and $N_{n,m}$ is its size, which increases rapidly with $n$ and $m$. The whole $k$-dependence of $\Gamma_k^{(n,m)}$ is carried by the dressings $\lambda_{n,m}$, while the tensor basis is typically kept $k$-independent. Note also that in \labelcref{eq:GammanmTensors}, we have used the short-hand notations \labelcref{eq:GnmShort} in order to keep things simple. For a given $m$ there are many different $\Gamma^{(n,m)}$ with different bases and different tensors and dressings, 
\begin{align}
	{\cal T}^{(i)}_{n,m}&\to {\cal T}^{(i)}_{n,\phi_{s_1}\cdots \phi_{s_m}}(\boldsymbol{x}_{n+m})\,,
	& 
	\lambda^{(i)}_{n,m}&\to \lambda^{(i)}_{n,\phi_{s_1}\cdots \phi_{s_m}}(\boldsymbol{x}_{n+m})\,. 
	\label{eq:T+lambdafull}
\end{align}
In general, the tensors and dressings in \labelcref{eq:T+lambdafull} depend on the chosen background metric $\bar g$ and for  $\bar\phi\neq 0$, also on the chosen fluctuation field.  For $\bar\phi=0$ and $\bar g=\eta$, the Fourier transformation can be readily performed, leading to 
\begin{align}
	&{\cal T}^{(i)}_{n,\phi_{s_1}\cdots \phi_{s_m}}(\boldsymbol{p}_{n+m})\,,
	& 
	&\lambda^{(i)}_{n,\phi_{s_1}\cdots \phi_{s_m}}(\boldsymbol{p}_{n+m})\,.
	\label{eq:T+lambdafullp}
\end{align}
The tensor basis is typically chosen such that the tensors are regular in momentum space. Importantly, this allows us to distinguish parametric singularities in the dressings from physical ones. While this is seemingly relatively trivial, physical irregularities in the dressings can have far-reaching consequences, a prominent example being QCD, where irregular vertices are required for the presence of confinement in a covariant gauge such as also used in quantum gravity, see e.g.~\cite{Cyrol:2016tym} and literature therein. In short, the choice of a well-defined tensor basis may be of eminent importance for unravelling the underlying physics of the system at hand.

The functional flow equations, \labelcref{eq:flow,eq:RenFlow}, allow us to compute the scale-dependence of the effective action in terms of the correlation functions $\Gamma_k^{(n,m)}$ or rather the dressings $\lambda^{(i)}_{n,m}$. To begin with, the flow of the $\Gamma_k^{(n,m)}$ is obtained from the flow of $\Gamma_k[\bar g,\phi]$ by applying $n$ derivatives with respect to $\bar g$ and $m$ derivatives with respect to $\phi$. This leads us to   
\begin{align}
	\partial_t \Gamma_k^{(n,m)}[\bar g,\phi] = \frac{\delta^{n+m}\partial_t \Gamma_k[\bar g,\phi]}{\delta^n\bar g\delta^m\phi}\equiv\textrm{Flow}^{(n,m)}[\bar g,\phi]\,, 
	\label{eq:flownm} 
\end{align} 
where we have used the $k$-independence of the fields. $\textrm{Flow}^{(n,m)}$ stands for the diagrams generated from the field derivatives on the right-hand side of \labelcref{eq:flow}.

\begin{figure}[t]
	\centering
	\includegraphics[width=0.9\linewidth]{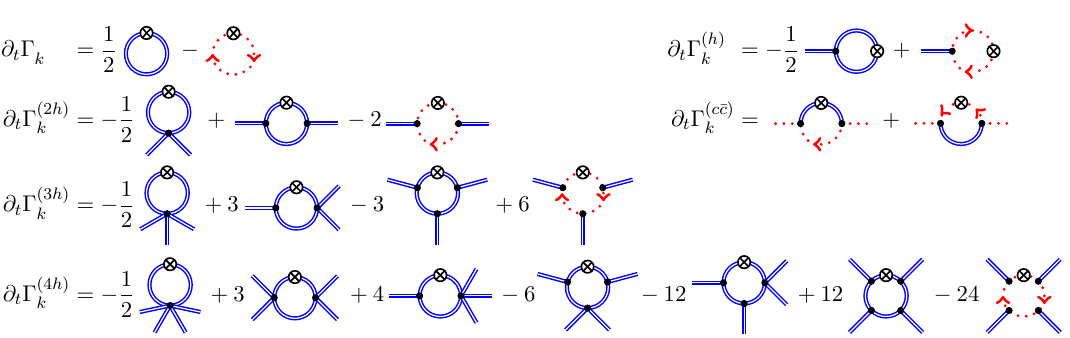}
	\caption{We display the diagrammatic flow of the $n$-point correlation functions. Blue double lines represent the fluctuation graviton propagators, red dotted lines stand for ghost propagators, while filled and crossed circles display dressed vertices and regulator insertions, respectively.
	}
	\label{fig:Diagrams}
\end{figure}

\Cref{eq:flownm} is a functional equation that holds for general background metrics and fluctuation fields. It hence offers a simple way to generate the diagrams on the right-hand side of the flow equation by successively applying field derivatives and using the rules 
\begin{subequations} 
	\label{eq:QMeSRules}
	\begin{align}
		\frac{\delta}{\delta \phi_b} \Gamma^{\phi_{a_1}\cdots \phi_{a_m}} = \Gamma^{\phi_b \phi_{a_1}\cdots \phi_{a_m}} \,, 
	\end{align}
	and 
	\begin{align}
		\frac{\delta}{\delta \phi_b} G_{\phi_{a_1}\phi_ {a_2}} = (-1)^{a_1 b} G_{\phi_{a_1}\phi_ {c}} \Gamma^{\phi_c \phi_b\phi_d} G_{\phi_{d}\phi_ {a_2}} \,,
	\end{align}	
\end{subequations} 
where $(-1)^{a_1 b}=-1$ for $\phi_{a_1}$ and $\phi_{b}$ both being fermionic and $(-1)^{a_1 b}=1$ for all other cases. As an explicit example, we show in \Cref{fig:Diagrams} the diagrammatic flows of $\Gamma^{(0,n)}$ for the graviton and ghost fields up to $n=4$, which can be readily derived from the rules \labelcref{eq:QMeSRules}. \Cref{eq:QMeSRules} extends readily to derivatives with respect to the background metric and allows us to iteratively derive the flow equations \labelcref{eq:flownm}.  

The right-hand side of the flows of general correlation functions \labelcref{eq:flownm} only contains correlation functions with at least two fluctuation fields, 
\begin{align}
	\partial_t \Gamma_k^{(n,m)}=\textrm{Flow}^{(n,m)}\left[\bar g,\left\{\Gamma_k^{(i,j)}\right\}\right]\,,\qquad \textrm{with}\quad 0\leq i\leq n\quad \textrm{and}\quad 2\leq j \leq m+2\,,
	\label{eq:Hierarchy}
\end{align}
see also \Cref{fig:Diagrams} as explicit example. This entails that correlation functions with higher orders in $\bar g$ depend on lower order ones in $\bar g$. Moreover, the flow of a correlation function $\Gamma_k^{(n,m)}$ depends on the correlation functions $\Gamma_k^{(i,m+1)}, \Gamma_k^{(i,m+2)}$ with $i\leq n$: the system does not close on a given order in the fluctuation field while it closes on a given order of the background field. Note that the origin of these properties is the dynamic nature of the fluctuation field and the auxiliary nature of the background field. In short, the correlation functions of the background field can only be computed, if that of the fluctuation field are known. 

In turn, the fluctuation correlation functions $\Gamma^{(0,n)}[\bar g\,,\,\bar\phi]$ constitute a closed system of flows for all backgrounds $\bar g$ and evaluation points $\bar \phi$ of the fluctuation fields, 
\begin{align}
	\partial_t \Gamma_k^{(0,m)}[\bar g,\bar \phi]=\textrm{Flow}^{(0,m)}\left[\bar g\,,\,\left\{\Gamma_k^{(0,j)}[\bar g,\bar \phi]\right\}\right]\,,\qquad \textrm{with}\quad 0\leq j \leq m+2\,.
	\label{eq:FluctuationHierarchy}
\end{align}
This leaves us with the following hierarchy: the only closed systems of flow equations are that of the fluctuation correlation functions $\Gamma^{(0,m)}$, and the computation of any other correlation functions requires this set as input. In short, the computation of the fluctuation correlation functions is at the core of any computation in the fRG approach to gauge-fixed metric quantum gravity.

All these properties are also visible in the example in \Cref{fig:Diagrams}: The flow of the graviton two-point function contains a tadpole diagram containing the graviton four-point function. In turn, the flow of the graviton four-point function contains a tadpole diagram containing the graviton six-point function. This exemplifies the hierarchy given in \labelcref{eq:FluctuationHierarchy}. By taking derivatives with respect to the background field, we only add external background-field legs to the diagrams but never any internal background propagator and therefore the hierarchy in \labelcref{eq:Hierarchy} holds.

In a final step, the flows \labelcref{eq:flownm} of the correlation functions have to be projected on the flow of the dressing functions $\lambda_{n,m}^{(i)}$ in \labelcref{eq:GammanmTensors} in a given expansion of the effective action. The projection is done by contracting the flows \labelcref{eq:flownm} with all the tensors ${\cal T}^{(i)}_{(n,m)}$ in a given tensor basis. In most cases such a basis is not diagonal: the basis elements overlap. Then, a further diagonalisation step is required. For the  $\lambda_{0,m}^{(i)}$, which is the expansion used in the fluctuation approach, this leads to the coupled flows 
\begin{align}
	\partial_t \lambda^{(i)}_{0,m} = \textrm{Flow}_{\lambda^{(i)}_{0,m}}[\{\lambda_{0,j}\} ]\,, \qquad  \textrm{with}\quad 0\leq j \leq m+2\,,  
	\label{eq:lambdaFluctuationHierarchy}
\end{align}
where we have used the hierarchy \labelcref{eq:FluctuationHierarchy}. Evidently, \labelcref{eq:lambdaFluctuationHierarchy} constitutes a closed set of flow equations. It is at the core of the flow equations of the mixed couplings including the dressings, $\lambda^{(i)}_{n,0}$, of the background effective action \labelcref{eq:BackGamma}. With \labelcref{eq:Hierarchy}, we are led to 
\begin{align}
	\partial_t \lambda^{(i)}_{n,m} = \textrm{Flow}_{\lambda^{(i)}_{n,m}}[\{\lambda_{l,j}\} ]\,,\qquad \textrm{with}\quad 0\leq l\leq n\quad \textrm{and}\quad 2\leq j \leq m+2\,.
	\label{eq:lambdaHierarchy}
\end{align}
Trivially, the hierarchy \labelcref{eq:lambdaFluctuationHierarchy,eq:lambdaHierarchy} of the flows of the dressings $\lambda_{n,m}$ carries the same properties as that of the full correlation functions: the system of flows for fluctuation dressings $\lambda_{0,m}$ is closed and is the core of the flows of the mixed dressings. We also remark, that the steps for the computation of the flows \labelcref{eq:lambdaFluctuationHierarchy,eq:lambdaHierarchy} can be implemented with the respective computer algebra packages for deriving the flows as well as contracting them: DoFun \cite{Huber:2011qr, Huber:2019dkb}, QMeS \cite{Pawlowski:2021tkk}, FormTracer \cite{Cyrol:2016zqb, github:FormTracer}, and VertExpand \cite{github:VertExpand}. 

The momentum-dependent dressings $\lambda^{(i)}_{n,m}(\boldsymbol{p}_{n+m})$ contain the full information of the quantum field theory. It is nonetheless useful to study their relation to curvature invariants. For example, the tensor structure of the transverse-traceless graviton two-point function, $\mathcal{T}_{0,h_{tt} h_{tt}}$, only has overlap with three tensor structures in an expansion about the flat background
\begin{align}
	\mathcal{T}_{0,h_{tt} h_{tt}} \sim  \sqrt{\pm g}, \, \sqrt{\pm g} R, \, \sqrt{\pm g} C_{\mu\nu\rho\sigma} f_C(\nabla^2) C^{\mu\nu\rho\sigma}\,.
\end{align}
Similarly, the tensor structure of the scalar graviton two-point function, $\mathcal{T}_{0,h_{s} h_{s}}$, also only has overlap with three tensor structures
\begin{align}
	\mathcal{T}_{0,h_{s} h_{s}} \sim  \sqrt{\pm g}, \, \sqrt{\pm g} R, \, \sqrt{\pm g} R f_R(\nabla^2) R\,.
\end{align}
The functions $f_R$ and $f_C$ have been called form factors \cite{Bosma:2019aiu, Knorr:2019atm, Draper:2020bop, Draper:2020knh, Knorr:2022lzn, Knorr:2022dsx}. These functions can be computed from solutions of the dressings $\lambda_{0,h_{tt} h_{tt}}$ and $\lambda_{0,h_{s} h_{s}}$, which has been done at the fixed point \cite{Christiansen:2012rx, Christiansen:2014raa, Christiansen:2015rva, Denz:2016qks, Knorr:2021niv} and recently also at vanishing cutoff scales \cite{Bonanno:2021squ, Fehre:2021eob} in Euclidean and Lorentzian signature, which will be discussed in detail in \Cref{sec:pure-grav,sec:LorentzianGravity}. Many expansion schemes focus on the transverse-traceless graviton mode, which is typically the dominant contribution. In this expansion, the tensor structure $\sqrt{\pm g} R f_R(\nabla^2) R$ can only be accessed order $h_{tt}^4$. This was used in \cite{Denz:2016qks} to show that $R^2$ fluctuations are relevant while $C^2$ fluctuations are irrelevant. 

One tensor structure only has overlap with finitely many curvature invariants in an expansion about the flat background. Together with the momentum dependence, this allows disentangling the running of different curvature operators, for example, the running of the universal one-loop beta functions of $R^2$, $C^2$, and $E$ \cite{Christiansen:2016sjn, HD-in-prep}. In the other direction, any curvature invariant has overlap with infinitely many $\mathcal{T}^{(i)}_{n,m}$. This creates a redundancy since they are related by diffeomorphism invariance, or more precisely by BRST invariance. How this redundancy is resolved is discussed in the next section.

\section{Background independence and symmetry identities}
\label{sec:symmetry-identities}

With the flow equations for the dressings,  \labelcref{eq:lambdaFluctuationHierarchy,eq:lambdaHierarchy}, we can compute the correlation functions within a given approximation to the full hierarchy of flows. However, the correlation functions or rather some of the dressings $\lambda_{n,m}^{(i)}$ are related by diffeomorphism symmetry constraints, and hence the set of all dressings carries diffeomorphism redundancies. These constraints provide non-trivial relations between background $\Gamma_k^{(n,0)}$ and fluctuation correlation functions $\Gamma_k^{(0,m)}$. These relations guarantee the background independence as well as the diffeomorphism invariance of the theory. In this chapter, we briefly review the symmetry identities that constrain the set of scale-dependent correlation functions  $\{\Gamma_k^{(n,m)}\}$. We shortly describe the origin and provide the final form of the symmetry identities. For a more detailed discussion of symmetry identities, we refer to \cite{Pawlowski:2020qer}. 

The quantum field theory approach to metric quantum gravity is based on the underlying symmetry of diffeomorphism invariance. With a linear metric split, $g_{\mu\nu} =  \bar g_{\mu\nu} + h_{\mu\nu}$, see \labelcref{eq:LinSplit}, this invariance is encoded in the quantum diffeomorphism transformation 
\begin{align}
	\label{eq:quantum-diff}
	h_{\mu\nu} &\longrightarrow h_{\mu\nu} + \mathcal{L}_{\omega} ( \bar g_{\mu\nu} + h_{\mu\nu} ) \,,
	&
	\bar g_{\mu\nu} &\longrightarrow \bar g_{\mu\nu} \,.
\end{align}
Here $\mathcal{L}_{\omega}$ is the Lie derivative with respect to some vector field $\omega_\mu$, which reads for a rank-two tensor 
\begin{align}\label{eq:Lie}
	\mathcal{L}_{\omega} T_{\mu\nu} = \omega_\rho \bar \nabla ^\rho T_{\mu\nu} + T_{\mu\rho} \bar \nabla^\rho \omega_\nu + T_{\nu\rho} \bar \nabla^\rho \omega_\mu \,,
\end{align} 
with the background covariant derivative $\bar \nabla$. The presence of the background metric triggers an auxiliary symmetry, the background diffeomorphism invariance given by the transformation
\begin{align}
	\label{eq:backgr-diff}
	h_{\mu\nu} &\longrightarrow h_{\mu\nu} + \mathcal{L}_{\omega} h_{\mu\nu} \,,
	&
	\bar g_{\mu\nu} &\longrightarrow \bar g_{\mu\nu} + \mathcal{L}_{\omega} \bar g_{\mu\nu} \,.
\end{align}
Both transformations, \labelcref{eq:quantum-diff,eq:backgr-diff}, imply the same transformation for the full metric $g_{\mu\nu}$ as expected from a diffeomorphism transformation. While \labelcref{eq:backgr-diff} is at first only an auxiliary symmetry, it is still related to the dynamical quantum symmetry via the split symmetry of the metric split. The split symmetry includes all transformations that leave the full metric invariant and is given by
\begin{align}
	\label{eq:metric-trafo}
	g( \bar g, h) \longrightarrow g( \bar g + \delta \bar g, h+ \delta h) = g( \bar g, h)\,.
\end{align}
For example in the linear split \labelcref{eq:LinSplit}, we have $\delta \bar g = -\delta h$. This symmetry encodes the background independence of the theory.

\subsection{Background independence}
\label{sec:background-independence}

Without the regulator function, the background independence of the theory is encoded in the Nielsen identity (NI) \cite{Nielsen:1975fs, Fukuda:1975di}
\begin{align}
	\text{NI}&\equiv\frac{\delta\Gamma}{\delta \bar g_{\mu\nu}} -\frac{\delta\Gamma}{\delta h_{\mu\nu}}
	-\llangle \!\left[\frac{\delta}{\delta \bar g_{\mu\nu}}-\frac{\delta}{\delta \hat h_{\mu\nu}} \right] \!( S_\text{gf}+S_\text{gh})\!\rrangle =0\,,
	\label{eq:linNielsen}
\end{align}
where $h_{\mu\nu} = \langle \hat h_{\mu\nu} \rangle$. For this identity, we have used the linear split. This identity has to hold at vanishing cut-off scales after the integrating-out of all quantum fluctuations. On the quantum equations of motions, the last term in \labelcref{eq:linNielsen} vanishes, which implies that a solution to the quantum equation of motion is also a solution to the background equation of motion, $\delta \Gamma/\delta h = \delta \Gamma/\delta \bar g$.

At finite cut-off scales, the NIs are modified by the presence of the regulator, leading to the modified NIs (mNIs)
\begin{align}
	\label{eq:mNIlin}
	\text{mNI}_k \equiv \text{NI}_k - \frac12 \text{Tr}\,\frac{\delta \sqrt{\bar g} R_k}{\sqrt{\bar g}\,\delta \bar g_{\mu\nu}} G_k  =0\,.
\end{align}
In terms of flowing correlation functions $\Gamma^{(n,m)}$ and the corresponding dressing functions of the tensor structures $\lambda^{(i)}_{n,m}$, see \labelcref{eq:lambdaHierarchy}, the mNI relates $\Gamma^{(n+1,m)}$ with $\Gamma^{(n,m+1)}$ where the last fluctuation derivative is with respect to the fluctuation metric $h$. For example, the background propagator $\Gamma^{(2,0)}$ can be determined with the knowledge of the fluctuation propagator $\Gamma^{(0,hh)}$ and vice versa, see \cite{Eichhorn:2018akn} for first results in that direction. Note however that the relation is non-trivial due to the expectation value in \labelcref{eq:linNielsen} and the trace in \labelcref{eq:mNIlin}. The background propagator can also be computed directly from the flow for a given fluctuation propagator but not the other way around due to the dynamical nature of the fluctuation field and the auxiliary nature of the background metric, see \labelcref{eq:Hierarchy}. mNIs have been discussed in detail in gravity, gauge theories, and in scalar theories, see \cite{Reuter:1996cp, Reuter:1997gx, Freire:2000bq, Litim:2002ce, Litim:2002hj, Pawlowski:2003sk, Pawlowski:2005xe, Manrique:2009uh, Manrique:2010mq, Donkin:2012ud, Bridle:2013sra, Becker:2014qya, Dietz:2015owa, Safari:2015dva, Safari:2016dwj, Safari:2016gtj, Labus:2016lkh, Morris:2016spn, Percacci:2016arh, Nieto:2017ddk, Ohta:2017dsq, Eichhorn:2018akn, Lippoldt:2018wvi, Reichert:2020mja}.

\subsection{Diffeomorphism invariance}
\label{sec:diff-invariance}
The auxiliary background diffeomorphism invariance \labelcref{eq:backgr-diff} remains unbroken by gauge-fixing and regularisation. The physical quantum diffeomorphism invariance \labelcref{eq:quantum-diff} turns into a BRST symmetry due to the gauge fixing, which is then further modified by the regulator. The related symmetry identities are called (modified) Slavnov-Taylor identities ((m)STI) \cite{Taylor:1971ff, Slavnov:1972fg}. They encode physical diffeomorphism invariance. 

In case of the linear gauge-fixing condition \labelcref{eq:gf-condition}, the generator of BRST-transformation (or BRST-operator) denoted by $\mathfrak{s}$, including the Nakanishi-Lautrup field $\lambda_\mu$, is given by 
\begin{align}
	\label{eq:brst}
	\mathfrak{s} (\bar g_{\mu\nu},h_{\mu\nu},c_\mu, \bar c_\mu, \lambda_\mu, \phi_\text{mat})
	&= ( 0,\mathcal{L}_{c}(\bar g_{\mu\nu} + h_{\mu\nu}),c_\rho \bar \nabla^\rho c_\mu, \lambda_\mu, 0, \mathfrak{s} \phi_\text{mat})\,. 
\end{align}
Here, the vector field $\omega_\mu$ in the Lie derivative \labelcref{eq:Lie} is given by the ghost field, $\omega_\mu=c_\mu$, for more details on the setup and the condensed notation used below see \cite{Pawlowski:2005xe}. The Nakanishi-Lautrup field $\lambda_\mu$ transforms trivially under the BRST transformation, $\mathfrak{s}\lambda_\mu=0$. The classical gauge-fixed action including the gauge-fixing and the ghost action is invariant under this transformation, $\mathfrak{s}(S_\text{grav}+ S_\text{gf}+S_\text{gh})=0$. Furthermore, $\mathfrak{s}$ is a nilpotent operator with $\mathfrak{s}^2=0$.

For the derivation of the STI, we include a source term $Q^a \mathfrak{s} \hat\phi_a$ for the BRST variations of the fields in the generating functional.  Expressed in terms of the effective action, see \cite{Pawlowski:2005xe}, the STI in the absence of the cutoff term reads
\begin{align}
	\label{eq:STI}
	\text{STI} \equiv\int\!\mathrm d^4 x \sqrt{\bar g} \,\frac{\delta \Gamma}{\delta \phi_a} 
	\frac{\delta \Gamma}{\delta Q^a} = 0 \,.
\end{align}
This equation is known as the quantum-master equation. The BRST variation of the effective action is given by $\delta \Gamma/\delta Q^a = \langle \mathfrak{s} \hat \phi_a \rangle$. These variations can be interpreted as generalised vertices of the theory.

\Cref{eq:STI} encodes diffeomorphism invariance at $k=0$ where the regulator vanishes. At finite cutoff scales, an additional regulator contribution has to be taken into account, and we are led to the mSTI, 
\begin{align}
	\label{eq:mSTI}
	\text{mSTI} \equiv \text{STI} - \,\text{Tr}\,R_k \,\frac{\delta^2 \Gamma_k}{\delta Q \delta \phi} \,G_k = 0 \,.
\end{align}
In terms of flowing correlation functions $\Gamma^{(n,m)}$, the mSTI relates different fluctuation correlation functions with each other. Some of the properties of the mSTI are theory-dependent, but most of them are generic: mSTIs in the presence and absence of background fields in gravity and gauge theories have been discussed in detail in \cite{Bonini:1993sj, Bonini:1994kp, Bonini:1994dz, Bonini:1995tx, Ellwanger:1994iz, DAttanasio:1996tzp, Reuter:1997gx, Litim:1998yc, Freire:2000bq, Igarashi:1999rm, Igarashi:2000vf, Igarashi:2001mf, Igarashi:2001ey, Igarashi:2001cv, Litim:2002ce, Pawlowski:2003sk, Pawlowski:2005xe, Gies:2006wv, Igarashi:2007fw, Igarashi:2008bb, Igarashi:2009tj, Donkin:2012ud, Sonoda:2013dwa, Safari:2015dva, Igarashi:2016gcf, Asnafi:2018pre, Morris:2018axr, Igarashi:2019gkm, Reichert:2020mja, Pawlowski:2021tkk}.

\subsection{Wrap up}
\label{sec:WrapUpSym}
The investigations of the last chapters leave us with the following structure of correlation functions and flows in functional approaches to metric quantum gravity: The required regularisation and gauge fixing introduces inevitably a background metric $\bar g_{\mu\nu}$, and hence two sets of fields, the background fields $\bar \phi$ and the dynamical fluctuation fields $\phi$. The flows of the $\Gamma^{(n,m)}$ can be derived, see \labelcref{eq:Hierarchy}, and it is the dynamical fluctuation correlation functions $\Gamma^{(0,m)}$ that drive all of these flows. Additionally to the flow equations, we have symmetry relations between the correlation functions. The mNI, see \labelcref{eq:mNIlin}, relates background correlation function $\Gamma^{(n,0)}$ with fluctuation ones $\Gamma^{(0,m)}$ and ensures background independence. The mSTI, see \labelcref{eq:mSTI}, relates different fluctuation correlation functions and ensures the BRST symmetry of the diffeomorphism invariance. 

In principle, it would be a great simplification if the dependence on the background fields could be eliminated.  However, the non-linear nature of the diffeomorphism transformation does not allow us to derive the effective action in a separable form such as 
\begin{align} 
	\Gamma[\bar \phi,\phi]\approx \Gamma[\bar \phi+\phi]+S_\textrm{gf}[\bar g,h]+S_\textrm{gh}[\bar g, h,c]\,,  
	\label{eq:BackgroundApprox}
\end{align}
where we have now included $\bar g$ in the expansion point $\bar \phi$ of the fluctuation field for the sake of convenience. With this notation, the commonly used expansion point is $\bar\phi=(\bar g_{\mu\nu},0,...,0)$.   
\Cref{eq:BackgroundApprox} is known as the background-field approximation, and is commonly used in computations in the fRG approach to metric quantum gravity. It is readily seen that the flow equation \labelcref{eq:flow} is diffeomorphism-invariant at $h=0$ and reduces to a flow of $\Gamma[\bar \phi,0]$. Note however, that the flow \labelcref{eq:flow} with $\phi\neq 0$ does not sustain the separability given in \labelcref{eq:BackgroundApprox} due to both, the gauge-fixing sector  $S_\textrm{gf}+S_\textrm{gh}$ and the cutoff term. 

Hence, while \labelcref{eq:BackgroundApprox} has the appealing feature of diffeomorphism invariance (up to the classical gauge-fixing term) and leads to a seemingly gauge-invariant flow equation, its diffeomorphism invariance rests on an approximation that explicitly breaks diffeomorphism invariance. This is readily seen by inserting  \labelcref{eq:BackgroundApprox} into the STIs. This leads us to the seemingly self-contradictory situation, that the diffeomorphism invariance of \labelcref{eq:BackgroundApprox} signals its breaking. Indeed, the fluctuation correlation functions are \textit{required} to be \textit{not} diffeomorphism-covariant for $\Gamma[\bar\phi]$ carrying \textit{physical} diffeomorphism invariance~\cite{IlGattorpardo}. 

There have been many attempts over the past decades to enforce diffeomorphism-covariance, or more generally gauge invariance of general gauge theories, of fluctuation correlation functions in a gauge-fixed approach via non-linear transformations. The latter are either directly applied or indirectly implemented with diffeomorphism constraints. For an (incomplete) list of respective references within the fRG approach see \cite{Branchina:2003ek, Pawlowski:2003sk, Donkin:2012ud, Morris:2016nda, Wetterich:2017aoy, Asnafi:2018pre, Falls:2020tmj, Mandric:2022dte}. Loosely speaking the price to pay for diffeomorphism covariance of fluctuation correlation functions in a gauge-fixed approach is non-locality. We emphasise in this context, that while commonly dropped in the discussions of general gauge theory, it is gauge invariance \textit{together} with locality, which is required, for a more detailed discussion see \cite{Dupuis:2020fhh}. In summary, while there is remarkable progress, it is fair to say that the diffeomorphism-covariant functional approaches are still in their infancy concerning the evaluation of their overall consistency and structure. 

The fluctuation approach discussed in this contribution aims at solving the closed set of flow equations for fluctuation correlation functions, see \labelcref{eq:FluctuationHierarchy}. The flows have to be solved with the boundary condition given by the STI at vanishing cutoff scale. The NI can then be used for the computation of the background correlation functions. This is an ambitious goal but it can be achieved for given orders in a systematic expansion scheme that does not spoil diffeomorphism invariance from the outset.

\section{Fluctuation approach}
\label{sec:fluctuation-approach}

In \Cref{sec:fRG-QG,sec:symmetry-identities}, we have discussed the general fRG approach to asymptotically safe metric quantum gravity. The goal is to compute the effective action $\Gamma[\bar g, \phi]$, \labelcref{eq:Gamma}, and the derived diffeomorphism-invariant background effective action $\bar\Gamma[g,\phi_\textrm{mat}]$, \labelcref{eq:BackGamma}. This task is tantamount to that of the respective correlation functions $\Gamma^{(n,m)}$, \labelcref{eq:def-derivates}, whose flows are governed by \labelcref{eq:flow}. As discussed in \Cref{sec:fRG-QG,sec:symmetry-identities}, both tasks require the computation of the fluctuation correlation functions in a given background at the chosen expansion point $\phi=\bar \phi$. In general, this reads 
\begin{align}
	\Gamma^{\phi_{a_1} \cdots\phi_{a_n}}[\bar g,\bar \phi](\boldsymbol{x}_n)	= \left.\frac{\delta^n \Gamma[\bar g, \phi]}{\delta\phi_{a_1}(x_1)\cdots \delta\phi_{a_n}(x_n)}\right|_{\phi=\bar\phi}\,. 
	\label{eq:CompFlucs}
\end{align}
The results for \labelcref{eq:CompFlucs} are the expansion coefficients of the effective action in an expansion about a generic expansion point $\bar\phi$, for $\bar\phi=0$ see \labelcref{eq:GammaExpand}. For $\bar\phi\neq 0$, we write schematically 
\begin{align}
	\Gamma[\bar g,\phi] = \sum_{n=0}^\infty \int_{\boldsymbol{x}_n}\!\Gamma^{\phi_{a_1} \cdots\phi_{a_n}}[\bar g,\bar\phi](\boldsymbol{x}_{n})\, \Biggl[(\phi-\bar\phi)_{a_1}(x_1)\cdots (\phi-\bar\phi)_{a_n}(x_n)\Biggr]\,, 
	\label{eq:GammaExpandbarphi}
\end{align}
The fluctuation correlation functions \labelcref{eq:CompFlucs} are then used to compute the complete system of correlation functions $\Gamma^{(n,m)}$, which is required for determining the background effective action $\bar\Gamma[g,\phi_\textrm{mat}]$.

\subsection{Expansion schemes and RG-invariant vertex dressings} 
\label{sec:ExpansionSchemes} 

The computation of the whole system of correlation functions without relying on the background approximation has been baptised \textit{fluctuation approach} for the reason that the core of any systematic computation is the fluctuation correlation functions \labelcref{eq:CompFlucs}. Evidently, one needs a systematic approximation scheme for any computation, but the fluctuation approach only rests on such a scheme and no further approximation.  

\Cref{eq:CompFlucs} also comprises the background correlation functions and the vacuum correlation function $\Gamma^{(0,0)}[\bar g,\bar \phi]$ at $h,c,\bar c=0$, which is nothing but the background effective action \labelcref{eq:BackGamma}. However, this is more a conceptual remark rather than one with computational significance, as already the computation of fluctuation correlation functions in symmetric background metrics is a very challenging task. Therefore, in most computations, we resort to a double expansion scheme both in terms of fluctuation fields about a simple background $\bar \phi$ as well as an expansion about a simple background metric $\bar g_{\mu\nu}$. We emphasise in this context that any expansion of the flow equation is only a tool for computing correlation functions, the expansion point itself is not required to have any physical significance such as the solution of the quantum equations of motion $g_\textrm{EoM}\,,\,\phi_\textrm{EoM}$ with 
\begin{align}
	\frac{ \delta \Gamma}{ \delta \bar g}=0= \frac{\delta \Gamma}{\delta \phi}\,, \qquad \textrm{for}\qquad (\bar g,\phi)=(g_\textrm{EoM},\phi_\textrm{EoM})\,.
	\label{eq:EoM}
\end{align}
Still, the further away the expansion point is from the solution to the equations of motion, the more the stability of the expansion scheme is at stake. 

In summary, the fluctuation approach is the general fRG approach to metric quantum gravity without resorting to further approximations. In explicit computations we have to specify the expansion scheme chosen: in almost all computational applications to date, the background metric has been set to the flat metric, either Euclidean or Minkowski. We emphasise that such a choice is related to a curvature expansion in the background effective action, the latter also being an expansion about the flat background. However, in the fluctuation approach with $\bar g=\eta$ typically momentum dependences are computed that give access to covariant momentum dependences in the background effective action.

The above discussion specifies the background and hence the expansion in \labelcref{eq:GammaExpandbarphi} used for most computations, but it does not specify the split between the background metric $\bar g$ and the fluctuation field $h$. While the approach works with any split, in this contribution we focus on the linear split \labelcref{eq:LinSplit} with a non-trivial prefactor. A very convenient choice for the linear metric split is given by 
\begin{align}
	g_{\mu\nu} =\bar g_{\mu\nu} + \sqrt{Z_h G_\text{N} }\, h_{\mu\nu}\,, 
	\label{eq:LinSplitGamma} 
\end{align} 
with the wave function $Z_h$ and the Newton coupling $G_\text{N}$. While the split \labelcref{eq:LinSplitGamma} works generically, it is natural for asymptotic safety based on the IR Einstein-Hilbert action \labelcref{eq:EH-Action}. This is already indicated by the occurrence of the Newton coupling leading to dimension 1 fluctuation fields $h_{\mu\nu}$. While it could be substituted by any dimension $-2$ coupling, the choice \labelcref{eq:LinSplitGamma} removes the $G_\text{N}$ prefactors from $S^{(2)}_{\textrm{EH}}$. Note in this context, that specifying the classical action is tantamount to setting the classically propagating degrees of freedom: In the classical Einstein-Hilberg action, \labelcref{eq:EH-Action}, this is only the massless graviton while in the classical Stelle action, \labelcref{eq:Stelle-Action}, this also includes the massive scalar mode and the massive spin-two ghost. We emphasise that while it is important to set the initial propagating degrees of freedom, the quantum degrees of freedom can be different. Furthermore, using the Einstein-Hilbert action in the path integral is not the same as working in the Einstein-Hilbert truncation. In the latter, only terms related to the cosmological constant and the Ricci scalar are taken into account, while here all generated contributions are considered. Importantly there are finite higher-curvature contributions in the limit $k\to \infty$: the difference is that these contributions do not have an ultra-local part, see the next section, \Cref{sec:ApparentConvergence}, for more details. 

Note that on the quantum level, each mode carries a different wave function and $Z_h$ would be a matrix. For now, we discuss approximations with a uniform wave function $Z_h$ and discuss the general case later. The wave function is dimensionless and depends in the flat background on the momentum $Z_h=Z_h(p)$. The Newton coupling carries the mass dimension $[G_\text{N}]= -2$ and therefore the fluctuation field carries the mass dimension $[h_{\mu\nu}]=1$ which is that of a standard scalar field in four dimensions. The rescaling in \labelcref{eq:LinSplitGamma} guarantees a canonical form of the graviton propagator, for example for the transverse-traceless part 
\begin{align}
	G_{h_{tt}h_{tt}} \propto \frac1{Z_h(p) (p^2 + \mu)} \,,
	\label{eq:canonical-prop}
\end{align}
where $\mu$ is the graviton mass parameter, which is related to the cosmological constant through a mNI. \Cref{eq:canonical-prop} holds similarly for the other modes of the graviton. We emphasise that these rescalings are used to simplify the relation between fluctuation correlation functions and background correlation functions, they do not affect or change the expansion scheme.

Inserting \labelcref{eq:LinSplitGamma} on both sides of the functional flow \labelcref{eq:flow} leads to a hierarchy of coupled integral-differential equations for fluctuation correlation functions. A diagrammatic depiction of the first five flows for  $\Gamma^{(0,0)}, \Gamma^{(0,1)}\,,...,\,\Gamma^{(0,4)}$ is given in \Cref{fig:Diagrams}. This system of equations has been solved in a flat background \cite{Denz:2016qks} including the momentum dependence of the correlation functions. It still constitutes the most advanced computation of momentum-dependent fluctuation correlation functions in metric quantum gravity. The fluctuation correlation functions are related to $S$-matrix elements via their RG-invariant vertex cores $\bar \Gamma^{(0,n)}(\boldsymbol{p}_{n})$ with the general definition   
\begin{align}
	\bar \Gamma_k^{(n,\phi_{a_1}\cdots \phi_{a_m})}(\boldsymbol{p}_{n+m}) := \frac{\Gamma_k^{(n,\phi_{a_1}\cdots \phi_{a_m})}(\boldsymbol{p}_{n+m})}{\prod_{i=1}^n Z^{1/2} _{\bar g}(p_i)\, \prod_{j=1}^m Z^{1/2} _{\phi_{a_j}}(p_{n+j})}\,.
	\label{eq:barGamman}
\end{align}  
Here, the $Z_{\phi_{a_i}}(p_i)$ are the fully momentum-dependent wave functions of the fields $\phi_i$ with $\Gamma^{(0,2)} \simeq Z(p) S^{(0,2)}(p)$, see also \labelcref{eq:canonical-prop}. The correlation functions \labelcref{eq:barGamman} can be expanded in terms of the same tensor basis as the $\Gamma^{(n,m)}$ in  	\labelcref{eq:GammanmTensors}, to wit, 
\begin{align}
	\bar \Gamma_k^{(n,m)}[\bar g,0] =\sum_i^{N_{n,m}} \bar \lambda^{(i)}_{n,m} {\cal T}^{(i)}_{n,m} \,,
	\label{eq:barGammanmTensors}
\end{align}
with
\begin{align}
	\bar \lambda^{(i)}_{n,\phi_{a_1}\cdots \phi_{a_m}}(\boldsymbol{p}_{n+m}) = \frac{\lambda^{(i)}_{n,\phi_{a_1}\cdots \phi_{a_m}}(\boldsymbol{p}_{n+m})}{\prod_{i=1}^n Z^{1/2} _{\bar g}(p_i)\, \prod_{j=1}^m Z^{1/2} _{\phi_{a_j}}(p_{n+j})}\,. 
	\label{eq:barlambda}
\end{align}
The correlation functions \labelcref{eq:barGamman} and their dressings \labelcref{eq:barlambda} are invariant under standard RG-transformations that constitute in their most general form reparametrisations of the theory and are typically done via the variation of an RG-scale $\mu$, for a detailed discussion see \cite{Pawlowski:2005xe}. Such a reparametrisation has to be contrasted with the IR cutoff scaling in the fRG approach, which entails the successive integration of momentum modes. The homogeneous RG-equation for the $\bar\Gamma^{(n,m)} $ reads 
\begin{align}
	\mu \frac{\mathrm d}{\mathrm d \mu}  \bar \Gamma^{(n,m)}(\boldsymbol{p}_{n+m})=0\,, \qquad \rightarrow\qquad \mu \frac{\mathrm d}{\mathrm d \mu}\bar \lambda^{(i)}_{n,m} (\boldsymbol{p}_{n+m})=0\,. 
	\label{eq:RGnm}
\end{align}
The $S$-matrix is built from the tree-level diagrams with the vertices \labelcref{eq:barGamman} and the fluctuation propagators. Hence,  the scalar dressings $\bar\lambda_{0,n}$ of $\bar \Gamma^{(0,n)}(\boldsymbol{p}_{n})$ are the non-diagrammatic part of the form factor of the respective $S$-matrix element. Note however, that in gauge theories neither \labelcref{eq:barGamman} nor its trivial extension to \labelcref{eq:BackGamma} provides the form factors of the $S$-matrix, but provide RG-invariant momentum-dependent scattering couplings, that are the core objects in the fluctuation approach. Note also, that the dressings are called form factors themselves and are at the root of the form factor approach to metric quantum gravity reviewed in \cite{Knorr:2022dsx}. 

It is worth mentioning that the \textit{essential} RG, developed in \cite{Baldazzi:2021ydj} and used in asymptotically safe gravity in \cite{Baldazzi:2021orb, Knorr:2022ilz}, leads to the definition of rescaled fields that absorb the wave functions and automatically generates the vertices of the type of \labelcref{eq:barGammanmTensors}. This also involves the field dependence of the wave functions and other terms, and hence includes general field-dependent reparametrisations, the \textit{flowing fields}. These reparametrisations are included within the the general fRG for the effective action derived in \cite{Pawlowski:2005xe}, based on \cite{Wegner_1974}. For further applications see \cite{Ihssen:2023nqd}, where the fRG with flowing fields is used to obtain classical dispersions for the full theory, which optimises the vertex expansion scheme. 

In the remainder of this chapter, we discuss the stability and convergence of such an expansion, the intricacies of stability investigations and relevant directions at a fixed point for over-determined systems such as present in gauge theories, as well as the relations to other approaches.  

The convergence of the expansion scheme in terms of correlation functions $\Gamma^{(0,n)}$ hinges on the sub-dominance of higher-order correlation functions in the flow of lower-order ones. Put differently, one studies the convergence of the results for a low-order correlation function such as the propagator in dependence on the full resolution of successively higher-order correlation functions. This is called \textit{apparent} convergence. Certainly, it is short of a proof of true convergence which however is absent for any other non-perturbative approach to metric quantum gravity or any other non-perturbative quantum field theory such as QCD for that matter. Note also that even in perturbation theory apparent convergence is used. However, in the latter case, the expansion at least relies on a small parameter, and we discuss such a parameter in the current expansion scheme in the next section. This discussion is based on an expansion about a flat background, which is the common choice for the background. However, we emphasise that the structural analysis does not depend on this choice. Moreover, in more sophisticated applications, in particular in gravity-matter systems, one typically resorts to a mixed expansion scheme: the effective action is expanded in terms of momentum-dependent correlation functions, but this expansion is augmented by full field-dependent functionals such as the effective potential $V_\textrm{eff}(\phi)$ of some matter field or even the fluctuating graviton, see e.g.~\cite{Henz:2013oxa, Henz:2016aoh, Pastor-Gutierrez:2022nki}. Such approximations have been commonly used specifically in QCD, and the current analysis draws a lot from results obtained there, for a discussion see \cite{Dupuis:2020fhh}.

\subsection{Apparent convergence}
\label{sec:ApparentConvergence} 

At the root of convergence or rather apparent convergence of the vertex expansion scheme is the phase space suppression of diagrams with higher-order vertices in the hierarchy of flows in \Cref{fig:Diagrams}. In a standard renormalisable local quantum field theory with polynomial classical interactions the vertices can be ordered in terms of a few, typically primitively divergent, vertices  $\Gamma^{(0,n)}_\textrm{cl}$ or rather the respective dressings $\lambda^{\textrm{(i)}}_{0,n}=\lambda^{\textrm{(cl)}}_{0,n}$ with a pointlike classical interaction core, and quantum vertices $\Gamma^{(n)}_\textrm{qu}$ that are only generated by quantum corrections and hence loop diagrams. The dressings of the classical tensor structures can be parametrised as 
\begin{align}
	\lambda^{\textrm{(cl)}}_{0,n}(\boldsymbol{x}_{n})\propto \lambda^\textrm{(cl)}_n \delta(x_1-x_2)\cdots \delta(x_{n-1}-x_n) +\lambda^{\textrm{(cl,dis)}}_{0,n}(\boldsymbol{x}_{n})\, ,
	\label{eq:point+dis}
\end{align} 
where $\lambda^{\textrm{(cl,dis)}}$ carries the part of the dressing without point-like contributions: hence it is a finite distribution function. Note that its different parts may still contain $\delta$-functions, but not the full product of $\delta$-functions in all positions as in the first term. In momentum space, the first term in \labelcref{eq:point+dis} gives rise to a constant term in the vertex for $p_i^2\to \infty$ for all $i=1,...,n$, while $\lambda^{\textrm{(cl,dis)}}$ decays if all momenta are taken to infinity, 
\begin{align}
	\left.\lim_{p\to\infty} \lambda^{\textrm{(cl)}}_{0,n}(\boldsymbol{p}_n)\right|_{p_i^2=p^2} &=(2 \pi)^{4n}  \lambda^\textrm{(cl)}_n \delta(\boldsymbol{p}_n)\,,
	&
	 \left.\lim_{p\to\infty} \lambda^{\textrm{(cl,dis)}}_{0,n}(\boldsymbol{p}_n)\right|_{p_i^2=p^2} &=0\,, 
	\label{eq:MomentumLocality1}
\end{align}
with $\boldsymbol{p}_n$ as defined in \labelcref{eq:xnpn}. We emphasise that it does not suffice to only take the limit $\boldsymbol{p}_n^2\to\infty$: if this limit is accomplished by $p_i^2\to \infty$ for some of the momenta, while others are kept finite, $\lambda^{\textrm{(cl,dis)}}_{0,n}$ does not necessarily vanish. 

A prominent non-perturbative example for the above properties is QCD where $\Gamma^{(0,n)}_\textrm{cl}$ with $n=2,3,4$ is given by the inverse propagator and three- and four-point vertex parts that carry the classical tensor structure $S^{(n)}$. The quantum vertices are all other verticesincluding the parts of the full three- and four-point vertices $\Gamma^{(0,3)}$ and $\Gamma^{(0,4)}$ in 	\labelcref{eq:barGammanmTensors} with non-classical tensor structures. We extend this distinction to metric quantum gravity and write in analogy, 
\begin{align}
	\left\{\Gamma^{(0,n)}(\boldsymbol{x}_n)\right\} = \left\{\Gamma_\textrm{cl}^{(0,n)}(\boldsymbol{x}_n)\,,\, \Gamma_\textrm{qu}^{(0,n)}(\boldsymbol{x}_n)\right\} \,.
	\label{eq:ApparentConvergenceOrdering}
\end{align}
In contradistinction to QCD, in metric quantum gravity, each $n$-point function of gravitons contains a classical part proportional to $S_\textrm{EH}^{(n)}$. However, all these parts are directly related to each other and the vertex expansion underlying apparent convergence is one about the classical pieces  $\Gamma_\textrm{cl}^{(0,n)}(\boldsymbol{x}_n)$ of all vertices. While this leaves us seemingly with an infinite set of classical vertices, the dressings of all these vertices are related to each other. In the present example of the Einstein-Hilbert action, the RG-invariant dressings \labelcref{eq:barlambda} of the fluctuation graviton $n$-point functions, $\bar\lambda^\textrm{(EH)}_n$, produce \textit{avatars} of the Newton coupling $G_n(p)$,  
\begin{align}
	\bar\lambda^\textrm{(EH)}_n(p) = G_n(p)^\frac{n-2}{2}+ \bar\lambda^\textrm{(EH,dis)}_n(p)\,,
	\label{eq:Avatars} 
\end{align}
where $p$ indicates the evaluation at the symmetric point or another appropriate momentum configuration. Note that the separation on the right-hand side is rather difficult to perform explicitly and is not relevant for any computation: it is only the full vertex dressings that enter any flow. The existence of avatars is directly related to the non-polynomial form of the Einstein-Hilbert action or any other diffeomorphism action and hence is related to diffeomorphism invariance. This complicates the stability analysis, which requires more care in a diffeomorphism or gauge invariant theory, see \Cref{sec:StabilityAnalsis}. 

In summary, this leaves us with a vertex expansion in terms of the quantum corrections of the vertices  
\begin{align}
	\Gamma_\textrm{qu}^{(0,n)}(\boldsymbol{x}_{n})\,, 
	\label{eq:ApparentConvergenceOrdering1}	 
\end{align}
for all theories. We emphasise that the use of the Einstein-Hilbert action in the above argument was done for the sake of an explicit example. The underlying structure is universal. However, in the vicinity of a non-trivial fixed point such as the UV Reuter fixed point in metric quantum gravity, the ordering in \labelcref{eq:ApparentConvergenceOrdering} has to be accessed carefully as there we have 
\begin{align}
	\Gamma_\textrm{cl}^{(0,n)}(\boldsymbol{x}_{n})\to \Gamma_\textrm{fp}^{(0,n)}(\boldsymbol{x}_{n}) \,, 
	\label{eq:ApparentConvergenceOrderingUV}	 
\end{align}
where $\Gamma_\textrm{fp}^{(0,n)}$ are the expansion coefficients of the fixed-point action or an approximate fixed-point action with a few relevant terms. 

Let us illustrate the above classification with the example of the Litim-Sannino (LS) model \cite{Litim:2014uca, Litim:2015iea, Rischke:2015mea, Bond:2016dvk, Bond:2017tbw}. In its vanilla version, the LS model is a gauge-Yukawa model that contains gauge fields in the SU($N_c$) gauge group, $N_f$ fermions in the fundamental representation, and an uncharged complex scalar field, which forms an $N_f \times N_f$ matrix field. The theory is considered in the Veneziano limit where $N_f$ and $N_c$ are sent to infinity in a constant ratio. The beta function then develops an interacting UV fixed point that is proportional to the Veneziano parameter $\varepsilon = N_f/N_c -11/2$. Choosing an infinitesimal value of $\varepsilon$ allows to have a UV fixed point that is under full perturbative control.

In the LS model, the set of classical vertices is clearly defined: it contains all propagators generated by the kinetic terms and the Yang-Mills action, as well as the classical gluon and gluon-fermion vertices, the Yukawa vertex, and the two four-scalar vertices. It is important to note that higher-order interactions such as the four-Fermi vertex are not vanishing at the UV fixed point, but nonetheless, these vertices are part of $\Gamma_\text{qu}$ since they belong to irrelevant operators at the UV fixed point. While the classification of relevant and irrelevant vertices is straightforward in the LS, the situation in gravity is far more intricate.

We proceed with the discussion of the sub-dominance of higher-order vertices $\Gamma^{(0,n)}_\textrm{qu}$ that is tantamount to extracting the small parameter of the expansion scheme. First, we remark that the correlation functions $\Gamma^{(0,n)}_\textrm{qu}$ are \textit{local} in position space. However, they are \textit{not} point-like (ultralocal). Hence, generic quantum vertices carry the same 
space-time and momentum dependences as the distribution parts 
\begin{align}
	\lambda^{(\textrm{cl,dis})}\to \lambda^{(\textrm{fp,dis})}\,,
	\label{eq:qudis}
\end{align}
of the dressing of the fixed point-parts of the effective action, see \labelcref{eq:point+dis,eq:MomentumLocality1}. This is not surprising as the distribution part belongs to the quantum part of the dressing of a relevant fixed-point tensor structure. Hence, \labelcref{eq:MomentumLocality1} generalises to all quantum dressings, 
\begin{align}
	\left.\lim_{p\to\infty} \lambda^{\textrm{(qu)}}_{0,n}(\boldsymbol{p}_n)\right|_{p_i^2=p^2} =0\,. 
	\label{eq:MomentumLocalityQuantum}
\end{align}
Moreover, we assume, that the dressings do not contain poles in momentum states or close massless resonances. This assumption entails that the dressings of the quantum parts of given vertices have a finite scattering length, 
\begin{align}
	\lim_{\|\boldsymbol{x}_n\|\to \infty} \lambda^{(i)}_{\phi_{a_1}\cdots \phi_{a_n}}(\boldsymbol{x}_n) \leq e^{- \Bigl(\xi^{(i)}_{\phi_{a_1}\cdots \phi_{a_n}} +\epsilon_+\Bigr)\|\boldsymbol{x}_n\|}\,, 
\end{align}
with a small positive $\epsilon_+\to 0$. 

The simplest example of classical and quantum tensor structures is at the hand of the graviton propagator, which contains 5 tensor structures of which 1 is transverse traceless. The Einstein-Hilbert part features two tensor structures, namely the transverse-traceless one and the physical scalar mode. The results in asymptotically safe gravity indicate that these two tensor structures are also that of the fixed-point action, and hence constitute $\Gamma_\text{fp}^{(0,2)}$, as the other potentially relevant directions in the effective action related to the $R^2$ and $R_{\mu\nu}^2$ tensor structure do not generate the additional three tensor structures. Accordingly, the $\Gamma_\text{qu}^{(0,2)}$ are triggered by the gauge fixing: they are only present for ensuring the invertibility of  $\Gamma^{(0,2)}$. Note however, that the dressings of $\Gamma_\text{fp}^{(0,2)}$ also contain quantum parts $\lambda^{(\textrm{tt,dis})}$ and $\lambda^{(\textrm{s,dis})}$  as defined in \labelcref{eq:point+dis,eq:MomentumLocality1,eq:qudis}. Here, 'tt' and 's' stand for the traceless-transverse and scalar part respectively. 

At the next order, the graviton three-point function contains 33 tensor structures of which 7 are completely transverse-traceless \cite{HD-in-prep}. One would need to diagonalise the tensor basis in order to determine how many of these tensor structures belong to $\Gamma_\text{cl}^{(0,3)}$. Moreover, the dressings of the seven transverse-traceless tensor structures are not all independent but are related by diffeomorphism constraints. This has to be taken into account within a relevance analysis, which is briefly discussed in \Cref{sec:StabilityAnalsis}.

\begin{figure}[t]
	\centering
	\includegraphics[width=.5\linewidth]{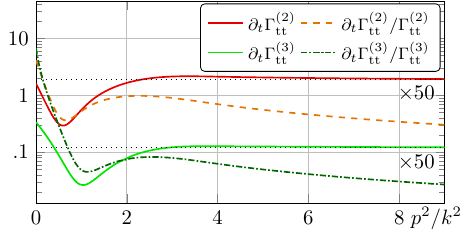}
	\caption{
		Flow of the transverse-traceless graviton two- and three-point function, $\partial_t \Gamma_\text{tt}^{(2)}$ and $\partial_t \Gamma_\text{tt}^{(3)}$, as well as the flows weighted by the $n$-point function itself $\partial_t \Gamma_\text{tt}^{(2)}/\Gamma_\text{tt}^{(2)}$ and $\partial_t \Gamma_\text{tt}^{(3)}/\Gamma_\text{tt}^{(3)}$. Both flows are momentum local. The graviton three-point function is displayed at the momentum symmetric point but the momentum locality holds at general non-singular momentum configurations. Figure adapted from \cite{Christiansen:2015rva}.
	}
	\label{fig:Locality}
\end{figure}

A key property for apparent convergence is \textit{momentum locality}. Momentum locality states that the flow of an $n$-point function decays faster than the $n$-point function itself in the limit of all momenta going to infinity
\begin{align}
	\lim_{p\to \infty}  \frac{|\partial_t \Gamma^{(n)} (\boldsymbol{p}_n)|}{|\Gamma^{(n)} (\boldsymbol{p}_n)| }=0 \,,
	\label{eq:MomLocal}
\end{align}
where the absolute value indicates the projection on some tensor structure. The limit should be understood such that all momenta $p_i$ go to infinity. In quantum gravity, this was first investigated in \cite{Christiansen:2012rx, Christiansen:2015rva} where	\labelcref{eq:MomLocal} was shown to hold for the transverse-traceless part of the graviton two- and three-point function, see also \Cref{fig:Locality}. Remarkably, the property holds for all gauge fixings and all momentum configurations as long as all external momenta are sent to infinity. The property does not hold for the transverse-traceless graviton four-point function and the scalar mode of the graviton \cite{Denz:2016qks, Pawlowski:2020qer, Knorr:2021niv}, which was associated with the overlap of the relevant $R^2$ direction. This intricate situation deserves a more systematic analysis. 

\subsection{Fixed point and critical exponents}
\label{sec:StabilityAnalsis} 
Metric quantum gravity in terms of an asymptotically safe quantum field theory requires an occurrence of a non-trivial UV fixed point, the Reuter fixed point, with a few relevant directions in coupling space $\boldsymbol{g}=(g_1,g_2,....)$, where all couplings are made dimensionless with the appropriate rescaling with powers of the RG-scale $k$. In the fluctuation approach, these couplings are typically expressed in terms of avatars of the cosmological constant, the Newton coupling, see \labelcref{eq:Avatars}, and other couplings. But also fixed points of entire correlation functions \cite{Christiansen:2012rx, Christiansen:2014raa, Christiansen:2015rva, Denz:2016qks, Bonanno:2021squ, Knorr:2021niv} and potentials \cite{Knorr:2017mhu} have been determined. 

In terms of avatars of the Newton coupling, the most extensive study was done in \cite{Eichhorn:2018ydy} where five avatars were considered: the three-graviton vertex $G_{h}$, the graviton-ghost vertex $G_{c}$, the graviton-scalar vertex $G_\varphi$, the graviton-fermion vertex $G_\psi$, and the graviton-gluon vertex $G_A$. All of these avatars were found to have very similar beta functions and fixed-point values,
\begin{align} \label{eq:UVFP-effUni}
	(G_h^* ,\, G_c^* ,\, G_\varphi^* ,\, G_\psi^* ,\, G_A^* )&= (0.58 ,\, 0.55 ,\, 0.74 ,\, 0.74 ,\, 0.84 ) \,. 
\end{align}
This is a remarkable feature since these avatars are by no means identical. Instead, they are related by non-trivial STIs as explained in \Cref{sec:symmetry-identities}. The fact that they display this similarity was called \emph{effective universality} \cite{Eichhorn:2018akn, Eichhorn:2018ydy} and indicates the near-perturbativeness of the interacting UV fixed point. 

In such a system with multiple avatars of one physical coupling, one needs to ask the question of how the mixing of the operators affects the relevant directions at the fixed point. In general, the relevant directions are obtained through the critical exponents, which are the (negative) eigenvalues of the stability matrix 
\begin{align}
	{\cal B}_{ij} = \frac{\partial \beta_i}{\partial g_j}\,,\qquad \textrm{with}\qquad \beta_i =\partial_t g_i\,.
\end{align}
This analysis has been done so far in terms of the couplings extracted from the vertex dressings of the renormalisation-group invariant vertices \labelcref{eq:barGamman} and specifically $\bar \Gamma_\textrm{cl}^{(0,n)}(p)$, evaluated at symmetric points $p$, see \cite{Denz:2016qks} for pure gravity, \cite{Eichhorn:2018ydy} for gravity-matter systems, and \cite{Pastor-Gutierrez:2022nki} for SM interactions.

Let us illustrate the general structure of a stability analysis in the presence of avatars within the simple toy example of a system with two couplings, $\boldsymbol{g}=(g_1, g_2)$, that resemble avatars of the Newton coupling. These couplings have the $\beta$-functions 
\begin{align} 
	\label{eq:beta-toy-example}
	\beta_1(\boldsymbol{g})&= 2 g_1 - (4 +2 \delta_1) g_1^2  +2 \delta_1 g_1 g_2\,, \notag \\
	\beta_2(\boldsymbol{g})& = 2 g_2 -  (4 + 2\delta_2) g_2^2 + 2\delta_2 g_1 g_2\,.
\end{align} 
Upon the identification $g_1=g_2$, both beta functions reduce to $\beta = 2g - 4 g^2$ with the simple fixed point $g^* =1/2$ and the critical exponent $\theta=-2$. In the combined system, the fixed points are still given by $\boldsymbol{g}^*=(1/2,1/2)$ and the stability matrix reads 
\begin{align}
	{\cal B}(\boldsymbol{g}^*)=\left(\begin{array}{cc} - 2 + \delta_1 & -\delta_1 \\ -\delta_2 & -2 + \delta_2 \end{array}\right)\,.
\end{align}
The eigenvalues are
\begin{align}
	\theta_{1,2} = (-2,\, -2 +\delta_1+\delta_2)\,.
\end{align}
The first eigenvalue is precisely the physical eigenvalue that was also found in the identified beta functions. The second eigenvalue is shifted by the parameters $\delta_1$ and $\delta_2$, and might generate a second spurious relevant direction. This scenario indeed closely resembles the situation in \cite{Eichhorn:2018ydy} where five avatars of the Newton coupling were investigated and six relevant directions were found at the fixed point. Presumably, five of these directions can be unified into one physically relevant direction after the usage of appropriate symmetry identities (mSTIs). Or put differently, the relevant directions can be used to fulfil the symmetry constraint by the mSTI.

\subsection{Relation to other approaches}
\label{sec:CombinedUse} 
The fluctuation approach within an expansion about the flat background is tantamount to an expansion about vanishing curvature and in particular a vanishing Ricci scalar $R=0$. As discussed before at the beginning of \Cref{sec:ExpansionSchemes}, this expansion scheme is explicit or implicit to most applications within the background approximation: few works go beyond the second order in curvature invariants and in those works one sees a rather quick convergence of the results in terms of the higher powers in the curvature scalar $R^n$ \cite{Falls:2013bv, Falls:2014tra, Falls:2016wsa, Falls:2018ylp}, or higher powers build out of the Riemann tensor, e.g., $(R_{\mu\nu\rho\sigma} R^{\mu\nu\rho\sigma})^n$ and $R(R_{\mu\nu\rho\sigma} R^{\mu\nu\rho\sigma})^n$ at even and odd orders of the curvature expansion \cite{Falls:2017lst, Kluth:2020bdv, Kluth:2022vnq}. In most of these computations the first three orders, e.g.~$R^0,R^1,R^2$, have an overlap with UV-relevant operators \cite{Falls:2020qhj, Knorr:2021slg}. This is in reassuring agreement with the findings in the fluctuation approach in pure quantum gravity, \cite{Christiansen:2012rx, Christiansen:2014raa, Christiansen:2015rva, Denz:2016qks, Christiansen:2017bsy, Burger:2019upn}.  In some works, a fourth relevant direction was found \cite{Kluth:2020bdv, Kluth:2022vnq} that is a combination of other higher-order curvature invariants. These findings in the Taylor expansion are sustained within computations with a full $f(R)$ potential. In the background field approximation this has been done in \cite{Dietz:2013sba, Benedetti:2013jk, Falls:2014tra, Demmel:2015oqa, Eichhorn:2015bna, Falls:2016wsa, Falls:2018ylp, Falls:2017lst, Kluth:2020bdv, Kluth:2022vnq, Gonzalez-Martin:2017gza, Mitchell:2021qjr, Morris:2022btf}, in the fluctuation approach with the additional inclusion of momentum dependences this has been studied in \cite{Christiansen:2017bsy, Burger:2019upn} and within a double expansion in momenta and background curvature in \cite{Knorr:2017fus}.

In summary, both schemes support and are based on an expansion in powers of curvature invariants. However, in direct comparison, the background approximation relies on yet another approximation, namely \labelcref{eq:BackgroundApprox}. This can be also clearly seen in the leading order of the curvature expansion: we can apply the momentum projection scheme used in the fluctuation approach for projecting on the running of the cosmological constant and the curvature scalar in the background approximation, leading to the same results as the heat-kernel techniques commonly used, see also \cite{Pawlowski:2020qer}. For higher-order curvature invariants the application of the momentum projection scheme requires the construction of a full tensor basis of local invariants including invariants that involve covariant derivatives. 

Of the approaches utilising the background-field approximation, the form-factor approach \cite{Bosma:2019aiu, Knorr:2019atm, Draper:2020bop, Draper:2020knh, Knorr:2022lzn, Knorr:2022dsx} is most closely related to fluctuation approach. Both approaches prioritise full momentum dependences. The form-factor approach expands in order of curvatures (not derivatives) and the fluctuation approach in the orders of the correlation functions. There are mappings between these expansions which are detailed below. The most significant difference is, that the form-factor approach utilises the background-field approximation. At finite $k$, the breaking of diffeomorphism invariance (or rather the BRST symmetry) by the regulator makes the comparison difficult since the background field approximation is at odds with the diffeomorphism symmetry as explained in \Cref{sec:WrapUpSym}. At vanishing cutoff scale $k\to0$, the relation to the form factor approach is more straightforward: The fluctuation approach uses a boundary condition that satisfies the STIs, and therefore the correlation functions can be mapped to a diffeomorphism or rather BRST invariant action. In terms of form factors, this action is written as
\begin{align}
	\Gamma[\bar g, h] = \int_x \left( \frac{2\Lambda - R}{16 \pi G_\text{N}} + R f_R(\Box) R + C_{\mu\nu\rho\sigma} f_C(\Box) C^{\mu\nu\rho\sigma} +\dots  \right) + S_\text{gf} + S_\text{gh}\,. 
\end{align}
Using the fluctuation approach, several correlation functions have been computed at vanishing cutoff scales, including the propagator \cite{Bonanno:2021squ, Fehre:2021eob} and the three-graviton vertex \cite{Bonanno:2021squ}. These can be matched in a one-to-one fashion to the form factors assuming that the diffeomorphism constraints from the STI, see \labelcref{eq:STI}, have been implemented as boundary conditions. For example, in an expansion about the flat background $g_{\mu\nu}=\eta_{\mu\nu} + \sqrt{G_\text{N}}\, h_{\mu\nu}$, the form factor $f_C$ is directly related to the wave-function renormalisation via the transverse-traceless graviton propagator
\begin{align}
	G_{h_{tt}h_{tt}} = \frac{32\pi}{Z_{h_\text{tt}}(p) p^2 } =  \frac{32 \pi}{p^2 + 32 \pi G_\text{N} p^4 f_C(p^2)} \,,
\end{align}
where we have used that on-shell the cosmological constant is vanishing on the flat background. In straight analogy, the form factor $f_R$ is expressed through the wave-function renormalisation of the physical scalar mode via its propagator
\begin{align}
	G_{h_{s}h_{s}} = \frac{-16 \pi}{  Z_{h_\text{s}}(p) p^2 } = \frac{-16 \pi}{p^2 - 96\pi G_\text{N} p^4 f_R(p^2)} \,.
\end{align}
The scalar mode is negative as expected. This does not imply that gravity has a ghost mode: the negative massless excitation is non-propagating since it cancels out with parts of the transverse-traceless mode. We also emphasise that the wave-function renormalisations $Z_{h_\text{tt}}$ and $Z_{h_\text{s}}$ produce form factors $f_R$ and $f_C$ that agree exactly with one-loop computations for small values of $p^2$ \cite{Bonanno:2021squ, Capper:1979ej}. Specifically, the prefactors of the $\log(p^2)$ terms in $f_R$ and $f_C$ are scheme independent but gauge dependent and these prefactors match exactly, see also \Cref{sec:LorentzianGravity}. The scheme independence of the contributions is seen in the fRG computation through the fact that they are related to a $p^4$ derivative, which results in dimensionless flows.

The transverse-traceless three-graviton vertex overlaps with the cosmological constant, the Newton coupling, and the $f_C$ form factor. These are however already fixed by lower-order correlation functions. The new overlap is with the Goroff-Sagnotti term and associated form factor
\begin{align}
	\Gamma_\text{GS}[\bar g, h] = \int_x  f_{C^3}(\nabla_1, \nabla_2, \nabla_3) C_{\mu\nu\rho\sigma} C^{\rho\sigma\kappa\lambda} {C_{\kappa\lambda}}^{\mu\nu} \,,
\end{align}
where $\nabla_i$ only acts on the $i$th Weyl tensor. The transverse-traceless graviton three-point function was so far computed at the momentum-symmetric point at $k=0$ \cite{Bonanno:2021squ}, which fixes the form factor $f_{C^3}$ at the momentum-symmetric point. This mapping can be systematically extended, see also \cite{Denz:2016qks} for an extensive discussion of overlaps between correlation functions and curvature operators.  

This concludes our discussion of the fundamentals of the fluctuation approach. We turn now to recently obtained results, and refer to \cite{Pawlowski:2020qer} for a more detailed overview.

\section{Momentum dependent correlation functions}
\label{sec:pure-grav}

In the present section, we discuss the computation of momentum-dependent correlation functions, which constitute the backbone of the fluctuation approach. In the past decade, significant progress was achieved and momentum dependencies up to the graviton four-point function, at vanishing cutoff scales, and even on Lorentzian backgrounds were computed, see \cite{Christiansen:2012rx, Christiansen:2014raa, Christiansen:2015rva, Denz:2016qks, Christiansen:2017cxa, Eichhorn:2018akn, Eichhorn:2018nda, Eichhorn:2018ydy, Knorr:2021niv,Bonanno:2021squ, Fehre:2021eob}.

\subsection{Momentum dependence, cutoff dependence, and diffeomorphism invariance}
Physical correlation functions are obtained at vanishing cutoff scale, $k\to0$, where all quantum fluctuations have been integrated out. At finite cutoff scales, the non-vanishing regulator may diffuse or even spoil important physics properties: while in many cases the cutoff dependence can be interpreted as an average momentum dependence at the symmetric point $p^2=k^2$, this analogy is sometimes deceiving. 

A prominent example in asymptotically safe gravity for the latter is the interpretation of the constant part of the cutoff-dependent effective action as the cosmological constant, which evidently does not have a momentum dependence, for a detailed discussion see \cite{Denz:2016qks}. Here we just mention that the cosmological constant is commonly defined by the $\int \!\mathrm d^4 x\,\sqrt{\bar g}$ part of the flow. However, this part of the flow is simply the normalisation of the functional integral of metric gravity just as the constant part in any ordinary quantum field theory. Hence, it does not relate to a physics observable even at $k=0$ in contrast to the flow of $\int \!\mathrm d^4 x\,\sqrt{g}$. In turn, the mass parameter of the graviton indeed is physical at $k=0$, and is not directly related to the normalisation of the functional integral. Moreover, the respective Nielsen identity derived from \labelcref{eq:linNielsen}, that relates these two parameters, has a dominant regulator dependence at finite cutoff scales.

It is therefore of paramount importance to disentangle momentum and RG scale dependence by integrating correlation functions from the UV fixed point to the IR while keeping full momentum dependence. The UV fixed point defines the starting point of the theory and the relevant perturbations around this fixed point define the trajectory along which quantum fluctuations are integrated out. In \cite{Denz:2016qks}, the system of coupled $n$-point functions has been solved up to the graviton four-point function. All $n$-point functions were evaluated at momentum-symmetric points with external transverse-traceless projections. A UV fixed point was found at
\begin{align} 
	\label{eq:UV-FP} 
	\left( \mu^*, \lambda_{3}^*, \lambda_{4}^*, g_{3}^*, g_{4}^* \right) 
	&= \left( -0.45,\, 0.12,\,0.028,\, 0.83,\, 0.57 \right) \,,
\end{align}
where $g_n$ and $\lambda_n$ are the dimensionless avatars of the Newton coupling and the cosmological constant, the latter corresponding momentum independent part of the graviton $n$-point function, for more details see \cite{Denz:2016qks}. The graviton mass parameter $\mu = - 2 \lambda_2$ is the momentum-independent part of the graviton two-point function. The critical exponents of the fixed point are given by
\begin{align}
	\theta_i &= (4.7,\, 2.0 \pm 3.1 i,\, -2.9,\, -8.0 ) \,,
	\label{eq:crit-exp}
\end{align}
where a positive sign corresponds to a UV-attractive direction. The three UV-attractive directions were associated with the operators or rather tensor structures $\sqrt{g}$, $\sqrt{g}R$, and $\sqrt{g}R^2$. In contrast, the operator $\sqrt{g}R_{\mu\nu}^2$ is not generated in the present approximation from the $\sqrt{g},\sqrt{g}R$ tensor structures. The latter property was inferred from the momentum dependence of the graviton three- and four-point functions.

From this UV fixed point, there is a three-dimensional critical hypersurface that leads to the IR where one wants to match general relativity with the Einstein-Hilbert action and hence all avatars of the Newton coupling and the cosmological constant agree,  
\begin{align}
	G_i = \bar G_i=G\,, \qquad \Lambda_i = \bar \Lambda_i=\Lambda\,,\qquad \forall i\in\mathbbm{N}\,. 
	\label{eq:GenRelativity}	
\end{align}
\Cref{eq:GenRelativity} should be obtained by first running to $k=0$ and then setting $p=0$. Evidently, \labelcref{eq:GenRelativity} satisfies the NI \labelcref{eq:linNielsen} and the STI \labelcref{eq:STI}, and hence any trajectory leading to \labelcref{eq:GenRelativity} 	 constitutes a diffeomorphism-invariant theory. Note that \labelcref{eq:GenRelativity} fixes two of the relevant parameters of the theory $G,\Lambda$ in terms of an IR renormalisation condition for the fluctuation couplings, whose flows provide the dynamical closed set of flow equations of the theory, see the discussion in \Cref{sec:flowCorFuncs} and specifically \labelcref{eq:lambdaFluctuationHierarchy}. The third relevant direction in coupling space is related to the $R^2$ term, and its IR value can be set to zero on the level of the fluctuation couplings with a further finetuning condition that concerns the $p^4$-part of the vertices.   

\begin{figure}[t]
	\centering
	\includegraphics[width=.5\linewidth]{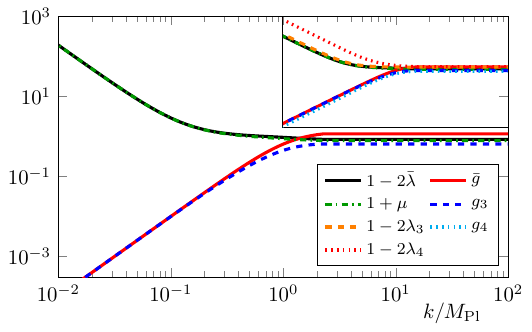}
	\caption{
		UV-IR trajectory from the UV fixed point \labelcref{eq:UV-FP} to the IR with general relativity. Here, the Newton couplings are fixed to their classical value and the cosmological constant is finite and negative. Other values of the cosmological constant can be chosen as well. The results are taken from \cite{Denz:2016qks}.
	}
	\label{fig:IR-trajectory}
\end{figure}

We emphasise that \labelcref{eq:GenRelativity} implies a fine-tuning condition of order  $n+m$ if $n$ \textit{fluctuation} avatars of $G$ and $m$ \textit{fluctuation} avatars of $\Lambda$ are considered in a given approximation: the initial conditions for these $n+m$ coupling parameters have to be tuned such that \labelcref{eq:GenRelativity} holds at $k=0$ (and $p=0$). If \labelcref{eq:GenRelativity} is obtained, the IR part of the full effective action (low momentum scales at $k=0$) is simply the Einstein-Hilbert action. Accordingly, solving these fine-tuning conditions at $k=0$ and $p=0$ is tantamount to solving the modified symmetry identities \labelcref{eq:GenRelativity,eq:linNielsen} in the given approximation. In turn, if \labelcref{eq:GenRelativity}	is not satisfied and other couplings and operators are absent in the IR, the theory is not diffeomorphism-invariant. A more detailed discussion can be found in \cite{Pawlowski:2020qer}. 

In \cite{Denz:2016qks}, many physically interesting UV-IR trajectories were found, including the example displayed in \Cref{fig:IR-trajectory}. On that trajectory, the Newton coupling $G(p)$ matches the physical value in the IR and the cosmological constant runs to a finite negative value. The fine-tuning condition for diffeomorphism invariance has been solved for the two avatars of the Newton coupling present in the approximation: $G=G_3\approx G_4$ and for two avatars of the cosmological constant, $\Lambda_2 \approx \Lambda_3$. The coupling $\lambda_4$ was not fine-tuned as the higher $\lambda_n$ are increasingly irrelevant and the fine-tuning effort increases substantially with the dimension of the fine-tuning manifold. 

With the diffeomorphism-consistent solution of the fluctuation system, the background couplings of the diffeomorphism invariant background effective action can be computed. In \Cref{fig:IR-trajectory}, the UV-IR trajectories of the background cosmological constant and background Newton coupling with $G_3=G$ and $\bar\Lambda= \Lambda$ are displayed. Such a choice can always be achieved easily since the background couplings are non-dynamical and higher-order background couplings do not feed back into the flow of the lower-order ones, see the discussion in \Cref{sec:flowCorFuncs} and specifically \labelcref{eq:lambdaHierarchy}. Hence there is no fine-tuning problem, whose difficulty is increasing with the dimension of the coupling manifold. 

This concludes our discussion of how to practically use the fluctuation approach for solving the flows of diffeomorphism invariant metric quantum gravity within given approximations. We add a few further technical remarks on the coupling choices, the Nielsen (NI) and Slavnov-Taylor (STI) identities in the limit $k\to 0$ and apparent convergence as discussed in \Cref{sec:ApparentConvergence}: \\[-1.5ex]

\emph{Choice of $\Lambda$:} The choice $\Lambda<0$ has been taken in order to avoid discussing the subtleties that come with the definition of the effective action as a Legendre transform: for non-convex classical action, the classical effective action is the convex hull of the classical action. In \cite{Denz:2016qks} also other values of the cosmological constant, including vanishing and positive values have been considered. \\[-2ex]

\emph{Physical limit $k\to 0$:}  In general the limits $k\to 0$ and $p\to 0$ do not commute, and the limit $k\to 0$ of $G_k(p=0)$ is in the regularised regime with non-trivial STIs and NIs even for $k=0$. In \cite{Denz:2016qks} we still have chosen this order of limits as the couplings have been obtained by taking into account the full momentum dependence in the regime $0\leq p^2 \lesssim k^2$. Hence we expect subleading effects of the non-trivial symmetry identities to be present which have been ignored there. \\[-2ex]

\emph{Apparent convergence:} Finally, it has been studied in \cite{Denz:2016qks}, how the results change if higher order couplings such as $g_4$ and $\lambda_4$ are computed instead of fixed to their Einstein-Hilbert counterparts $\lambda^{(\textrm{EH})}(p)$ as discussed in \Cref{sec:ApparentConvergence}. We have interpreted the respective minimal changes of the lower-order avatars of the Newton coupling and cosmological constant as the onset of apparent convergence. These promising signatures award further studies in more advanced approximations. \\[-1.5ex]

\begin{figure}[t]
	\includegraphics[width=.48\linewidth]{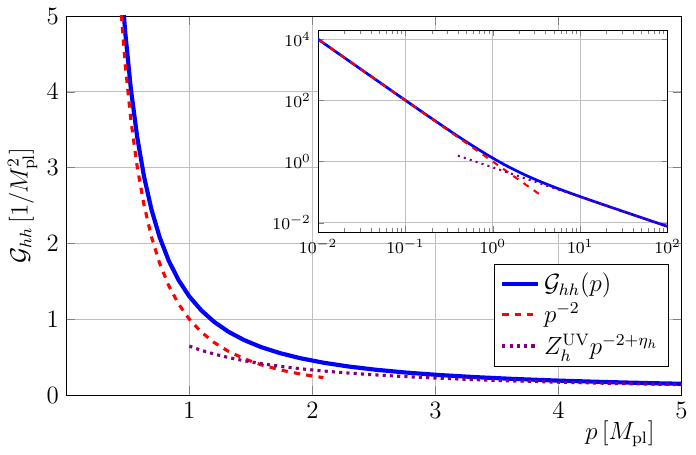}
	\hfill
	\includegraphics[width=.48\linewidth]{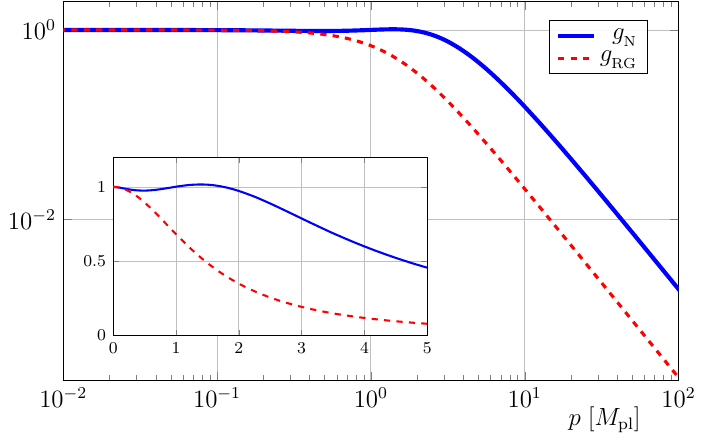}
	\caption{We show the physical momentum dependence of the propagator (left panel) and the Newton coupling (right panel). The propagator scales with $1/p^2$ in the IR and with $1/p^{2-\eta_h^*}$ in the UV where $\eta_h^*$ is the fixed-point anomalous dimension. The crossover regime is around the Planck scale. In the right panel, we show a comparison between the physical Newton coupling, $g_\text{N}(p) = G_\text{N}(p) M_\text{pl}^2$, and the RG improved Newton coupling, $g^{\ }_\text{RG}= g_{k=p}$, both in units of the Planck mass. While the functions agree qualitatively, the physical Newton coupling is needed for a quantitative understanding. Figure adapted from \cite{Bonanno:2021squ}.}
	\label{fig:Newton-coupling}
\end{figure}

The results in \cite{Denz:2016qks} obtained by the fine-tuning procedure described above include momentum-dependent correlation functions and hence form factors in disguise. On a given UV-IR trajectory, the momentum-dependent correlation functions or rather their momentum-dependent flows can be integrated, resulting in momentum-dependent physical correlation functions at $k=0$. This integration on the basis of the momentum-dependent flows in \cite{Denz:2016qks} was performed in \cite{Bonanno:2021squ} for the transverse-traceless propagator mode and the transverse-traceless three-graviton vertex at the symmetric momentum point. There, a UV-IR trajectory with a vanishing IR cosmological constant was chosen. The results are displayed in \Cref{fig:Newton-coupling}. The left panel shows the momentum dependence of the propagator in units of the Planck scale $M_\textrm{pl}$. Below the Planck scale, the propagator scales with $1/p^2$ and has a sub-leading logarithmic contribution that matches exactly the effective field theory results \cite{Bonanno:2021squ, Capper:1979ej}. This is discussed in more detail in \Cref{sec:LorentzianGravity}. There is a crossover regime around the Planck scale, and in the trans-Planckian regime for $p/M_\textrm{pl}\to\infty$ the propagator scales with $1/p^{2-\eta_h^*}$, where $\eta_h^*$ is the fixed-point anomalous dimension. In this truncation, the fixed-point anomalous dimension was found to be $\eta_h^* \approx 1.03$ implying that the fluctuation graviton propagator scales approximately with $1/p$ in the UV. This propagator was analytically continued and used to reconstruct the graviton spectral function from the displayed Euclidean data. More details are provided in \Cref{sec:LorentzianGravity}, where also results of the direct Lorentzian computations are presented. 

The right panel in \Cref{fig:Newton-coupling} shows the momentum dependence of the physical Newton coupling at vanishing cutoff, $g_\text{N}(p) = G_\text{N}(p) M_\text{pl}^2$, which was extracted from the momentum dependence of the graviton three-point function. It is compared to the RG improved Newton coupling where the RG scale is identified with the momentum, $g^{\ }_\text{RG}= g_{k=p}$. In the IR, both couplings are constant and agree exactly as they are both fixed to the classical value. Beyond the Planck scale, both couplings scale with $1/p^2$ which implies that their dimensionless counterpart takes the constant fixed-point value. Hence both couplings have qualitatively the same features while they quantitatively clearly differ. These results support that RG improvement can be used qualitatively, at least in a system with only one physical scale. The only physical scale here is the momentum of the vertex at the symmetric point. It is expected that the RG improvement works less well with multiple scales, e.g., away from the momentum symmetric point. 

\begin{figure}[b]
	\centering 
	\includegraphics[width=0.22\linewidth]{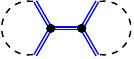}
	\caption{Tree-level 2-to-2 graviton scattering diagram in the tt-channel. The dashed lines indicate the transverse-traceless contraction of the external legs, see \labelcref{eq:Msym-channel}. Figure adapted from \cite{Bonanno:2021squ}.}
	\label{fig:graviton_scattering}
\end{figure}

\subsection{Scattering amplitudes}

We close this section with a discussion on the relation of the results on vertex dressings and the respective RG-invariant couplings or form factors to matrix elements of the $S$-matrix and the respective form factors of the $S$-matrix. To that end, we consider 2-to-2 graviton scattering in the tt-channel and explain its relation to the momentum-dependent Newton coupling $G_\text{N}(p)$. The lowest order tree-level diagram for this matrix element is depicted in \Cref{fig:graviton_scattering}, where the dashed lines indicate the transverse-traceless contraction of the external legs that is done with the tt-projection operator. The transition amplitude ${\cal M}_{gg\to gg}$ of this process is nothing but 
\begin{align}
	{\cal M}_{gg\to gg} &\simeq p^2 G_{h_bh_a}^{(tt)}(p) \left[ \Gamma^{(h_a h_ b h_c)}(p) G_{h_c h_d }(p)\Gamma^{(h_d h_e h_f )}(p)\right]  p^2 G^{(tt)}_{h_f h_e}(p)\,,
	\label{eq:Msym-channel}
\end{align}
where $a,...,f$ in \labelcref{eq:Msym-channel} are the indices of second rank Lorentz tensors, for example, $a=\mu\nu$, and hence all Lorentz indices in  \labelcref{eq:Msym-channel} are contracted. We remark that in \labelcref{eq:Msym-channel} we evaluated the transition amplitude at the symmetric point $p_i^2=p^2$ for a general $p$ for the sake of simplicity, and hence the external gravitons are not on-shell. The external dressings $p^2 G_{hh}$ accommodate the $tt$-projection as well as the wave function factors $1/Z_h^{1/2}(p_i^2)$ in the LSZ formula. At its core and in the approximation used here, \labelcref{eq:Msym-channel} is nothing but the renormalisation group invariant three-graviton vertex squared, $\bar\Gamma_{hhh}^{(3)}$ defined in \labelcref{eq:barGamman}. This leads us to 
\begin{align}
	{\cal M}_{gg\to gg} &\propto p^2 G_3(p)\,,  
	\label{eq:Msym-channel-G3}
\end{align}
which entails that the matrix element goes to a constant for large momenta and that the form factor of the transition amplitude is nothing but the form factor squared of the three-graviton vertex. Note that strictly speaking such a 2-to-2 graviton scattering process is not physical, as a state with $n$-gravitons is not gauge independent. In any case, \labelcref{eq:Msym-channel-G3} or similar scattering processes for (physical) matter fields which is transmitted by gravitons have to be evaluated on-shell. Then the external momenta 
$(p^{(\textrm{ext})}_i)^2=m_i^2$, where $m_i$ are the pole masses of the respective particles or fields. Then, the process \Cref{fig:graviton_scattering} constitutes an $s$-channel scattering process with $s=p^2$ and the coupling $G(p)$ in  \labelcref{eq:Msym-channel-G3}  generalises to a renormalisation group invariant form factor $\bar \lambda_{hhh}(p_1,p_2,p)\bar \lambda_{hhh}(-p,-p_3,-p_4)$, where all momenta are counted incoming, see also the discussion in \cite{Bonanno:2020bil}.

\subsection{Outlook}
\label{sec:pure-gravoutlook}

The results on momentum-dependent correlation functions reviewed in this section allow for the resolution of asymptotically safe gravity and asymptotically safe gravity-matter systems in a systematic expansion scheme. This scheme is controlled by  apparent convergence and allows for the resolution of the underlying Slavnov-Taylor identities at vanishing cutoff scales which guarantee physical diffeomorphism invariance.

\section{The asymptotically safe Standard Model}
\label{sec:ASSM}

The results obtained in the fluctuation approach for pure gravity, including that on apparent convergence discussed in \Cref{sec:pure-grav}, can be put to work in a general gravity-matter system. This line of study was initiated in \cite{Meibohm:2015twa} and by now includes the study of the fixed-point structure of general gravity-matter theories with scalars, fermions, and gauge fields. The general interplay of gravity and matter in the background field approximation, hybrid approaches and the fluctuation approach including studies beyond the minimal coupling approximation, see \cite{Dou:1997fg, Percacci:2002ie, Granda:2005nd, Granda:2005py, Dona:2012am, Biemans:2017zca, Alkofer:2018fxj, Alkofer:2018baq, Gies:2018jnv, Dona:2013qba, Dona:2015tnf, Meibohm:2015twa, Christiansen:2017cxa, Eichhorn:2018akn, Eichhorn:2018ydy, Eichhorn:2018nda, Burger:2019upn, Percacci:2003jz, Narain:2009fy, Zanusso:2009bs, Vacca:2010mj, Daum:2010bc, Folkerts:2011jz, Harst:2011zx, Eichhorn:2011pc, Eichhorn:2012va, Henz:2013oxa, Percacci:2015wwa, Labus:2015ska, Oda:2015sma, Eichhorn:2016esv, Henz:2016aoh, Eichhorn:2016vvy, Meibohm:2016mkp, Christiansen:2017gtg, Hamada:2017rvn, Eichhorn:2017sok, Christiansen:2017qca, Eichhorn:2017lry, Pawlowski:2018ixd, deBrito:2019epw, Held:2020kze, Daas:2020dyo, Eichhorn:2019dhg, Eichhorn:2017muy, Eichhorn:2021qet, deBrito:2021akp, deBrito:2021pyi}, was extensively reviewed in \cite{Eichhorn:2022gku}.

\subsection{Fundamentals of the asymptotically safe Standard Model}
In the present section, we focus on the recent milestone of full non-perturbative SM flows including quantum gravity within the fluctuation approach achieved in~\cite{Pastor-Gutierrez:2022nki}. There, the SM was embedded in asymptotically safe gravity within the advanced fluctuation approximation discussed in  \Cref{sec:pure-grav}. In \cite{Pastor-Gutierrez:2022nki} the flows of the wave functions \labelcref{eq:WaveASSM} of all SM fields and the graviton as well as the flows of the primitively divergent couplings \labelcref{eq:ApproxShort2} have been considered. The field content of the Asymptotically Safe Standard Model (ASSM) is given by  
\begin{align}	\label{eq:SuperfieldGravMat} 
	\phi_\textrm{grav}=&\,(h_{\mu\nu}, c_\mu,\bar c_\mu)\,, 
	&
	\phi_\textrm{SM}=&\,({\cal A}_\mu,\, {\cal C},\, \bar {\cal C}, \,l, \,\bar l,\, q,\,\bar q,\, \Phi)\,. 
\end{align}
The matter superfield $\phi_\textrm{SM}$ comprises the gauge and ghost fields of the SM gauge group  U(1)${}_\textrm{Y}\,\times\,$SU(2)${}_\textrm{L}\,\times\,$SU(3)${}_\textrm{C}$, 
\begin{align} \label{eq:SuperfieldGauge}
	{\cal A}_\mu&=( B_\mu,\, A_\mu^a,\, G_\mu^b)\,,
	&
	{\cal C}&= (c^a,\, c^b)\,, 
\end{align}
with the hypercharge gauge field $B_\mu$, the weak gauge fields $A_\mu^a$ with $a=1,2,3$, and the gluons $G_\mu^b$ with $b=1,...,8$. The field $\cal C$ contains the respective ghost fields of the weak and strong gauge groups. $\phi_\textrm{SM}$ also contains the three families of quarks $q$ and leptons $l$, 
\begin{align} \label{eq:SuperfieldFerms}
	q&=(d,\,u,\,s,\,c,\,b,\,t)\,,
	&
	l&= (e,\,\nu_e, \,\mu,\, \nu_\mu,\, \tau,\,\nu_\tau)\,.
\end{align}
The scalar field $\Phi$ is the Higgs doublet, which is parametrised with
\begin{align}\label{eq:Higgs}
	\Phi&= \frac{1}{\sqrt{2}}\begin{pmatrix} \mathcal{G}_1 +i \mathcal{G}_2 \\  v +H  +i \mathcal{G}_3  \end{pmatrix} \,, 
	&
	\rho &= \tr \Phi^\dagger\Phi\,, 
\end{align}
where $v$ is the flowing minimum, and $H$ is the (radial) fluctuation Higgs field with a vanishing expectation value $\langle H\rangle =0$, as the latter is explicitly carried by $v$.  The ${\cal G}_i$ with $i=1,2,3$ are the Goldstone modes.

We briefly discuss the approximation to the full effective action of the ASSM in terms of the flows taken into account in \cite{Pastor-Gutierrez:2022nki}. The flows have been computed self-consistently: all flowing parameters computed have been fed back into the flows. First of all, all fields have to be augmented with their cutoff-dependent wave functions, 
\begin{align}
	\label{eq:WaveASSM}
	Z_{\phi,k}=(Z_{{\cal A},k}\,,\, Z_{{\cal C},k}\,,\, Z_{l,k}\,,\, Z_{q,k}\,,\, Z_{\Phi,k}\,,\, Z_{h,k}\,,\, Z_{c,k})\,.
\end{align}
and the respective flows have been solved. Moreover, the flow of all primitively divergent couplings was considered, 
\begin{align}
	\vec \lambda_k=( 
	g_{1,k} \,,\,
	g_{2,k}\,,\,
	g_{3,k}\,,\,
	y_{q,k}\,,\,
	y_{l,k}\,,\,
	\lambda_{\Phi,k}\,,\,
	G_{k}\,,\,
	\Lambda_{k})\,. 
	\label{eq:ApproxShort2}\end{align}
These couplings have been obtained from the RG-invariant dressings \labelcref{eq:barlambda} by dividing the vertex dressings with the respective powers of the wave functions in \labelcref{eq:WaveASSM}. 

The $g_i$ with $i=1,2,3$ are related to the hypercharge gauge coupling with $g_1 \equiv \sqrt{5/3}\, g_Y $, the weak gauge coupling $g_2$, and the strong gauge coupling $g_3$. The couplings $y_q$ and $y_l$ are the Yukawa couplings to the quarks and leptons, and $\lambda_\Phi$ is the quartic interaction of the Higgs doublet. In the Higgs sector, also higher-order couplings were considered: the Higgs potential was expanded in a high-order Taylor expansion around the flowing minimum. On the gravity side, the running of the Newton coupling $G$ and the cosmological constant $\Lambda$ were taken into account.

The couplings in \labelcref{eq:ApproxShort2} have different avatars depending on the correlation function from which the flow was extracted. In \cite{Pastor-Gutierrez:2022nki}, all gauge couplings were extracted from the fermion-fermion--gauge vertex, and the Newton coupling from the three-graviton vertex, at the momentum symmetric point. For more details, we refer to \cite{Pastor-Gutierrez:2022nki}.

We proceed with the discussion of several points that need to be handled with care in order to obtain full SM flows if aiming for quantitative precision required for even a qualitative resolution of the phase structure of the ASSM:\\[-1.5ex]

\emph{Strong IR QCD sector:} In the IR regime for $k, p\lesssim 10$\,GeV, one needs an improved approximation for the gluonic sector for including confinement, as well as in the matter sector for including dynamical spontaneous chiral symmetry breaking, see \cite{Fu:2019hdw}. Therefore, results from functional approaches for 2 and 2+1 flavour computations were utilised that match lattice 2 and 2+1 flavour benchmarks in the IR regime of physical QCD \cite{Cyrol:2017ewj, Fu:2019hdw, Gao:2021wun}. With this external input, the computation is able to effectively describe the mass generation of quarks due to chiral symmetry breaking.\\[-2ex]

\emph{Matching of the top pole mass:} For most Yukawa couplings it is sufficient to fix their values in the IR with the Euclidean curvature mass, e.g., with the simple relation $y_{b,k=0}= \sqrt{2} M_{b,\text{pole}}/v$ for the bottom quark. For the top quark, the situation is different since the SM flows are very sensitive to the top mass. In \cite{Pastor-Gutierrez:2022nki}, the non-trivial relation between the pole and the Euclidean curvature mass of the top quark was taken into account by determining the pole in the complex momentum plane of the two-point function for a given Yukawa coupling. This leads to the relation
\begin{align}\label{eq:topMassCurvature}
	m_{t}&=165.4^{+0.9}_{-0.2}\,\textrm{GeV}
	&&\leftrightarrow & 
	y_{t}&=0.950^{+0.005}_{-0.001}\, . 
\end{align}
Determining the pole in the complex plane additionally provides a prediction for the decay width of the top quark, which is in agreement with experiments \cite{Pastor-Gutierrez:2022nki}. \\[-2ex]

\emph{Strong UV gravity sector:} Since the strong UV gravity sector has several important points we will address that in the next subsection. At first we will discuss the impact of matter fluctuations on the UV fixed point, which is key for the existence of a combined SM-gravity fixed point. Then we will look at the impact of gravity fluctuations on the matter sector. The matter sector in the SM contains UV Landau poles at high-energy scales which need to be avoided by gravity fluctuations.

\subsection{Gravity-matter interplay}
\label{sec:Gravity-Matter-Interplay}

We initiate our discussion of the phase structure of the ASSM with that of the impact of the matter fluctuations on the fixed point of the gravity couplings \labelcref{eq:UV-FP}. For simplicity, we consider at first only \textit{minimally coupled}  matter fields: matter fields where the coupling to gravity only comes via the kinetic term and no matter self-interactions are considered. Starting from the pure gravity fixed point \labelcref{eq:UV-FP}, one adds successively scalars, fermions, and gauge bosons. The current state-of-the-art computations can be found in \cite{Eichhorn:2018akn} (scalars), \cite{Eichhorn:2018nda} (fermions), and \cite{Christiansen:2017cxa} (gauge bosons), where the momentum dependent running of all two-point functions as well as the running of the three-graviton vertex and the graviton-matter vertex was taken into account. Therefore these systems contain several momentum-dependent anomalous dimensions: $\eta_h(p^2),\, \eta_c(p^2), \eta_\text{mat}(p^2)$, two avatars of the cosmological constant: the graviton mass parameter $\mu$ and the avatar from the three graviton vertex $\lambda_3$, as well as two avatars of the Newton coupling: $g_3$ from the three-graviton vertex and $g_\text{mat}$ from the matter-graviton vertex.

\begin{figure}[t]
	\includegraphics[width=\linewidth]{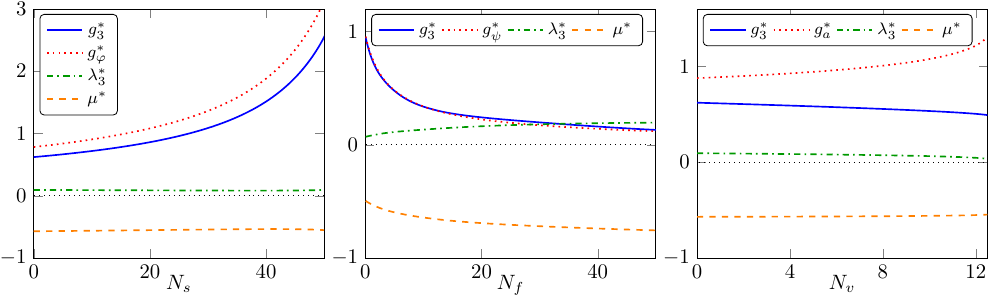}
	\caption{
		Fixed-point values of the fluctuation couplings as a function of the number of scalar (left), fermion (middle), and gauge fields (right). All truncations include the graviton two- and three-point function as well as the respective graviton-matter vertex. In the scalar case, the Newton couplings, $g_3$ and $g_\varphi$, are diverging at $N_s\approx 52 $. The fermionic case is stable for all $N_f$. In the gauge field case, the fixed point disappears in the complex plane at $N_v\approx 13$. It was explained in \cite{Christiansen:2017cxa} that the vanishing of the fixed point is an artefact of the truncation and how it can be lifted in the gauge-field case. In \cite{Burger:2019upn}, it was suggested that an expansion about a background that is a solution to the quantum equation of motions might remove the divergence in the scalar case. The result are taken from \cite{Eichhorn:2018akn} (scalar), \cite{Eichhorn:2018nda} (fermion), and \cite{Christiansen:2017cxa} (gauge).
	}
	\label{fig:matter-dependence}
\end{figure}

The results are displayed in \Cref{fig:matter-dependence} and consolidate the first fluctuation work in this direction \cite{Meibohm:2015twa}. The left panel of \Cref{fig:matter-dependence} shows the fixed-point values as a function of the number of scalar fields $N_s$. The Newton couplings diverge around $N_s\approx 52$, which is in a regime where the truncation cannot be trusted anymore. In \cite{Burger:2019upn}, it was suggested that an expansion about a background that is a solution to the quantum equation of motion might remove the divergence in the scalar case.

The results for minimally coupled fermions are displayed in the central panel in \Cref{fig:matter-dependence}. The Newton couplings decrease with the number of fermionic flavours, and arbitrarily many fermions can be included in the system. 

In the gauge field case, the fixed point is disappearing in the complex plane due to a fixed-point merger at $N_v\approx 13$. In \cite{Christiansen:2017cxa} it was shown that this strongly depends on the regulator: with a slightly different regulator arbitrarily many gauge bosons can be included in the system. This indicates that the merger is unphysical and an artefact of the truncation.

In all panels of \Cref{fig:matter-dependence}, we can observe that the Newton coupling from the three-graviton vertex has similar values and behaves similarly to the Newton coupling from the graviton-matter vertices. This effect was dubbed \emph{effective universality} \cite{Eichhorn:2018akn, Eichhorn:2018ydy} as already discussed in \Cref{sec:StabilityAnalsis}. In \cite{Eichhorn:2018ydy}, it was shown that this effect also extends to the ghost-graviton vertex and is only present in a small parameter space of the couplings, which is precisely where the UV fixed point is located. While the different Newton couplings $g_n$ are related by STIs, see \Cref{sec:symmetry-identities}, it is not expected that their behaviour is that similar in the presence of further non-trivial operators beyond the curvature scalar. In fact, we would expect large differences between the couplings in a highly non-perturbative setting as the $g_n$'s also have overlap with higher-curvature terms as well as terms with curvature and covariant momentum dependences, for pure gravity this was discussed in \cite{Denz:2016qks}. Therefore, the existence of effective universality was taken as evidence of a close-perturbative nature of the UV fixed point \cite{Eichhorn:2018ydy} with the relevant operators $R, \Lambda$ and $R^2$. In the right panel of \Cref{fig:matter-dependence}, we see that effective universality is increasingly violated with an increasing number of gauge bosons and the system runs into a fixed-point merger in this approximation. This is not the case if different regulators are chosen, see \cite{Christiansen:2017cxa}. Moreover, it was shown there, that the gauge boson-gravity systems in the minimally coupled approximation always have a Reuter-type fixed point as they can be mapped onto a pure gravity system with effective Newton constant and cosmological constant. In conclusion, the fixed-point merger signals a failure of the approximation to the flow in this minimally coupled system and not a disappearance of the fixed point. Different regulators or shifted relative cutoff scales are required to stabilise the approximation. Only then do the flows show the above feature of the system that it has a Reuter-type fixed point.

In \cite{Pastor-Gutierrez:2022nki}, the problem was circumvented with the observation that SM matter content can be easily accommodated with the approximation of effective universality, $g_3=g_a=g_\varphi= g_\psi=g_c$, leading to the SM fixed point
\begin{align}
	\label{eq:SM-FP}
	(g_3,\,\lambda_3,\,\mu)_\text{SM}^* = (0.17\,,0.15\,,-0.71)\,,
\end{align}
with the critical exponents
\begin{align}
	\label{eq:SM-crit-exp}
	\theta_i^\text{SM} = (-2.9 \pm 4.5 \imag ,\,10.2)\,.
\end{align}
We expect that the fixed-point values will be similar in an improved truncation that does not possess the fixed-point merger in $N_v$ due to the effective universality of the Newton couplings. Consequently, these values were utilised in the computation in \cite{Pastor-Gutierrez:2022nki}.

\begin{figure}[t]
	\includegraphics[width=.49\columnwidth ]{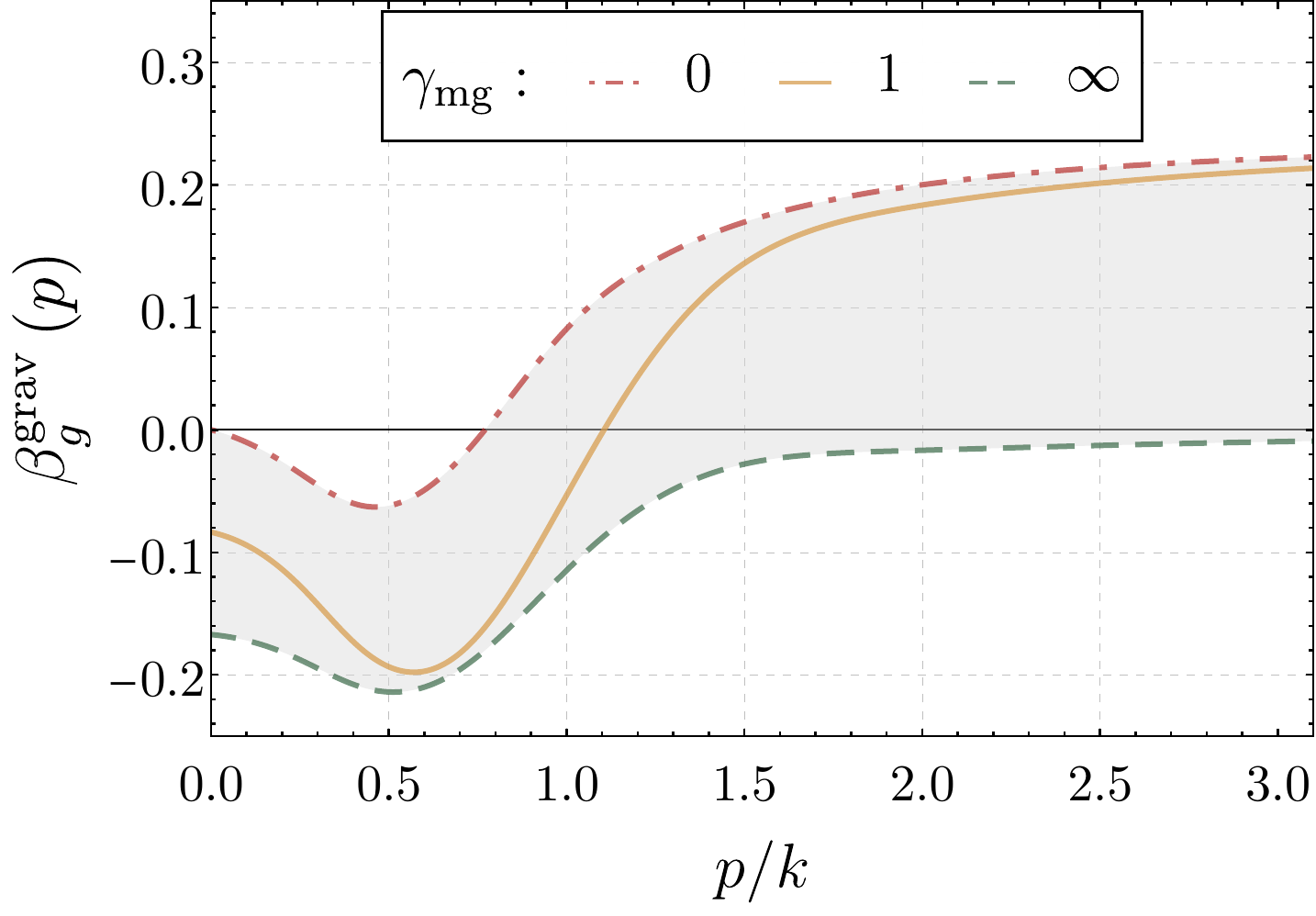}
	\hfill
	\includegraphics[width=.49\columnwidth ]{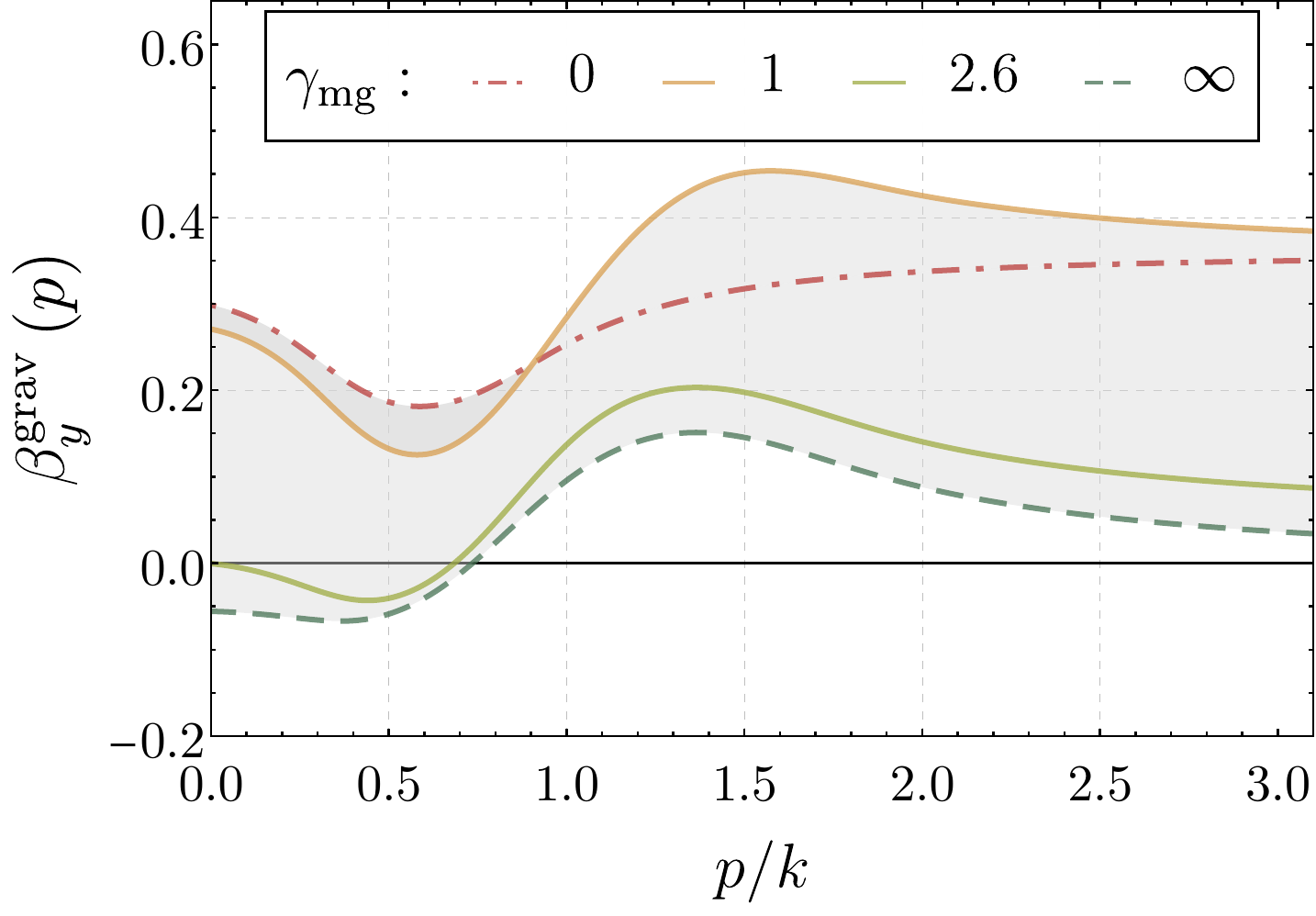}
	\caption{Matter-gravity contributions to the gauge beta functions $\beta_{g}^{\text{grav}}$ derived from the gauge-fermion vertex for $g=1$, $g_h=1$, $\mu_h=0$, and different values of $\gamma_\textrm{mg}$. For $\gamma_\textrm{mg}= 0$, $\beta_{g}^{\text{grav}}$ vanishes due to an exact cancellation between the fermion anomalous dimension and the three-point diagrams. Taken from \cite{Pastor-Gutierrez:2022nki}.}
	\label{fig:betagrav_gy}
\end{figure}

Next, we discuss the impact of graviton fluctuation on the marginal SM matter couplings. For example, the graviton contribution to the Yukawa coupling is generated by the diagrams
\begin{align}
	\label{eq:beta-grav-y}
	\parbox{.8\columnwidth}{\includegraphics[width=.8\columnwidth]{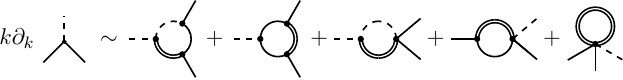}},
\end{align}
where double lines represent gravitons, dashed lines scalars, and solid lines fermions. We have suppressed the prefactors of the diagrams and the regulator insertions for simplicity. All diagrams in \labelcref{eq:beta-grav-y} are linear in the Yukawa coupling but higher-order contributions of the Yukawa coupling enter through the fRG resummation of higher-order interactions. It has been common practise to use the parametrisation \cite{Eichhorn:2017eht, Eichhorn:2017ylw, Eichhorn:2017lry, Eichhorn:2017als, Eichhorn:2019yzm, deBrito:2022vbr}, 
\begin{align}
	\label{eq:beta-grav-y-param}
	\partial_t y \bigg|_\text{gravity} = - f_y (G_\text{N},\mu,\dots) y\,,
\end{align}
where $f_y$ only depends on the gravity couplings. In the fixed-point regime beyond the Planck scale, $f_y$ is in a good approximation a constant while it vanishes below the Planck scale. Many applications used $f_y$ as a constant free parameter beyond the Planck scale for the fixed-point analysis of (beyond) the SM physics, and then utilised standard perturbative beta functions with vanish $f_y$ below the Planck scale \cite{Eichhorn:2018whv, Reichert:2019car, Alkofer:2020vtb, Eichhorn:2020sbo, deBrito:2020dta, Eichhorn:2020kca, Kowalska:2020gie, Kowalska:2020zve, Eichhorn:2021tsx, Kowalska:2022ypk, Chikkaballi:2022urc, Kotlarski:2023mmr}. The gravity contributions to the gauge and scalar quartic couplings are commonly parametrised in straight analogy
\begin{align}
	\partial_t g &= - f_g g \,,
	&
	\partial_t \lambda &= - f_\lambda \lambda\,.
\end{align}
The signs and the size of the gravity contributions to the marginal SM couplings are crucial for the compatibility of the SM with asymptotically safe gravity. They were analysed in with fRG techniques in \cite{Daum:2009dn, Harst:2011zx, Folkerts:2011jz, Christiansen:2017gtg, Christiansen:2017cxa, Eichhorn:2017lry, Hamada:2020vnf, Wetterich:2022bha} (gauge coupling), \cite{Rodigast:2009zj, Zanusso:2009bs, Oda:2015sma, Eichhorn:2016esv, Eichhorn:2017eht} (Yukawa coupling), and \cite{Percacci:2003jz, Rodigast:2009zj, Zanusso:2009bs, Narain:2009fy, Eichhorn:2017als, Pawlowski:2018ixd, Wetterich:2019zdo, Ohta:2021bkc} (quartic scalar coupling), as well as with perturbative techniques  \cite{Deser:1974xq, Pietrykowski:2006xy, Toms:2007sk, Ebert:2007gf, Anber:2010uj, Toms:2010vy, Bevilaqua:2021uzk, Souza:2022ovu, Souza:2023wzv}.

For the compatibility of the U(1) hypercharge coupling with asymptotic safety, $f_g \geq 9.8 \cdot 10^{-3}$ is needed \cite{Eichhorn:2017lry}. The Yukawa sector is more intricate and also depends on whether the hypercharge coupling becomes asymptotically safe or free. Assuming an asymptotically free hypercharge coupling,  $f_y \geq 10^{-4}$ is needed \cite{Eichhorn:2017ylw, Alkofer:2020vtb}. The quartic scalar coupling will be discussed below.

In \cite{Pastor-Gutierrez:2022nki}, the regulator and momentum dependence of the gravity contributions to the gauge and Yukawa coupling were investigated. The regulator dependence was assessed by integrating out gravity and matter fluctuations at different scales, $k_\text{grav}$ and $k_\text{mat}$. The parameter $\gamma_\text{mg} = k_\text{mat}/k_\text{grav}$ describes the ratio of these cutoff scales, see \cite{Pastor-Gutierrez:2022nki} for more details. A stable truncation should be insensitive to a change of cutoff scales of order one.

The results of \cite{Pastor-Gutierrez:2022nki} are displayed in \Cref{fig:betagrav_gy}. The left panel shows the gravity contribution to the gauge coupling, which was computed from the quark-gluon vertex. At vanishing momentum, the contribution has a definite sign which is compatible with asymptotic safety supporting asymptotic freedom. For $\gamma_\text{mg}=0$, the contribution is exactly vanishes due to a kinematic identity that was also found in other gauge correlation functions \cite{Folkerts:2011jz, Christiansen:2017cxa}.

The situation is qualitatively different in the Yukawa sector as shown in the right panel of  \Cref{fig:betagrav_gy}. Two different signs are possible at vanishing momentum and the crossover happens at $\gamma_\text{mg}= 2.6$ which is an order one change in the cutoff scales. For the naive choice of cutoff scales, $\gamma_\text{mg} = 1$, the gravity contributions drive the Yukawa coupling into a Landau pole and only for $\gamma_\text{mg}> 2.6$ the couplings become asymptotically safe or free. This computation emphasises the urgency to improve computations in the Yukawa sector. We note large negative values of the cosmological constant can stabilise the computation \cite{Eichhorn:2017ylw}, which is not realised in the SM fixed point \labelcref{eq:SM-FP}, or the inclusion of ultralocal higher-derivative interactions \cite{Eichhorn:2017eht}, i.e., working with the classical Stelle action \labelcref{eq:Stelle-Action}.

In case of the quartic scalar coupling, the sign of $f_\lambda$ was found to be negative \cite{Percacci:2003jz, Rodigast:2009zj, Zanusso:2009bs, Narain:2009fy, Eichhorn:2017als, Pawlowski:2018ixd, Wetterich:2019zdo, Ohta:2021bkc} leading to a irrelevant direction at the $\lambda=0$ fixed point. The irrelevant direction can be used to predict the ratio between Higgs and top mass, which led to the result $m_H = 125$\,GeV with one-loop beta functions \cite{Shaposhnikov:2009pv}. The result shifts to $m_H = 132$\,GeV with higher-order beta functions and a more accurate determination of the top-quark pole mass, see \cite{Pastor-Gutierrez:2022nki} and also below. Note however that this result is very sensitive to the value of the top mass, just as the metastability of the Higgs potential in perturbative computations \cite{Buttazzo:2013uya, Bezrukov:2014ina, Bezrukov:2014ipa}. Furthermore, the value is modified with beyond SM physics \cite{Kwapisz:2019wrl, Reichert:2019car, Eichhorn:2020kca, Eichhorn:2020sbo, Eichhorn:2021tsx}.

A natural question to ask is if there are other fixed points in the scalar sector that can connect to the SM in the IR, in particular, due to the small discrepancy between the measured value of the Higgs mass and the prediction by asymptotic safety for $\lambda^* =0$. In \cite{Pastor-Gutierrez:2022nki}, the Higgs fixed-point potential was investigated in a high-order Taylor expansion. A fixed point was identified that converges to the Gau\ss ian fixed point but contains a second relevant direction. For more details, we refer to \cite{Pastor-Gutierrez:2022nki}, and for more details on gravity-matter interactions, we refer to \cite{Eichhorn:2022gku}.

\begin{figure}[t]
	\centering
	\includegraphics[width=.9\linewidth]{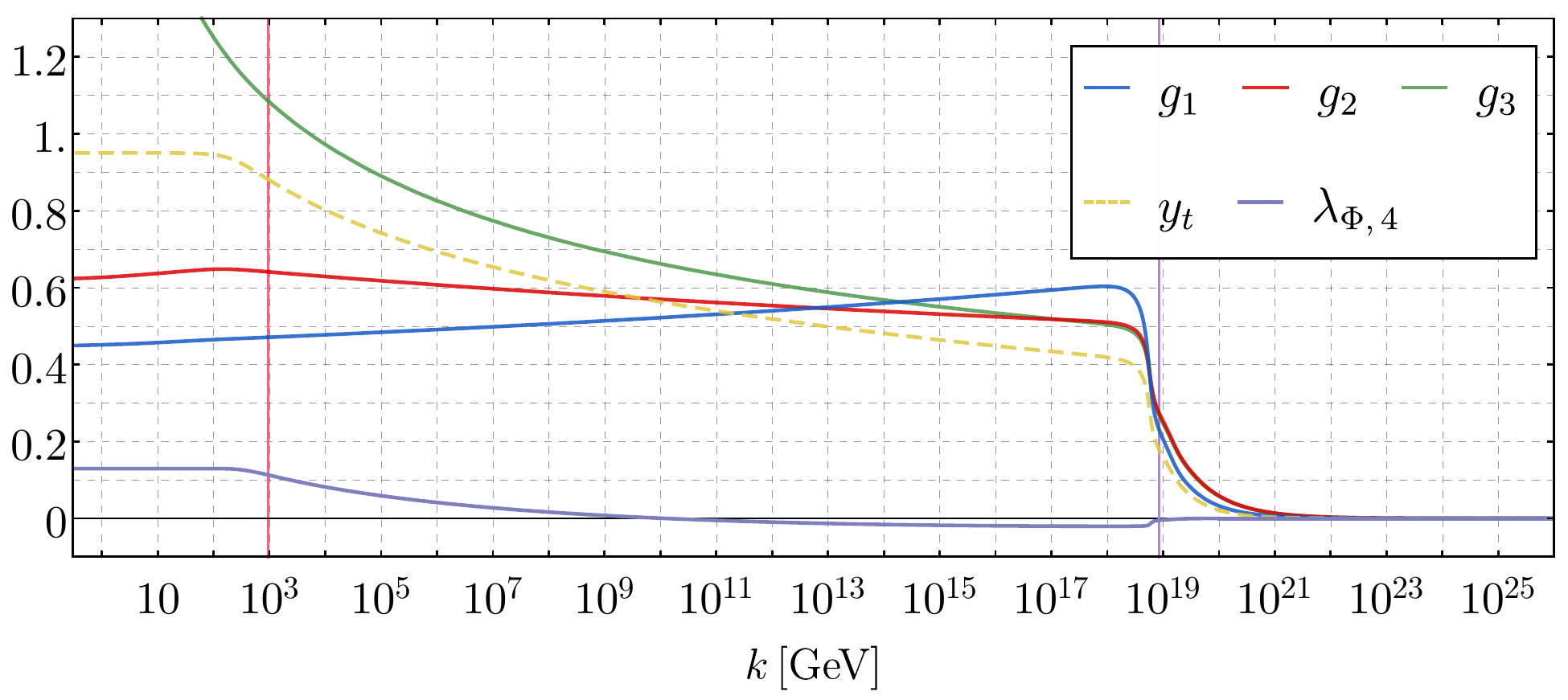}
	\caption{We show the RG running of the classically marginal SM couplings including quantum gravity and threshold effects. The threshold effects due to the masses of the particle cause couplings to freeze out in the IR, with the exception of the strong and the electro-magnetic coupling. In the UV, quantum gravity drives all couplings into the Gau\ss ian fixed point. Taken from \cite{Pastor-Gutierrez:2022nki}}
	\label{fig:ASSMtrajectory}
\end{figure}

\subsection{Phase structure of the asymptotically safe Standard Model}
With the approximations and details discussed in the last section, we are now ready to look at full UV-IR flows for all SM couplings. The result from \cite{Pastor-Gutierrez:2022nki} is shown in \Cref{fig:ASSMtrajectory}. All marginal SM couplings emerge from the Gau\ss ian fixed point in the UV and reach their physical values in the IR. The quartic Higgs coupling approaches the Gau\ss ian fixed point from below and turns negative around $k = 10^{10}$\,GeV, which is compatible with the metastability scale found in perturbative studies. Threshold effects in the IR and in the UV have been included which causes most couplings to freeze out below the electroweak symmetry breaking scale, which is around 1\,TeV in the RG scale $k$.

\begin{figure}[t]
	\centering
	\includegraphics[width=.45\linewidth]{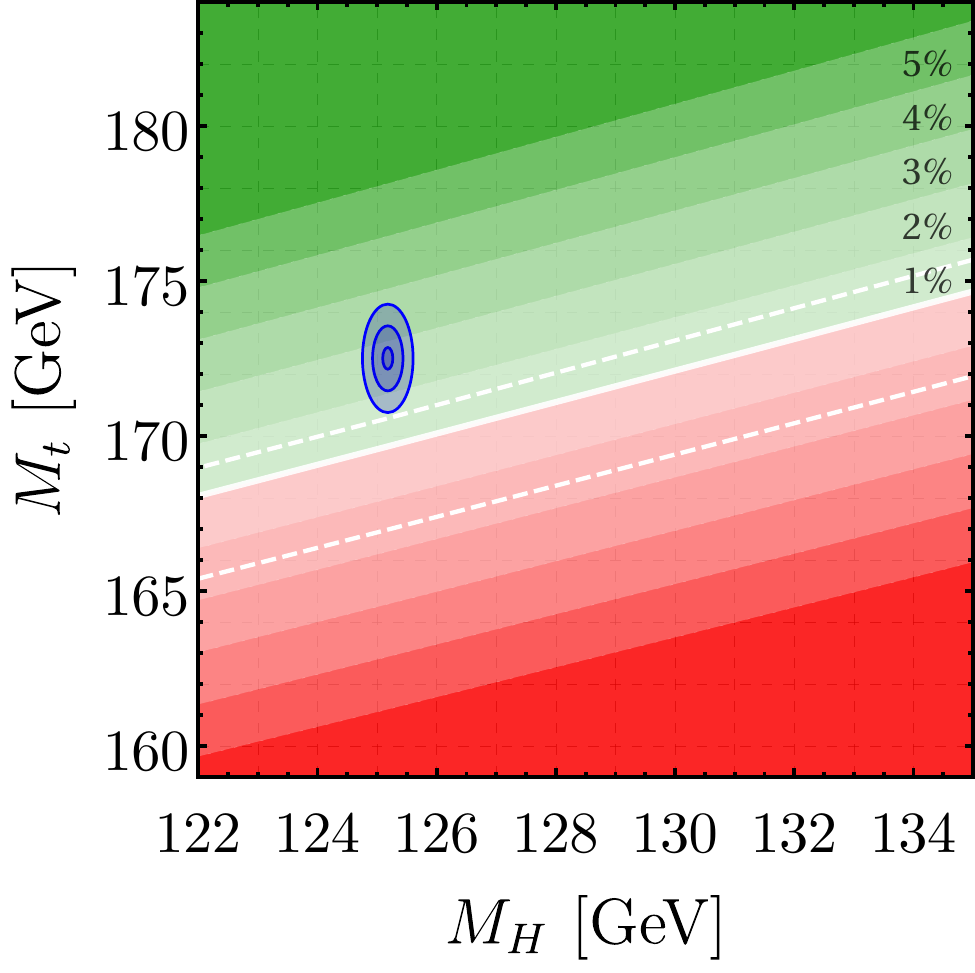}
	\caption{IR masses of the Higgs boson and top quark that can be reached from the asymptotically safe UV. The green region can be approached from the non-trivial Gau\ss ian fixed point with two relevant directions, the white line can be approached from the standard Gau\ss ian fixed point with one relevant direction, and the red region does not have a UV completion. The blue circles represent the central measured values including the 1$\sigma$, 3$\sigma$, and 5$\sigma$ uncertainty. Taken from \cite{Pastor-Gutierrez:2022nki}}
	\label{fig:mtop-mh}
\end{figure}

Not all values of the Higgs and top mass can be connected to the UV fixed point, which is displayed in \Cref{fig:mtop-mh}. The green area can be connected to the non-trivial Gau\ss ian fixed point. The area is bounded by the white line, which is precisely the line that connects to the predictive standard Gau\ss ian fixed point. On the other side of the white line, we have the red regime for which no UV completion was found.

The experimental values are shown in \Cref{fig:mtop-mh} with blue circles including the 1$\sigma$, 3$\sigma$, and 5$\sigma$ uncertainty. They lay in the green regime with roughly a $2.9$\,GeV distance between the central values and the white Gau\ss ian line. The combined 5$\sigma$ experimental and theoretical error are roughly of the same size implying that it is not excluded but unlikely that the pure SM lays in the red regime or on the white line. This can certainly change upon inclusion of beyond SM physics such as dark matter or right-handed neutrinos. For example, the asymptotically safe dark matter models in \cite{Reichert:2019car, Eichhorn:2020kca, Eichhorn:2021tsx} allow for a correct Higgs mass from the standard Gau\ss ian UV fixed point.

\subsection{Outlook}
\label{sec:ASSMtlook}

The results on the phase structure of the asymptotically safe Standard Model reviewed in this section constitute an important milestone in the quest for reliable predictions in asymptotically safe particle physics. It is the first \textit{self-consistent}  computation: all running coupling parameters are obtained within the same RG framework, which is of paramount importance for reliable predictions and the systematic improvement within the given vertex expansion scheme. It is based on  momentum-dependent correlation functions and hence allows the computation of on-shell observables such as pole masses, in contradistinction to other approaches. In summary, these results and their systematics in the fluctuation approach open a path towards reliable computations with increasingly small systematic errors, and hence usher in the era of quantitative predictivity in asymptotically safe particle physics.

\section{Lorentzian quantum gravity}
\label{sec:LorentzianGravity}

In this final chapter, we discuss a direct fRG approach to Lorentzian metric quantum gravity. It allows what can be considered the holy grail of any approach to quantum gravity: direct non-perturbative computations within a Lorentzian signature. Such an approach has been set up the first time in \cite{Fehre:2021eob}. The spectral fRG approach underlying \cite{Fehre:2021eob} has been discussed in detail in \cite{Braun:2022mgx}, and is based on the spectral functional approach put forward in \cite{Horak:2020eng}.

\subsection{Fundamentals and effective field theory}
We start this chapter with a brief overview of the state of the art concerning timelike correlation functions in asymptotically safe gravity or more generally Lorentzian quantum gravity. To date, most computations in asymptotically safe quantum gravity including those in the fluctuation approach reviewed so far have been performed with a Euclidean signature. The Euclidean signature allows for a simple ordering principle of the momentum fluctuations and removed poles and cuts from the integration domain. This property is lost with Lorentzian signature since the squared momentum can vanish, $p^2 =0$, while, for example, the spatial momentum is large. The ordering of momentum fluctuations is however important for the coarse gaining procedure of the fRG. Moreover, for numerical integrations of loop momenta, the absence of singularities in or close to the integration contour is essential. 

In principle, it is feasible to first perform the integrating out of all momentum fluctuations and then Wick rotate the results to Lorentzian signature, see \cite{Bonanno:2021squ} for an explicit example. However, the Wick rotation is already a subtle issue on flat Minkowski spacetimes and it is further complicated by the dynamical nature of the metric in quantum gravity. Moreover, the Wick rotation based on numerical data comes with an ill-conditioned or even ill-posed reconstruction problem, and the systematic error of the results has to be evaluated with great care \cite{Bonanno:2021squ}. For general considerations on the reconstruction problem see e.g.~\cite{Cyrol:2018xeq, Kades:2019wtd, Horak:2021syv, DelDebbio:2021whr,  Horak:2023xfb, Candido:2023nnb}. 

For all of the above reasons, approaches that allow for direct non-perturbative computations with Lorentzian signature are much wanted. First steps towards computations with Lorentzian signature have been reported in \cite{Manrique:2011jc, Rechenberger:2012dt, Demmel:2015zfa, Biemans:2016rvp, Houthoff:2017oam, Wetterich:2017ixo, Knorr:2018fdu, Baldazzi:2018mtl, Nagy:2019jef, Eichhorn:2019ybe, Platania:2020knd, Saueressig:2023tfy}, for attempts in other approaches to quantum gravity see e.g.~\cite{Ambjorn:1998xu, Ambjorn:2000dv, Ambjorn:2001cv, Engle:2007uq, Freidel:2007py, Feldbrugge:2017kzv, Asante:2021zzh, Asante:2021phx}.

Recently, several milestones have been achieved in the fluctuation approach: In \cite{Bonanno:2021squ}, the graviton propagator was reconstructed on the basis of numerical data for the Euclidean or spacelike propagator, resulting in a graviton spectral function and hence a graviton propagator for spacelike and timelike momenta. In \cite{Fehre:2021eob}, the first direct computation of asymptotically safe gravity in Lorentzian background was put forward. This allowed us to compute the graviton spectral function and hence the space- and timelike graviton propagator. The results of this direct computation corroborated that of the reconstruction procedure in \cite{Bonanno:2021squ}. Moreover, it is worth emphasising that the spectral approach to asymptotically safe gravity gains much from recent results on spectral properties of QCD and Yang-Mills theory. In some sense, spectral considerations in these theories face even more intricacies as in asymptotically safe gravity: while the latter enjoys effective universality and a close-perturbative behaviour also signalled by a relatively weak gravitational coupling even in the UV scaling regime, the former has a truly non-perturbative IR regime with emergent resonances and confinement, which manifests as a dynamical mass gap in the gluon propagator. For respective investigations in QCD see \cite{Horak:2020eng, Horak:2021pfr, Horak:2022myj, Horak:2022aza}, for the discussion of the conformal regime of QCD close to the Banks-Zaks fixed point highly relevant in the present context see \cite{Kluth:2022wgh}. 

The spectral functional approach is based on the spectral representation of correlation functions, and most prominently on that of the propagator, the K\"all\'en-Lehmann (KL) spectral representation \cite{Kallen:1952zz, Lehmann:1954xi} as already discussed in \Cref{sec:flowEffActLorentzian} around \labelcref{eq:KL-Rep}. The KL spectral representation is a powerful tool which allows the encoding of the propagator in the entire complex momentum plane in just one real function. It provides intuitive access to physics properties such as on-shell states and unitarity properties. If the KL representation of the propagator of a given field exists, the spectral function and the propagator are related via
\begin{align}
	G(p_0,|\vec{p}\,|) = \int_{0_-}^\infty \frac{\mathrm d\lambda}{2 \pi}\frac{\lambda\,\rho(\lambda,|\vec{p}\,|)}{\lambda^2+p_0^2} 
	= \int_{0_-}^\infty \frac{\mathrm d\lambda}{2 \pi}\frac{\lambda\, \rho(\lambda)}{\lambda^2+p^2}\,, \qquad \rho(\lambda)=\rho(\lambda,0)\,,
	\label{eq:KS-L} 
\end{align}
with the temporal and spatial momentum $p_0$ and $\vec{p}$ and the spectral values $\lambda$. For the sake of simplicity, we again restricted ourselves to the scalar case. The lower bound of the spectral integral, $0_-$, takes into account that for massless particles the spectral function contains a delta function at vanishing spectral values, $\delta(\lambda)$.

The second identity of \labelcref{eq:KS-L} follows in the presence of Lorentz symmetry. For example, at finite temperature or density, this step cannot be performed as the heat bath or the medium singles out a rest frame. Here, we consider Lorentzian quantum gravity in an expansion about the flat background with full Lorentz symmetry. Then, the propagator is a function of momentum squared,  $p^2$. 

Note that the KL representation \labelcref{eq:KS-L} does not necessarily hold for unphysical fields such as the graviton or the gluon, but the spectral function is defined in any case by the discontinuity of the propagator on the time-like axis with
\begin{align}
	\label{eq:rho-G}
	\rho(\lambda,|\vec{p}|) = \lim_{\varepsilon\to 0} 2\, \text{Im}\, G(p_0=-\imag (\lambda + \imag\,\varepsilon),|\vec{p}|) \,.
\end{align}
The spectral function acts as a linear response function of the two-point correlator, encoding the energy spectrum of the theory. For fields that present asymptotic states, the existence of the KL representation can be readily shown with a simple insertion of a complete set of states with masses $\lambda^2$ and propagators $1/(\lambda^2+p^2)$ into the two-point correlation function of the fields. Then, the spectral function can be understood as a probability density for the transition to an excited state with energy $\lambda$. Accordingly, in this case, the spectral function is positive semi-definite $\rho(\lambda) \geq 0$ and the total spectral weight can be normalised to unity, 
\begin{align}
	\int_{0_-}^\infty  \frac{ \mathrm d \lambda}{2 \pi}\,  \lambda \,\rho(\lambda) \mathrm= 1\,, \qquad \textrm{with}\qquad \rho(\lambda)>0\,.
	\label{eq:TotalWeight1}
\end{align}
\Cref{eq:TotalWeight1} holds true if the dressing of the Euclidean propagator is normalised to unity in the UV, $\lim_{p^2\to\infty} p^2 G(p^2)=1$, which encodes the canonical commutation relations. The total spectral weight as well as the analytic IR and UV limits have been discussed in detail in \cite{Cyrol:2018xeq, Bonanno:2021squ, Horak:2021pfr}. 

\begin{figure}[t]
	\includegraphics[width=.53\linewidth]{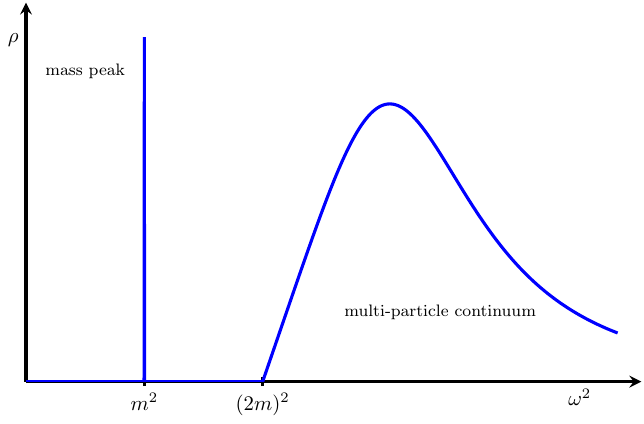}
	\hfill
	\includegraphics[width=.35\linewidth]{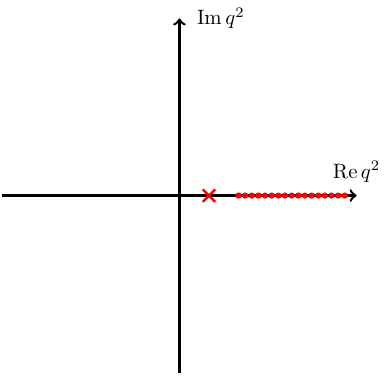}
	\caption{Schematic illustration of the spectral function (left) and the corresponding propagator in the complex momentum plane (right) for a stable physical particle of mass $m$. The on-shell mass peak in the spectral function corresponds to a singularity in the propagator (red cross), and the multi-particle continuum corresponds to a branch cut (red line).}
	\label{fig:spectral-schematic}
\end{figure}

A schematic example of a spectral function and the corresponding propagator are displayed in \Cref{fig:spectral-schematic}. In the left panel, we see the one-particle delta-peak at the spectral value $\omega^2 = m^2$, and the onset of the multi-particle continuum at $\omega^2 = (2m)^2$. Correspondingly, we show the singularity structure of the propagator in the complex momentum plane in the right panel of  \Cref{fig:spectral-schematic}: the propagator has a singularity at $q^2=m^2$, which corresponds to the delta-peak in the spectral function, and a branch cut starting at $q^2 = (2m)^2$ corresponding to the multi-particle continuum.

The above considerations also extend to the spectral representation of higher correlation functions. However, they are far more complicated and even their closed form for higher-order correlation functions such as four-point functions is not completely worked out. Still, the approximations used so far in asymptotically safe gravity are well covered with the known spectral representations of correlation functions in the spectral approach.

\subsection{The graviton spectral function}
The explicit computations in the spectral fRG approach have so far been done in an expansion about the flat Minkowski background $\eta_{\mu\nu}$ with \labelcref{eq:LinSplitGamma}, 
\begin{align}
	g_{\mu\nu}= \eta_{\mu\nu}+\sqrt{Z_h  G_\text{N}} \,h_{\mu\nu}\,,\qquad \textrm{with}\qquad  \eta=\text{diag}(1,-\boldsymbol{1})\,.
\end{align}
In the following, we present results for the gauge-fixing condition \labelcref{eq:gf-condition} with $\alpha=\beta=1$ (harmonic Feynman gauge).  The graviton propagator in the flat  background has the parametrisation  
\begin{align}
	G_{hh,\mu\nu\rho\sigma}(p) =  32\pi	G_{hh}^{(tt)}(p) \Pi^{(tt)}_{\mu\nu\rho\sigma}(p) - 64\pi G_{hh}^{(s)}(p) \Pi^{(s)}_{\mu\nu\rho\sigma}(p) +\cdots\,, 
	\label{eq:PropGravFlat}
\end{align}
where $	G_{hh}^{(tt/s)}(p)$ are the scalar dressings of the tt-part with five degrees of freedom and the physical scalar part with one degree of freedom respectively. Further terms are included in the dots and comprise the vector and other scalar modes that are gauge degrees of freedom. In the following, we assume uniform scalar dressing functions which are determined by the dominant tt-mode. This leads to the approximation 
\begin{align}
	G_{hh}^{(tt)}(p) &= G_{hh}(p) \,,
	&
	G_{hh}^{(s)}(p)&=	 G_{hh}(p)\,,
	\label{eq:UniformGraviton}
\end{align}
and similarly for the other modes. In the approximation \labelcref{eq:UniformGraviton}, the graviton propagator has a uniform scalar dressing $G_{hh}$, which can be written in terms of the KL-representation \labelcref{eq:KS-L}. 

Before we discuss the full non-perturbative computation and results for the spectral function, we proceed with a discussion of the spectral function for different effective theories of gravity. This will allow us to embed and interpret the full quantum results. We start with the classical Einstein-Hilbert action \labelcref{eq:EH-Action} with a vanishing cosmological constant, $\Lambda=0$. Then, the flat background is a solution of the equations of motion and the scalar propagator $G_{hh}$ is simply $1/p^2$ and the corresponding spectral function is a $\delta$-function at vanishing spectral value, to wit, 
\begin{align}
	G_{hh}(p) &= \frac{1}{p^2}\,,
	&
	\rho_h(p)&= \frac{2 \pi}{\lambda}\delta(\lambda)\,.
	\label{eq:EHprop+rho}
\end{align}
Conversely, starting from the higher-derivative Stelle action \labelcref{eq:Stelle-Action}, we find that after a partial fraction decomposition the transverse-traceless graviton propagator and the corresponding spectral function are given by 
\begin{align}
	G_{hh}(p) &=  \frac{1}{p^2}- \frac{1}{p^2+M_\text{ghost}^2}\,,
	&
	\rho_h(p)&= 2 \pi\left[ \frac{1}{\lambda} \delta(\lambda) -  \frac{1}{M_\text{ghost}} \delta(M_\text{ghost} - \lambda) \right]\,.
	\label{eq:Stelleprop+rho}
\end{align}
The second delta function in the spectral function in \labelcref{eq:Stelleprop+rho} comes with a negative sign indicating that this is a ghost state. The mass of the ghost is given by $M_\text{ghost} = M_\text{pl}/\sqrt{32\pi b}$, which is of the order of the Planck mass. Depending on the sign of $b$, the state can also be tachyonic. This is a reflection of the unitarity problems in higher-derivative gravity, commonly known as Ostrogradsky instabilities. Note that only the prefactor of the Weyl-squared tensor enters in the propagator since we are only looking at the transverse-traceless part. The Ricci-scalar--squared term would enter in the physical scalar mode giving rise to a massive scalar mode that can be used for inflation \`a la Starobinsky \cite{Starobinsky:1982ee}, for reviews see e.g.~\cite{Kehagias:2013mya, deHaro:2021swo, Cruces:2022imf} and for applications in asymptotically safe gravity \cite{Platania:2020lqb, Copeland:2013vva, Liu:2018hno, Platania:2019qvo, Bonanno:2017pkg, Bonanno:2015fga}.

\begin{figure}[t]
	\includegraphics[width=.48\linewidth]{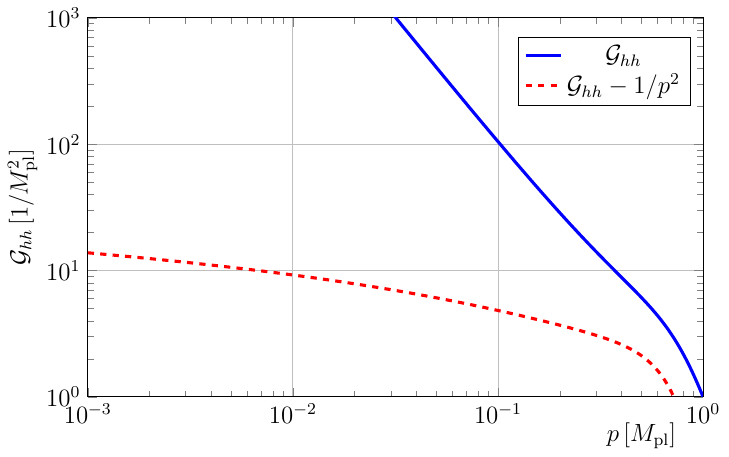}
	\hfill
	\includegraphics[width=.48\linewidth]{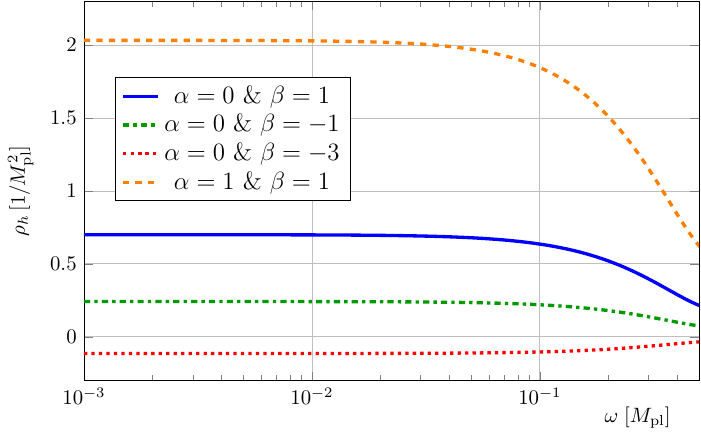}
	\caption{The graviton propagator (left) and spectral function (right) below the Planck scale computed from one-loop effective field theory. The spectral function contains a gauge-invariant delta peak at vanishing frequencies (not shown in the plot) corresponding to $1/p^2$ behaviour in the propagator. The one-loop contribution induces a sub-leading logarithm in the propagator corresponding to a constant part in the spectral function below the Planck scale.}
	\label{fig:graviton-in-EFT}
\end{figure}

We can go beyond the classical graviton spectral function by treating gravity as an effective field theory with the Planck scale as cutoff. The one-loop effective action reads
\begin{align}
	\label{eq:one-loop-eff-action}
	\Gamma_\text{1-loop} = \int_x \sqrt{g} \left(  \frac{R-2 \Lambda}{16 \pi G_\text{N}} + c_1 R \ln(\Box/M_\text{pl}^2)R + c_2 \, C_{\mu\nu\rho\sigma}  \ln(\Box/M_\text{pl}^2)  C^{\mu\nu\rho\sigma} + \ldots\right),
\end{align}
where $\Box= - g^{\mu\nu}\nabla_\mu \nabla_\nu$ is the d'Alembertian. The prefactors $c_1$ and $c_2$ can be determined analytically and they depend on the two gauge-fixing parameters $\alpha$ and $\beta$ \cite{Capper:1979ej, Bonanno:2021squ}. The coefficient $c_2$ enters in the transverse-traceless mode, while $c_1$ enters in the scalar mode. The explicit form of $c_2$ is, for example, displayed in equation (F3) in \cite{Bonanno:2021squ}. With this one-loop effective action, the inverse graviton propagator takes the form
\begin{align}
	\mathcal{G}_{hh} (p)^{-1} \sim p^2 + \tilde c_2 \ln(p^2) p^4 + \ldots\,,
	\label{eq:prop-one-loop}
\end{align}
with $\tilde c_2 = 32 \pi G_\text{N} c_2 $  and the corresponding spectral function
\begin{align}
	\rho_h (\lambda)\sim \frac1{\lambda} \delta(\lambda) + \tilde c_2 +\dots\,.
	\label{eq:spec-one-loop}
\end{align}
The graviton propagator and spectral function are displayed in \Cref{fig:graviton-in-EFT}. The left panel shows the propagator. The dominant contribution is the standard $1/p^2$, and the first sub-leading contribution is the \nobreakdash{logarithmic} $\ln p^2$ contribution. The latter leads to a constant part in the spectral function, displayed in the right panel of \Cref{fig:graviton-in-EFT}. The magnitude of the constant part is directly given by the coefficient $c_2$ from the one-loop effective action \labelcref{eq:one-loop-eff-action}. The coefficient $c_2$ is gauge dependent, as explained before. We display the result for commonly used values for the gauge-fixing parameters. We intentionally also include the values $\alpha = 0$ and $\beta = -3$ where the coefficient $c_2$ is negative and consequently the spectral function has a negative part. This emphasises that even an un-truncated result of the graviton spectral function is gauge-dependent, and for some values of the gauge-fixing parameter, the result will contain negative parts. The one-loop EFT graviton propagator contains additional cuts and poles in complex momentum plane beyond the Planck scale. These cuts and poles are outside of the validity regime of the effective field theory.

We turn now to the computation of the full non-perturbative spectral function from \cite{Fehre:2021eob} based on the spectral renormalisation group \labelcref{eq:RenFlow} with the Callan-Symanzik regulator reviewed in  \Cref{sec:flowEffActLorentzian}. For the graviton propagator, the scalar part of the CS-regulator, \labelcref{eq:CS-Cutoff}, reads 
\begin{align} 
	R_h= Z_h k^2\,,
\end{align}
where the inverse wave function $1/Z_h$ is the amplitude of the pole contribution of the graviton spectral function, 
\begin{align}\label{eq:rhoh-para}
	\rho_h(\lambda)= \frac{1}{Z_h}\!\Big[ 2 \pi \delta(\lambda^2-m^2_h) +\theta(\lambda^2- 4 m_h^2)  f_h(\lambda)\Big],
\end{align}
with $Z_h\equiv Z_h(p^2=-m_h^2)$, $m_h^2= k^2(1+\mu)$ is the graviton mass parameter, and $f_h$ is the multi-graviton continuum. 

In summary, we are led to the flow of the graviton spectral function, 
\begin{align}\label{eq:dotrho-dotlambdahh}
	\partial_t \rho_h &= -2\,\text{Im}\, G_{hh}^2 \left(\partial_t \Gamma^{(hh)}_\text{TT} + \partial_t R_h\right)\,,
\end{align}
and the respective one of the ghosts. These flows are supplemented with a flow equation for the Newton coupling from the graviton three-point function at vanishing momentum. This leads to the Lorentzian asymptotically safe UV fixed point
\begin{align} 
	(g,\, \eta_h,\,\mu)\big|_*=(1.06,\,0.96,\,-0.34)\,.
	\label{eq:FP-Lorentzian}
\end{align}
The scaling exponents $\theta= 2.49 \pm 3.17\,\imag$ compare well with those found in Euclidean studies. From the UV fixed point \labelcref{eq:FP-Lorentzian}, a trajectory was found that connects to standard GR behaviour in the IR. The counter-term action in \labelcref{eq:RenFlow} reabsorbs divergences of the flow. We have to introduce a renormalisation condition for each divergence, just as in perturbation theory, and make sure that there are only finitely many renormalisation conditions. This analysis was done in \cite{Braun:2022mgx} and it was shown that if the Newton couplings become asymptotically safe, corresponding to a $G_\text{N} (p) \to 1/p^2$ in the UV, then there are finitely many renormalisation conditions and the theory is predictive. In the absence of a UV fixed point, this analysis would fall back onto the standard non-predictivity found in perturbation theory.  

\begin{figure}[t]
	\includegraphics[width=.48\linewidth]{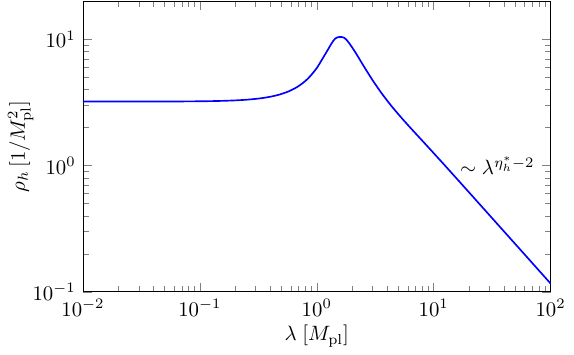}
	\hfill
	\includegraphics[width=.48\linewidth]{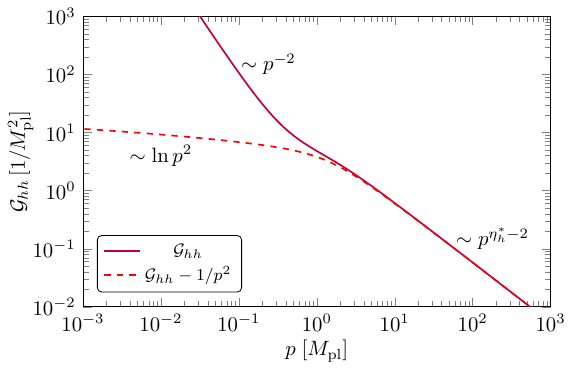}
	\caption{The full graviton spectral function (left) and propagator (right). Below the Planck scale, they match effective field theory, while they display an asymptotically safe scaling above the Planck scale. Taken from \cite{Fehre:2021eob}}
	\label{fig:full-graviton}
\end{figure}

On this trajectory, the multi-graviton continuum can be integrated. The result of the computation is displayed in \Cref{fig:full-graviton}. The graviton spectral function, displayed in the left panel of \Cref{fig:full-graviton}, has a delta-peak at vanishing frequencies corresponding to the massless graviton, and a positive ensuing multi-graviton continuum. In the IR, the spectral function is constant as expected from the EFT considerations, and in the UV the spectral function decays with an asymptotically safe power law $\lambda^{\eta_h^*-2}$. The corresponding Euclidean propagator is shown in the right panel of \Cref{fig:full-graviton}. In the IR, the leading order behaviour is the standard GR $1/p^2$ behaviour, with a subleading logarithmic $\ln(p^2)$ behaviour from the EFT corrections. In the UV, we encounter the same asymptotically safe scaling  $p^{\eta_h^*-2}$. The constant part of the spectral function in the IR is universal (regulator-independent) but gauge-dependent \cite{Kallosh:1978wt, Bonanno:2021squ}. It can be determined within effective theory, giving $61/30\approx 2.03$. In \cite{Fehre:2021eob}, the value $70 \pi/(9\sqrt{3})-11\approx 3.11$ was found, since the feedback of $f_h$ onto the flow was neglected. In comparison, in the analytic continuation in \cite{Bonanno:2021squ}, the exact EFT value was found: $7/10$ in the gauge $\alpha=0$ and $\beta=1$.

We know from the discussion of the EFT results, see \Cref{fig:graviton-in-EFT}, that the graviton propagator and the graviton spectral function are gauge dependent. \Cref{fig:full-graviton} shows the result in the gauge $\alpha= \beta=1$. Despite the gauge dependence, the result is very encouraging: the existence of a KL spectral representation in any gauge is already non-trivial. Furthermore, the propagator does not have any tachyonic and ghost-like instabilities and is therefore compatible with all requirements from unitarity, see also \cite{Platania:2022gtt} for a discussion. The latter is expected to be a gauge-independent feature.

\begin{figure*}[t]
	\includegraphics[width=.45\linewidth]{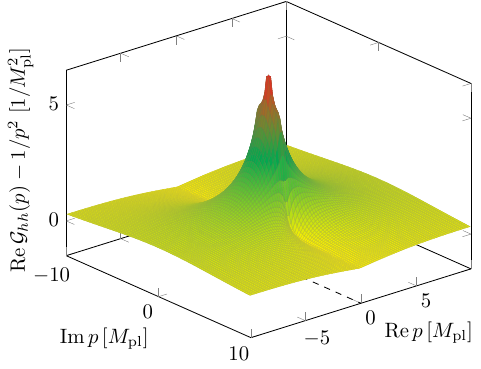}
	\qquad 
	\includegraphics[width=.45\linewidth]{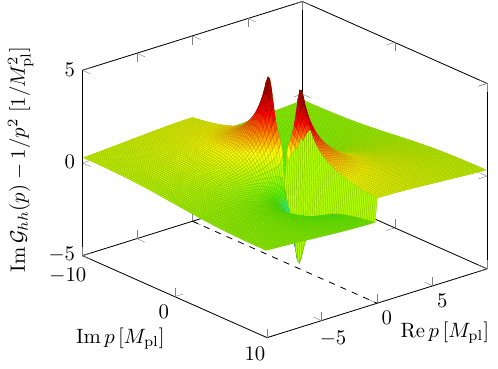}
	\caption{Real and imaginary part of the graviton propagator in the complex plane. The dashed line indicates the timelike axis. Taken from \cite{Fehre:2021eob}.
	}
	\label{fig:3d-prop}
\end{figure*}

Finally, the spectral function shown in \Cref{fig:full-graviton} can be inserted in the spectral representation of the graviton,  \labelcref{eq:KS-L}, giving access to the propagator in the whole complex momentum plane. The real and imaginary parts of the propagator are depicted in \Cref{fig:3d-prop}, where we excluded the pole contribution in the real part.  Both parts vanish for asymptotically large $p$. The real part displays a unique pole at vanishing $p$ not shown in \Cref{fig:3d-prop}, while the imaginary part shows a branch cut along the timelike axis. 

\subsection{Towards curved backgrounds}
We close this section with a discussion of Lorentzian quantum gravity with a non-vanishing cosmological constant, $\Lambda\neq 0$. This amounts to the IR values 
\begin{align}\label{eq:boundary}
	(G_\text{N}(k),Z_h(k),k^2\mu(k))\big|_{k\to0}=(G_\text{N},1,-2\Lambda)\,,
\end{align}
for the dimensionful Newton constant and cosmological constant respectively. The value of the wave function or residue value, $Z_h$, is normalised to unity. 

The following results have been obtained within the expansion about the flat Minkowski background used for vanishing cosmological constant, $\Lambda = 0$, for which the flat background is a solution to the Einstein equations. 
For a non-vanishing cosmological constant, the flat background is not a solution to the Einstein equations and should be understood as a background field such as in quantum electrodynamics (QED) with a background electric or magnetic field. It is well-known that QED in such a background does not represent a closed system and hence unitarity and probability conservation is potentially lost. 

We continue our discussion, bearing the above in mind. On de~Sitter (dS) or anti-de~Sitter (AdS) backgrounds, the classical graviton and ghost are still massless, while the graviton vertices are deformed in comparison with flat backgrounds. We expect to find modifications of the spectral function at small spectral values since alterations of the geometry are relevant for large spatial distances and are negligible at small distances. As discussed above, we still expand about the flat Minkowski background, and hence our setup is an off-shell expansion at $\Lambda\neq 0$. For simplified trajectories
\begin{align}
	\label{eq:gk-simple}
	G_\text{N}(k) = \frac{g^*}{k^2+g^* M_\text{pl}^2}\,,
\end{align}
the spectral flows admit analytic solutions. We also neglect the ghost contributions and in combination this facilitates the present qualitative discussion. 

The UV fixed point value $g^*$ of the Newton constant in \labelcref{eq:gk-simple} is a free parameter, and the respective UV fixed point is given by
\begin{align}
	\label{eq:FP-mu-simple}
	\mu^*&=\frac{-g^* }{c_\mu  + g^*}\,,
	&
	\eta_{h}^*&=\frac{2 g^* }{2 c_\eta  + g^*}\,,
\end{align}
with $(c_\mu,\, c_\eta) = ( 1.77,\, 0.49)$ known analytically, for more details see \cite{Fehre:2021eob}. Using $g^*= 1.06$ from \labelcref{eq:FP-Lorentzian}, we find $\mu^*= -0.38$ and $\eta_h^* =1.04$. Both values are approximately 10\% off, see \labelcref{eq:FP-Lorentzian}, which supports the present qualitative approximation. In particular, it indicates that the ghost contributions are indeed subleading. Moreover, the anomalous dimension of the fluctuating graviton is positive and takes values in the range $\eta_h^* \in (0,2)$. This allows for a positive spectral function for the fluctuating graviton in the presence of a non-vanishing cosmological constant as for $\Lambda=0$. 

The flow is readily integrated analytically with the IR boundary conditions \labelcref{eq:boundary},
\begin{align}
	\label{eq:mu-sol}
	Z_{h}(k) &= \left( 1 + \frac{1}{c_\eta\,\eta_h^*} \frac{k^2}{M_\text{pl}^2} \right)^{\!\!-\frac12 {\eta_h^*}}, 
	&
	\mu(k)&= \mu^* -\frac{2 \Lambda}{k^2} + \frac{c_1 M_\text{pl}^2-2\Lambda}{k^2} \left[Z_h(k)^{- c_2}-1\right],
\end{align}
with  $c_1=2.17\, g^* /(1.77  + g^*)$ and $c_2 =0.45$, again we refer to \cite{Fehre:2021eob} for the details. 
\begin{figure}[t]
	\centering
	\includegraphics[width=.5\linewidth]{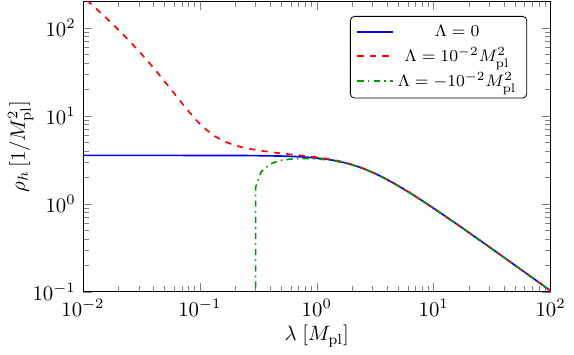}
	\caption{Enhancement (or suppression) of the spectral function due to a positive (or negative) cosmological constant. Taken from \cite{Fehre:2021eob}.}
	\label{fig:finite-Lambda}
\end{figure}
The UV tail of the solution \labelcref{eq:mu-sol} has the same qualitative behaviour as the full solution for $\Lambda=0$ displayed in \Cref{fig:full-graviton}, and the differences can be traced back to the additional approximations used here.  Remarkably, the short distance mass parameter is constrained within the narrow range $\mu^*\in (-1,0)$ and only takes negative values. Evidently, the mass parameter $\mu(k)$ interpolates smoothly between $\mu^*$ in the UV  and the cosmological constant $-2\Lambda/k^2$ in the IR. In summary, \labelcref{eq:gk-simple,eq:mu-sol} are viable approximate solutions interpolating between an asymptotically safe fixed point and general relativity with a cosmological constant.

The results for the spectral function in a de~Sitter or anti-de~Sitter background with $\Lambda\neq 0$ are depicted in  \Cref{fig:finite-Lambda}. We find that a cosmological constant $\Lambda\neq 0$ does not affect the spectral function for spectral values above $\lambda \gtrsim \sqrt{|8\Lambda|}$. This is in line with our expectation that the geometry should not leave an imprint for larger spectral values. In turn, for smaller spectral values, the geometry leaves such an imprint. For AdS backgrounds, the cosmological constant acts like a mass term which leads to a suppression. Conversely, the spectral function is enhanced for dS backgrounds because $\Lambda>0$ acts like a negative mass-squared term.

The off-shell effects due to the cosmological constant become even more pronounced if the ghost contributions are retained. The ghost remains on-shell at $k^2$ compared to the off-shell graviton at $m_h^2=k^2(1+\mu)$. We find that for AdS backgrounds (at $\mu= 3$), off-shell gravitons can directly scatter into the on-shell multi-ghost continuum and the flow of $f_h$ diverges, while it stays finite for dS backgrounds. As already mentioned before, in this off-shell computation, the flat Minkowski background bears similarities to an external electric or magnetic field in QED. External backgrounds or boundary conditions can introduce driving forces or friction that constantly feed or suppress scattering processes, which then destroy unitarity much like in open quantum systems. This analogy allows for a heuristic interpretation of the AdS singularity in the flow: there the off-shell background serves as a driving force for graviton scattering processes. We expect that full on-shell AdS flows with ghost contributions remain finite. Then, graviton and ghost are both on-shell massless, and it is the off-shell shift of mass scales that triggers the divergence. 

In summary, we consider the result in the presence of a non-vanishing cosmological constant as very promising. However, in our opinion, a fully conclusive analysis requires the expansion about a solution of the (quantum) Einstein equation and hence the expansion about a non-trivial background such as done in the Euclidean setting in \cite{Christiansen:2017bsy}.  

\subsection{Outlook}
We have reviewed the first direct approach to asymptotically safe gravity in spacetimes with Lorentzian signatures. This is based on the novel spectral renormalisation group which allows for non-perturbative computations on spacetimes with Lorentzian signature. This approach has been used to compute the graviton spectral function, which matches expectations from effective field theory at small momenta and from asymptotic safety at large momenta. The propagator has no ghost or tachyonic instabilities and thus there are no indications of unitarity violation, which is an encouraging first result. Next steps concern the computation of spectral representations of vertices and the direct computation of scattering amplitudes in asymptotically safe quantum gravity.

\section{Summary}
\label{sec:summary}
In this contribution, we have reviewed the state of the art of the fluctuation approach to quantum gravity. This approach is based on the path integral of metric quantum gravity, which we introduced in \Cref{sec:qft-approach}. The key element is to disentangle the correlation functions of the dynamical graviton fluctuation field $h_{\mu\nu}$ with that of the background field $\bar g_{\mu\nu}$ within a systematic vertex expansion.  For these correlation functions, we set up flow equations on spacetimes with Euclidean and Lorentzian signatures, and it is the running fluctuation correlation functions that build the core of the renormalisation group flow, see \Cref{sec:fRG-QG}. The distinction of fluctuation and background field is strictly necessary to fulfil the diffeomorphism symmetry identities, as discussed in \Cref{sec:symmetry-identities}. The technical foundations including the underlying concept for the convergence of the expansion scheme are presented in \Cref{sec:fluctuation-approach}.

We have highlighted some remarkable recent results in this approach, for example, the computation of momentum-dependent correlation functions at vanishing cutoff scales, see \Cref{sec:pure-grav}, relevant for the evaluation of S-matrix elements. In  \Cref{sec:ASSM}, we have discussed the results for the UV-IR phase structure of the asymptotically safe Standard Model, and in \Cref{sec:LorentzianGravity} we have reviewed the first direct Lorentzian computation leading to the graviton spectral function. These sections come with their own conclusions and in summary the fluctuation approach with its systematic vertex expansion is by now entering a stage with access to quantitative results for observables in asymptotically safe particle physics including spacetimes with Lorentzian signature.   

\smallskip
\noindent  {\bf Acknowledgements}\\
\noindent We thank G.~Assant, A.~Bonanno, B.~B\"urger, N.~Christiansen, T.~Denz, A.~Eichhorn, K.~Falls, J.~Fehre, H.~Gies, A.~Held, Y.~Kluth, B.~Knorr, P.~Labus, S.~Lippoldt, D.~Litim, J.~Meibohm, A.~Pastor-Gutiérrez, A.~Pereira, A.~Platania, A.~Rodigast, B.-J.~Schaefer, M.~Schiffer,  F.~Saueressig, J.~Smirnov, M.~Yamada, and C.~Wetterich for discussions and work on the subjects reported on. This work is funded by the Deutsche Forschungsgemeinschaft (DFG, German Research Foundation) under Germany’s Excellence Strategy EXC 2181/1 - 390900948 (the Heidelberg STRUCTURES Excellence Cluster) and the Collaborative Research Centre SFB 1225 - 273811115 (ISOQUANT),  and by the Science and Technology Research Council (STFC) under the Consolidated Grant ST/T00102X/1.

\bibliographystyle{apsrev4-1_custom}
\bibliography{GravityStatus}

\end{document}